%% file: main.tex
\documentclass[paper]{JFM-FLM_Au}
\pdfpagewidth=\paperwidth
\pdfpageheight=\paperheight

\usepackage{mathtools}
\usepackage{amssymb}

\usepackage{graphicx}
\usepackage{subcaption}
\usepackage{booktabs}
\usepackage{multirow}

\usepackage{siunitx}
\usepackage{placeins}

\usepackage{tikz}
\usepackage{pgfplots}
\usepackage{standalone}
\pgfplotsset{compat=1.18}

\usepackage{hyperref}
\usepackage{cleveref}

\crefname{figure}{Fig.}{Figs.}
\Crefname{figure}{Figure}{Figures}

\newcommand{\plotwidth}{0.9\textwidth}
\newcommand{\plotwidths}{1.2\textwidth}
\newcommand{\ceq}{\!=\!}
\hypersetup{
  pdftitle={A Priori Assessment of Tensor-Network Encoding for Isotropic Turbulent Flows},
  pdfauthor={Massen Esmaeili et al.},
  pdfkeywords={tensor networks, matrix product states, isotropic turbulence, passive scalar, reduced-order modeling}
}
\pgfplotsset{
  every axis/.append style={
    tick style={thick, line width=1pt},
    tick align=inside,
    tick pos=left,
    axis line style={line width=1pt}
  }
}

\lefttitle{Esmaeili et al.}
\righttitle{Tensor-network encoding of isotropic turbulence}

\title{\textit{A priori} Assessment of Tensor-Network Encoding for Isotropic Turbulent Flows}

\author{
Massen Esmaeili\aff{1},
Hirad Alipanah\aff{1},
Robert Pinkston\aff{1},
Peyman Givi\aff{1,2},
Daniel Livescu\aff{3},
Andrew J. Daley\aff{4},
Dieter Jaksch\aff{4,5},
\and Juan José Mendoza-Arenas\aff{1,6}
}

\affiliation{
\aff{1}Mechanical Engineering and Materials Science, University of Pittsburgh, Pittsburgh, PA, USA\\
\aff{2}Chemical and Petroleum Engineering, University of Pittsburgh, Pittsburgh, PA, USA\\
\aff{3}Los Alamos National Laboratory, Los Alamos, NM, USA\\
\aff{4}Clarendon Laboratory, University of Oxford, Oxford, UK\\
\aff{5}Institute for Quantum Physics, University of Hamburg, Hamburg, Germany\\
\aff{6}Physics and Astronomy, University of Pittsburgh, Pittsburgh, PA, USA
}

\corresau{Juan José Mendoza-Arenas, \email{jum151@pitt.edu}}  

\usepackage[authoryear]{natbib}
\begin{document}

\maketitle

\begin{abstract}
Tensor networks (TNs), originally developed for simulating many-body quantum systems, provide a systematic framework for approximating high-dimensional fields. This is achieved by factorizing the field into interconnected tensors with small bond dimensions, thereby restricting the strength of the correlations that can be captured across bipartitions of the field. Belonging to the family of TNs, the matrix product state (MPS) ansatz is utilized here as a reduced-order modeling framework to construct truncated representations of isotropic turbulent flow data. Two direct
numerical simulation (DNS) datasets are considered: the hydrodynamic field of an
incompressible three-dimensional flow, and a conserved Fickian scalar in a
similar flow. Each field is encoded as an MPS through a sequence of singular value decompositions (SVDs) in which small singular values are discarded. The
truncated representation is contracted back to the full grid, and the
resulting \emph{reconstructed field} is compared against DNS. An interleaved ordering
of the spatial tensor indices of the transport variables is applied prior to decomposition in order to
localize the dominant inter-tensor correlations. Velocity reconstructions achieve
$99.8\%$ fidelity using only $5\%$ of the original DNS memory, while the scalar
field reaches the same fidelity at $15\%$ memory usage. A wide range of lower- and higher-order
statistics, including velocity gradients, dissipation, and structure functions,
are systematically examined. At these compression levels, the total kinetic
energy and the scalar energy are both recovered within $0.2\%$ relative error,
while the mean dissipation and mean scalar dissipation remain within
approximately $10\%$ of the DNS generated values. These findings support the suitability of MPS for scalable reduced-order analysis of complex turbulent datasets and
motivate further exploration of TN-based methods in computational turbulence.
\end{abstract}

\section{Introduction}

Direct numerical simulation of turbulent flows is notoriously expensive \citep{moin2024fundamentals}. The need to resolve the full spectrum of turbulent transport, from the largest energy-containing scales to the smallest dissipative scales, requires extremely fine spatial grids and correspondingly small time steps. As the Reynolds number increases, the required number of grid points and the complexity of the flow increase dramatically \citep{pope2000turbulent}. This  leads to memory and processing requirements that quickly surpass the capabilities of even the largest supercomputers \citep{Livescu2020Modeling,GDMM20}. Moreover, accurately capturing transient phenomena and intermittent bursts of vorticity requires long simulation times and frequent data sampling, which increases the computational cost even further. There is a similar complexity in simulations of quantum many-body problems, as the Hilbert space describing lattice systems of many interacting particles also grows exponentially with the number of particles \citep{coleman2015introduction}. This fundamental challenge has led to the development of TN methods, which provide a powerful tool to efficiently represent and simulate interacting many-body systems by restricting the quantum state description to a small, physically relevant subspace of the full Hilbert space, particularly in low dimensions \citep{vidal2007entanglement, schollwock2011density, orus2014practical,cirac2021rmp}. This representation is typically sufficient for states that do not exhibit strong long-range correlations. This favorable representation has motivated research aimed at extending TN-based methods to problems beyond many-body physics. In  recent years, there has been an emerging interest in employing an important family of TNs, known as MPS, to simulate a variety of other problems such as machine learning \citep{huggins2019towards, sengupta2022tensor, liu2023tensor} and image processing \citep{zhang2015qsobel, jiang2019improved, jiang2015quantum, caraiman2014histogram, nakaji2021quantum}. Furthermore, MPS methods exhibit strong potential in reduced-order modeling for numerical simulations of partial differential equations (PDEs) \citep{khoromskij2011qtt, gelss2022solving, benner2021regularization, manzini2023tensor, truong2024tensor, Siegl_2026}. Recent applications are found in computational fluid dynamics (CFD) 
\citep{gourianov2022quantum, gourianov2025tensor, peddinti2024quantum,
danis2025tensor, adak2024tensor, pinkston2025matrix,
vanhulst2026quantuminspiredsimulation2dturbulent}, quantum fluids \citep{connor2025tensornetworkmethodsgrosspitaevskii,niedermeier2025solvinggrosspitaevskiiequationmultiple,boucomas2025quanticstensortrainsolving,chen2025solvinggrosspitaevskiiequationquantic,gomezlozada2025simulatingquantumturbulencematrix}, plasma physics \citep{ye2022quantum, ye2024quantized}, and phonon transport in crystals \citep{sangyeop2026peierlsboltzmann}.

Despite the increasing use of TN methods in scientific computing, their applicability to characterize fully developed turbulent flows remains largely unexplored. In particular, there is a lack of systematic {\it a priori} assessments quantifying how MPS-based compression impacts turbulence statistics across spatial scales. Existing reduced-order modeling approaches typically rely on modal decompositions or projection-based techniques \citep{Berkooz1993,Schmid2010,Rowley2017}, with limited attention to tensor factorizations and their ability to preserve multiscale correlations inherent in isotropic turbulence. A recent study by \cite{pisoni2026} applied MPS compression to a \(1024^3\) DNS dataset and examined selected quantities such as the energy spectrum, flatness, and probability density functions (PDFs) of velocity increments (i.e. velocity differences across a separation distance $r$). The present work extends this line of investigation by providing a broader and more systematic assessment of MPS encoding of DNS data of two turbulent flows, quantifying the compression-accuracy trade-off across a wide range of low- and high-order statistics. These include large-scale mean quantities (e.g. the total kinetic energy), PDFs of velocity components and their gradients, joint PDF of the invariants such as \(Q\)-\(R\), and higher-order structure functions. They probe both large-scale dynamics and fine-scale intermittency.

The subjects under consideration are three-dimensional incompressible homogeneous isotropic turbulence and the mixing of a passive scalar. The hydrodynamical fields offer a statistically uniform flow without a mean shear or boundary effects, allowing the isolation of the intrinsic multiscale dynamics of turbulence. Isotropic turbulence serves as a fundamental reference case in this setting, since sufficiently small scales are expected to recover local isotropy \citep{pope2000turbulent}. Assessing MPS encoding in this setting provides a baseline for understanding its applicability to more complex flows. Investigating the scalar field provides a more demanding test of the MPS encoding, as passive scalars typically develop sharp gradients and fine structures. Some scalar statistics also depend directly on spatial derivatives of the field. Since differentiation amplifies high-wavenumber components, these quantities are generally more sensitive to the loss of small scale information. Consequently, a close agreement in scalar statistics indicates that the TN representation preserves not only large-scale structures, but also the finer scales associated with turbulent mixing \citep{BRODKEY197549,JABERI1997}. Some of these statistics are sensitive to intermittent extreme events; only a high-fidelity surrogate can reproduce them.

The manuscript is organized as follows. Section~\ref{sec:mps_formalism} introduces the MPS formalism together with the compression and reconstruction procedures. Section~\ref{sec:hydro} presents the results for MPS-truncated velocity fields, including error metrics, Schmidt spectra, and turbulence statistics computed from velocity data \citep{isotropic1024coarse}. Section~\ref{sec:scalar} examines the corresponding truncation analysis of DNS-generated passive-scalar data \citep{Daniel_2018, LIVESCU_JABERI_MADNIA_2000}. Section~\ref{sec:conclusions} summarizes the main findings and suggests some directions for future work.

\section{MPS Formalism}\label{sec:mps_formalism}

\begin{figure}[t!]
    \centering    \includegraphics[width=0.9\textwidth]{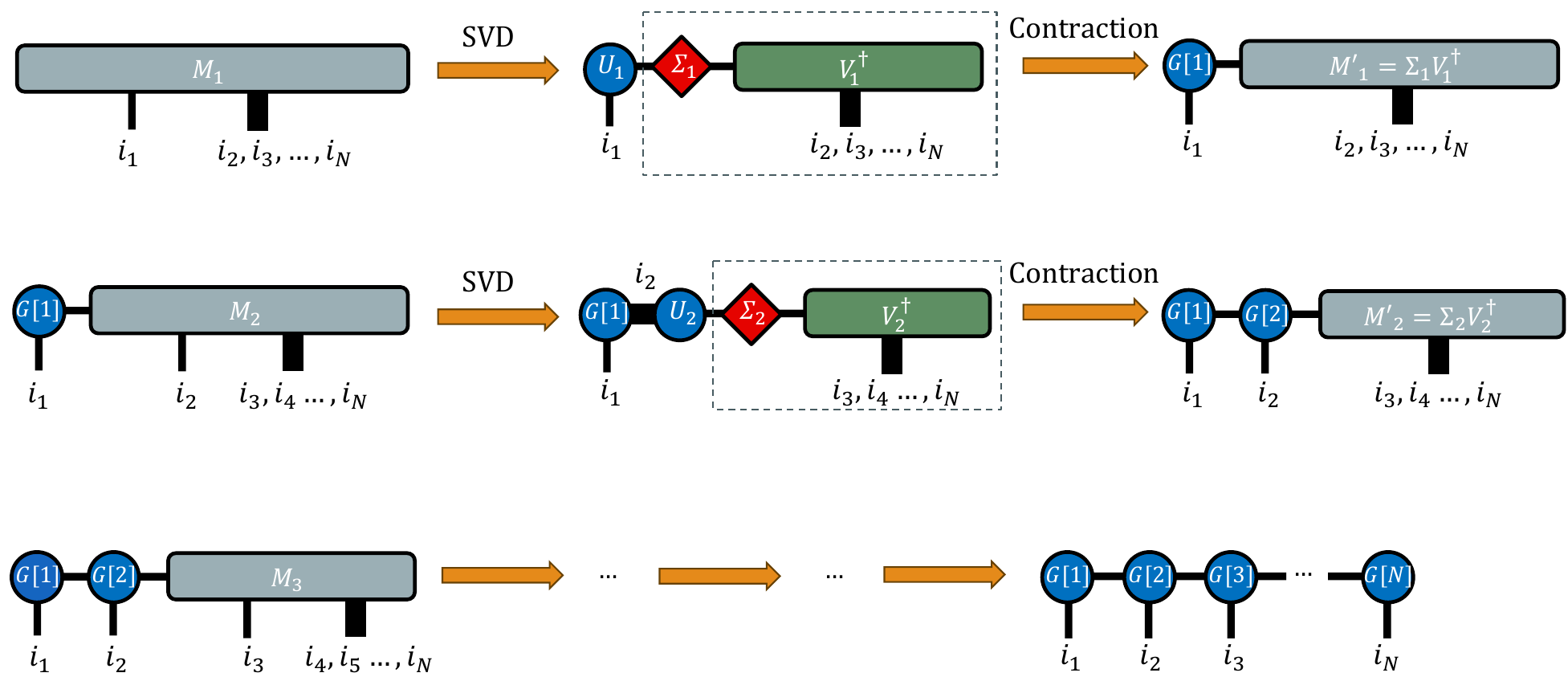}
    \caption{Schematic of the iterative MPS construction consisting of repeated cycles of SVD, contraction, and reshaping.}
    \label{fig:mps_construction}
\end{figure}

The MPS is a one-dimensional TN. Consider an unstructured tensor \(T\in \mathbb{R}^{d \times d \times \cdots \times d}\) of order \(N\). This order also corresponds to the number of tensors in the resulting MPS representation (see~\cref{fig:mps_construction}). The total number of elements, $n$, in the tensor grows exponentially with the order: \(n = d^{N}\). An element of tensor \(T\) is indexed by the multi-index \(i_1 i_2 \cdots i_N\), where each physical index satisfies \(i_k \in \{0,1\}\) for the qubit-based encoding employed in this work and encodes one scale along a spatial direction. Thus, the local physical dimension is uniform and fixed to \(d = 2\).

In the MPS representation, the tensor \(T\) is factorized into a chain of site tensors connected by auxiliary indices, commonly referred to as \emph{bond} indices. These bond indices characterize the inter-tensor correlations that emerge between different parts of the MPS. This structured decomposition, when combined with the truncation of small singular values (described later), enables a compact representation of the original tensor. Specifically, the MPS factorizes \(T\) into tensors \(\{G[k]\}_{k=1}^N\) connected by bond indices \(\{\alpha_k\}_{k=1}^{N-1}\) as

\begin{equation} \label{mps_eq}
    T_{i_1 i_2 \cdots i_N} \approx \quad
    \mathclap{\sum_{\alpha_1,\ldots,\alpha_{N-1}}}
    \quad \:
    G[1]^{\,i_1}_{\alpha_1}
    \,G[2]^{\,i_2}_{\alpha_1\alpha_2}
    \cdots
    G[N-1]^{\,i_{N-1}}_{\alpha_{N-2}\alpha_{N-1}}
    \,G[N]^{\,i_N}_{\alpha_{N-1}} .
\end{equation}

\noindent Here, \(\alpha_k \in \{1,\ldots,\chi_k\}\) is the bond index connecting sites \(k\) and \(k+1\), and \(\chi_k\) is referred to as the bond dimension. Consequently, the tensor shapes are
\(G[1]\!\in\!\mathbb{R}^{d\times \chi_1}\),
\(G[k]\!\in\!\mathbb{R}^{\chi_{k-1}\times d \times \chi_k}\) for \(2\!\le\!k\!\le\!N\!-\!1\),
and \(G[N]\!\in\!\mathbb{R}^{\chi_{N-1}\times d}\).
The bond dimensions \(\chi_k\) control how much inter-tensor correlation can be captured at each bond. Their precise meaning will become clear once the MPS construction process is described. 

The MPS is constructed through a sequence of SVDs. Before the first SVD, the tensor \(T\) is reshaped into a matrix by separating the first physical index from the remaining ones, which yields a matrix.
\begin{equation}
T\in \mathbb{R}^{\overbrace{d \times d \times \cdots \times d}^{N}}{} \xrightarrow{\text{reshape}} M_1 \in \mathbb{R}^{d \times d^{N-1}},
\end{equation}
where the first physical index forms the row dimension and the remaining \(N-1\) physical indices are grouped into the column dimension. An SVD of \(M_1\) is then performed:
\begin{equation}
M_1 = U_1 \Sigma_1 V_1^\dagger,
\end{equation}
where the dagger ($\dagger$) denotes the conjugate transpose (Hermitian adjoint). The matrices \(U_1 \in \mathbb{R}^{d \times d}\) and \(V_1 \in \mathbb{R}^{d^{N-1} \times d}\) contain the left and right singular vectors and satisfy the orthonormality conditions \(U_1^\dagger U_1 = I\) and \(V_1^\dagger V_1 = I\)\footnote{A matrix whose columns satisfy such an orthonormality condition is said to be left-orthonormal; analogously, a matrix whose rows are orthonormal is termed right-orthonormal.}. The diagonal matrix \(\Sigma_1\) contains the singular values in descending order. The matrix \(U_1\) is the first core tensor \(G[1]\), and the product \(M'_1 = \Sigma_1 V_1^\dagger\) is carried forward to the next step.

The procedure is repeated along the chain. At step \(k\), the tensor passed forward from the previous step, \(M'_{k-1} = \Sigma_{k-1} V_{k-1}^\dagger\), is reshaped by absorbing one additional physical index into the row dimension:
\begin{equation}
M'_{k-1} \in \mathbb{R}^{\chi_{k-1} \times d^{N-(k-1)}} \xrightarrow{\text{reshape}} M_k \in \mathbb{R}^{\chi_{k-1}d \times d^{N-k}}.
\end{equation}

\noindent An SVD \(M_k = U_k \Sigma_k V_k^\dagger\) is then performed, \(U_k\) is reshaped into the core tensor \(G[k]\), and the product \(M'_k = \Sigma_k V_k^\dagger\) defines the tensor passed to the next step. The bond dimension \(\chi_k\) corresponds to the number of singular values retained from the SVD at bond $k$. If all singular values produced by every SVD are retained, the factorization is exact and the approximation in Eq.~\eqref{mps_eq} becomes equality. However, for tensors with a low-rank structure, an exact representation can be achieved with fewer retained singular values. Retaining fewer singular values than required for an exact factorization yields a truncated MPS that approximates the original tensor \(T\). The procedure ends once the last physical index is isolated, producing the full MPS representation depicted in \cref{fig:mps_construction}.

The truncation of singular values is controlled by a user-defined cutoff point which sets a threshold for the retained singular values. This cutoff is measured by \(10^{-C}\), with $C$ a positive integer specifying the order of magnitude of the threshold. At bond \(k\), \(r_k\) denotes the total number of singular values resulting from the SVD. The minimum number \(\chi_k\) of singular values is retained so that
\begin{equation} \label{cutoff_eq}
 \frac{\sum_{j=\chi_k+1}^{r_k} \sigma_{j,k}^2}
      {\sum_{j=1}^{r_k} \sigma_{j,k}^2}
 < 10^{-C},
\end{equation}
thereby bounding the truncation error by the cutoff. The approximation error satisfies the following:
\begin{equation}
    \| M_k - M_k^{(\chi_k)} \|_F^2
    =
    \sum_{j=\chi_k+1}^{r_k} \sigma_{j,k}^2,
\end{equation}
that is, the truncation error measured in the squared Frobenius norm equals the sum of the squared discarded singular values. This truncation is optimal in the \(L^2\) (Frobenius) sense \citep{Eckart_Young_1936}, no rank-\(\chi_k\) approximation of \(M_k\) achieves a smaller Frobenius norm error \citep{schollwock2005density, gourianov2022quantum}. A lower cutoff will retain more singular values and reduce the truncation error, at the expense of increased computational cost and memory usage, whereas larger cutoffs yield better compression but may remove fine-scale features. The cutoff selects the minimum \(\chi_k\) needed for each bond independently, so the dimension of the bond can vary throughout the chain. It will be larger at bonds where the inter-tensor correlations are stronger. In this sense the construction is \emph{adaptive}: the cutoff is the same everywhere, but the resulting bond dimensions adapt to the local correlations. Denoting the maximum bond dimension by \(\chi_{\max}\), the storage required for the MPS representation of the tensor is reduced from \(\mathcal{O}(d^N)\) to \(\mathcal{O}(N d \chi_{\max}^2)\) \citep{schollwock2011density,orus2014practical}.

\begin{figure}[b!]
    \centering
    \includegraphics[width=\textwidth]{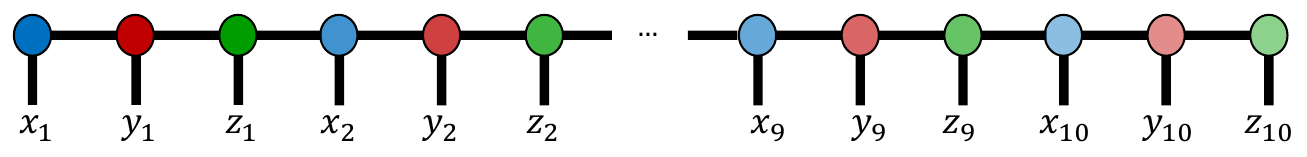}
    \caption{Schematic of the interleaved tensor ordering used for MPS encoding.}
    \label{fig:MPS_plots}
\end{figure}

The same SVD-based construction admits a complementary interpretation through the \emph{Schmidt decomposition}, a tool from quantum information. It expresses a state in terms of its correlations across a bipartition of the system. The decomposition is obtained by performing a single SVD of the tensor \(T\) reshaped in the chosen bipartition. The resulting singular values quantify the degree to which the two sides of the bipartition are correlated. To make this connection precise, the notion of mixed-canonical form is introduced: an MPS is in mixed-canonical form at bond \(k\) when all tensors to the left of site \(k\) are left-orthonormal and all those to the right are right-orthonormal. Bringing the MPS into this form at the bond \(k\) corresponds to performing an SVD of the reshaped tensor across the bipartition 
\((i_1,\ldots,i_k)\,|\, (i_{k+1},\ldots,i_N)\). That is, reshaping the full tensor \(T\) into the same matrix \(M_k \in \mathbb{R}^{d^k \times d^{N-k}}\) introduced in the MPS construction above, and performing the SVD \(M_k = U_k \Sigma_k V_k^\dagger\), yields the Schmidt decomposition \citep{Wilde_2016}. For a concise and compact presentation, Dirac's bra-ket notation is adopted: a ket \(|v\rangle\) denotes a (column) vector and the bra \(\langle v|\) its conjugate transpose, so that \(\langle u|v\rangle\) is the inner product. With this notation, the tensor \(T\) is regarded as the vector \(|T\rangle\) whose elements are \(T_{i_1 i_2 \cdots i_N}\), and the Schmidt decomposition reads
\begin{equation}
|T\rangle
=
\sum_{\alpha_k=1}^{r_k}
\sigma_{\alpha_k}
\,
|L_{\alpha_k}\rangle
\,
|R_{\alpha_k}\rangle,
\label{eq:schmidt_decomposition}
\end{equation}
where \(\{\sigma_{\alpha_k}\}\) are the singular values of \(M_k\), and 
\(|L_{\alpha_k}\rangle\), \(|R_{\alpha_k}\rangle\) are the left and right Schmidt vectors formed from the columns of \(U_k\) and \(V_k\), respectively. The Schmidt vectors satisfy the orthonormality relations:
\begin{equation}
\langle L_\alpha | L_\beta \rangle = \delta_{\alpha\beta},
\qquad
\langle R_\alpha | R_\beta \rangle = \delta_{\alpha\beta},
\label{eq:schmidt_orthonormality}
\end{equation}

\noindent where \(\delta_{\alpha\beta}\) is the Kronecker delta. The set \(\{\sigma_{\alpha_k}\}\) constitutes the \emph{Schmidt spectrum} at the bond \(k\). The rate of decay of these singular values characterizes the strength of correlations across the bipartition: rapid decay indicates weak coupling between the left and right subsystems, whereas slow decay signals stronger correlations, requiring more singular values to be retained to capture them. The MPS formalism is applied to represent the DNS generated fields. Therefore, the total number \(n\) of elements corresponds to the total number of grid points within the computational domain: \(n = n_x n_y n_z\) with \(n_x\), \(n_y\), and \(n_z\) denoting the number of grid points in the directions \(x\)-, \(y\)- and \(z\), respectively. To introduce the MPS representation, the velocity or the scalar field is first reshaped into an unstructured tensor of order \(N = \log_d(n)\) under the qubit encoding ($d\,$ = $\,2$). In this construction, each physical index \(i_k\) corresponds to a length scale along a given spatial direction \citep{gourianov2022quantum}. The MPS decomposition is obtained by the sequence of SVDs shown in \cref{fig:mps_construction}. This process reorganizes the spatially discretized turbulent field into a chain structure, allowing compression while preserving correlations across spatial scales. The MPS construction corresponds to a particular ordering of the physical indices along the chain. For each spatial direction, the binary indices are ordered from the coarsest to the finest spatial scale. Thus, \(x_1\) represents the largest scale in the \(x\)-direction, whereas \(x_{N_x}\) represents the smallest scale, where \(N_x=\log_2(n_x)\). The indices in the \(y\)- and \(z\)-directions are defined analogously. These indices may be freely permuted before the SVD sequence is performed, and any such permutation defines a different ordering scheme with potentially different truncation behavior. The number of possible permutations is obviously very large; here only three of the forms are considered: split, interleaved, and staircase. These span representative ways of grouping length scales across spatial directions. In the \emph{split} ordering, all indices of one spatial direction are placed consecutively, \(x_1, x_2, \ldots, x_{N_x}, y_1, y_2, \ldots, y_{N_y}, z_1, z_2, \ldots, z_{N_z}\). The \emph{staircase} ordering is a variant in which the \(y\) indices are reversed. In \emph{interleaved} ordering, shown in \cref{fig:MPS_plots}, the indices corresponding to the same length scale in different directions are placed adjacent to each other, i.e., \(x_1, y_1, z_1, x_2, y_2, z_2, \ldots\).
Based on several preliminary tests, the  interleaved format yielded the  best overall performance; so it was used in all subsequent analyses. For isotropic turbulence, the dominant correlations exist between length scales with comparable size across the three spatial directions. A length scale of size \(\ell\) along \(x\) is statistically related to length scales of size \(\ell\) along \(y\) and \(z\). Interleaved ordering places such same scale indices adjacent in the MPS chain. The dominant correlations are then captured by short-range bond indices and incur a small truncation error. In the split scheme, all the \(x\) indices precede all \(y\) and \(z\) indices, so the same correlations must instead be propagated over long stretches of the chain. This requires larger bond dimensions to retain them and therefore increases the truncation error at a fixed cutoff. The optimal ordering is thus expected to depend on the correlation structure of the underlying flow. A different scheme may be preferred for anisotropic flows. All MPS construction and truncation processes are implemented via the ITensor library \citep{itensor, itensor-r0.3}.

\subsection{Implementation}

\begin{figure}[t!]
\vspace{0.5em}
    \centering
    \includegraphics[width=0.5\textwidth]{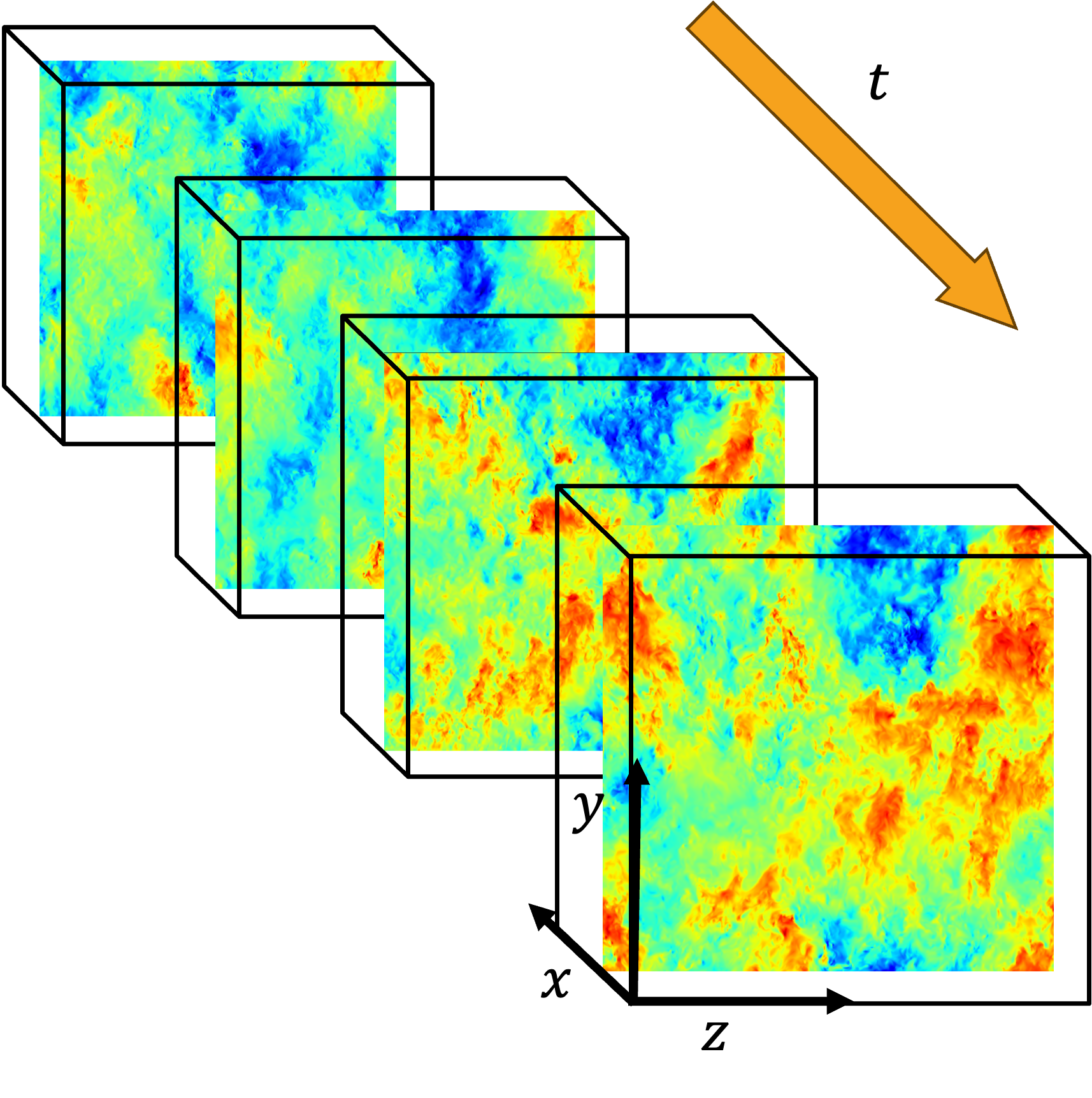}
    \caption{Contours of the \(u\)-velocity component from the isotropic turbulence dataset \citep{isotropic1024coarse}.}
    \label{fig:3d_field}
\end{figure}

The MPS is applied to DNS-generated three-dimensional isotropic turbulent flow  datasets at uniformly spaced grid points \(1024^3\). Two datasets are considered. The first is from the Johns Hopkins turbulence databases (JHTDB) \citep{Li_2008, isotropic1024coarse, Perlman_2007}, and corresponds to an incompressible Newtonian flow with Taylor microscale Reynolds number of \(Re_\lambda \approx 433\) (\cref{fig:3d_field}). The flow is statistically stationary, maintaining approximately constant turbulent kinetic energy over several large-eddy turnover times. The second pertains to a passive Fickian scalar (denoted by \(\phi\)) \citep{Daniel_2018, LIVESCU_JABERI_MADNIA_2000} transported by a statistically stationary turbulent velocity field with \(Re_\lambda \approx 180\) and molecular Schmidt number \(Sc\ceq1\).\footnote{Here the Schmidt number refers to the fluid-dynamical quantity defined as the ratio of momentum diffusivity to mass diffusivity, \(Sc = \nu/D\), where \(\nu\) is the kinematic viscosity and \(D\) is the scalar diffusivity. This parameter is not related to the Schmidt decomposition used in the TN formulation.} The velocity forcing method for the second dataset prescribes the Kolmogorov length scale \(\eta\) at the start of the simulation and maintains it close to the target value, satisfying \(\eta \kappa_{\max}=1.5\). Here, \(\kappa_{\max} = \sqrt{2}\,n_x/3\) denotes the maximum resolved wavenumber on the cubic grid \(n_x = n_y = n_z\). The scalar field has a zero mean value with  no external scalar forcing and is initialized after the turbulence reaches a stationary state. Statistics are taken later when the scalar field develops an approximately Gaussian distribution with variance \(6\times 10^{-4}\). The JHTDB data are available in single precision, and the passive scalar dataset in double precision. Multiple snapshots from both flows are examined.

To quantitatively assess the quality of the MPS approximation, both compression and accuracy metrics are required. The compression ratio (CR) measures the reduction in storage relative to the DNS representation and is defined as
\begin{equation}
\text{CR} =
\frac{\displaystyle 2^{N}}
{\displaystyle \sum_{k=1}^{N} 2\, \chi_{k-1}\, \chi_k},
\end{equation}
where the numerator is the total number of elements in DNS, and \(\chi_k\) is the bond dimension between sites \(k-1\) and \(k\) (with the convention \(\chi_0=\chi_N=1\)). The higher CR values correspond to better compressions \citep{Holtz2012}. The maximum bond dimension, \(\chi_{\max}=\max_k (\chi_k)\), characterizes the largest number of singular values retained across all bonds. As the cutoff point decreases, \(\chi_{\max}\) increases, reflecting the additional correlations the MPS must retain to represent the data with higher precision. The compression considering all components of the velocity field is quantified using an average compression ratio defined as
\(\mathrm{CR}_{\mathrm{Avg}} = \tfrac{1}{3}(\mathrm{CR}_u + \mathrm{CR}_v + \mathrm{CR}_w)\), where the subscripts \(u\), \(v\), and \(w\) denote the velocity components in the \(x\)-, \(y\)-, and \(z\)-directions, respectively. From this quantity, an equivalent DNS resolution is estimated as:
\begin{equation}
\text{Equivalent DNS} = \frac{n}{\text{CR}_{\text{Avg}}}.
\end{equation} 
This quantity gives a cubic grid resolution that would produce a storage comparable to that of the compressed field. The precision is quantified using the fidelity \citep{Wilde_2016} between the reconstructed field and the DNS field. For example, for the  $u$-component of the velocity:
\begin{equation}
\mathcal{F}(u_{\text{MPS}}, u_{\text{DNS}}) =
\frac{|\langle u_{\text{MPS}}, u_{\text{DNS}} \rangle|^2}
{\langle u_{\text{MPS}}, u_{\text{MPS}} \rangle \, \langle u_{\text{DNS}}, u_{\text{DNS}} \rangle},
\qquad
\mathcal{I} = 1 - \mathcal{F},
\end{equation}
where \(u_{\text{DNS}}\) is the \(u\)-component of the velocity field obtained from DNS, and \(u_{\text{MPS}}\) is the corresponding component of the compressed velocity field. Although the MPS representation stores the field in a smaller number of parameters, contracting the MPS back to the original grid yields a reconstructed field of size \(1024^3\) on which the inner product is evaluated. Here, \(\langle \cdot , \cdot \rangle\) denotes the Euclidean inner product over all grid points. Infidelity \(\mathcal{I}\) therefore provides a normalized measure of the error between the MPS approximation and DNS.

\section{MPS representation of the hydrodynamic field}\label{sec:hydro}

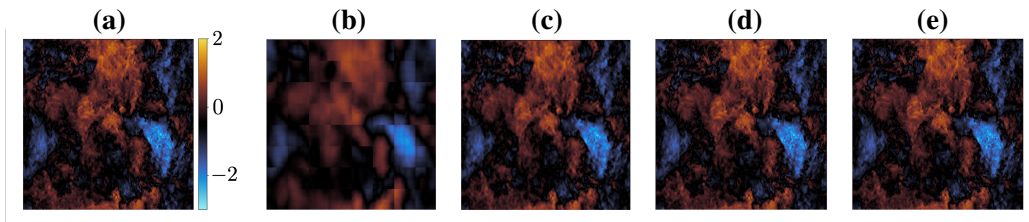
\begin{figure}[b!]
\centering
\begin{tikzpicture}
  \node[anchor=south west, inner sep=0] (img) 
    {\includegraphics[width=1.02\textwidth]{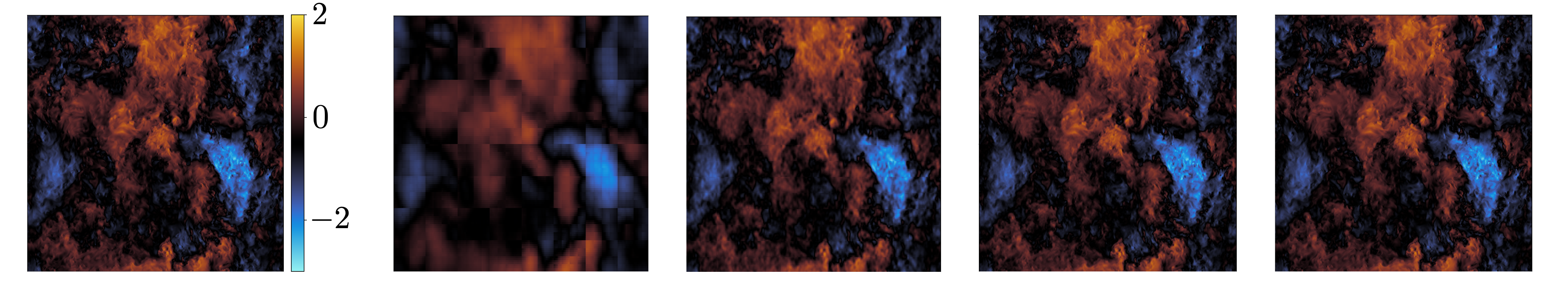}};
  \begin{scope}[x={(img.south east)}, y={(img.north west)}]
    \node at (0.10,1.04) {\textbf{(a)}};
    \node at (0.33,1.04) {\textbf{(b)}};
    \node at (0.52,1.04) {\textbf{(c)}};
    \node at (0.71,1.04) {\textbf{(d)}};
    \node at (0.89,1.04) {\textbf{(e)}};
  \end{scope}
\end{tikzpicture}
\caption{Contour plot of \(u\)-velocity in the \(xy\)-plane at \(z=\pi\). Left to right: the original DNS field (a) and MPS reconstructions for cutoffs \(10^{-C}\) with $C\ceq2$ to $5$ (b--e).}
\label{fig:velocity_slice_dns}
\end{figure}

\begin{figure}[t!]
    \centering
    \includegraphics[width=0.8\textwidth]
    {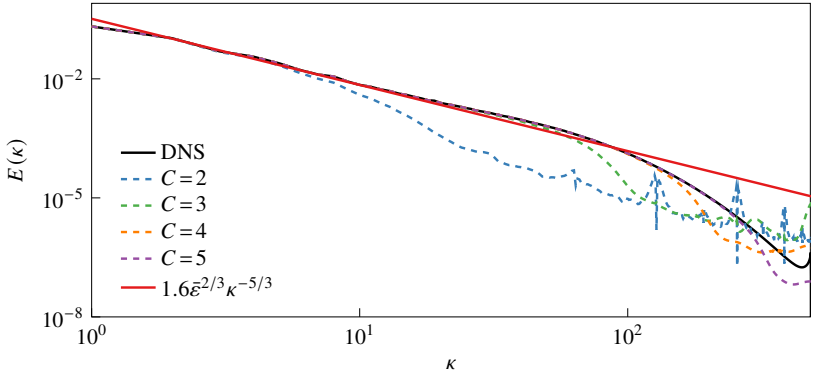}
    \caption{Energy spectrum comparison between DNS and MPS-reconstructed data for $C\ceq2$ to $5$.
    The red line shows the Kolmogorov inertial-range scaling.}
    \label{fig:energy_spectrum}
\end{figure}

Compression is implemented in the three components of the velocity field \(\mathbf{U}=(u,v,w)\), using four cutoff values \(10^{-C}\) with \(C\in\{2,3,4,5\}\). For each cutoff, the truncated MPS representations of \(u\), \(v\), and \(w\) are contracted back to the full grid to obtain the reconstructed velocity field. A truncation of \(10^{-C}\) removes singular values below the threshold indicated by Eq.~\eqref{cutoff_eq}, thus filtering those singular values with small magnitudes. The behavior of the truncation across scales is examined in detail via the Schmidt spectra. Large-scale energetic structures are typically associated with large singular values and are expected to be well preserved by the truncation. fine-scale features and gradients are expected to be more sensitive to truncation, as will be quantified by the statistics. Consequently, low-order moments should remain accurate even at moderate truncation levels, whereas quantities depending on spatial derivatives, such as dissipation or velocity-gradients, are expected to exhibit larger deviations and require smaller truncation thresholds for accurate reconstruction. The reconstructed fields are compared with the original DNS data, both instantaneously and statistically. For flow visualization, \cref{fig:velocity_slice_dns} shows a slice of the $u$-velocity component at $z=L/2=\pi$.
Because the turbulence field is statistically isotropic, the  statistics are equivalent in all spatial directions; therefore, some of the results are presented only in a single direction (\(x\)). The reconstruction for $C\ceq2$ appears noticeably blurred and exhibits horizontal and vertical blocking artifacts, reflecting the loss of small scale structures. At $C\ceq3$, more detail is retained, but the field remains blurred compared to DNS. For $C\ceq4$ and $C\ceq5$, the visual differences relative to DNS become minimal. A more quantitative comparison is provided by the energy spectrum \(E(\kappa)\) in \cref{fig:energy_spectrum}, where \(\kappa\) here denotes the wavenumber. The solid red line indicates the Kolmogorov inertial-range scaling \(E(\kappa) = C_K\,\bar{\varepsilon}^{2/3}\,\kappa^{-5/3}\), with \(C_K \approx 1.6\), and \(\bar{\varepsilon}\) is the mean dissipation rate. At low wavenumbers, all reconstructed spectra agree closely with DNS, indicating an accurate approximation of the dominant energy-containing scales. At higher wavenumbers, the cutoff level controls how far into the inertial range the agreement extends. For \(C\ceq3\), \(4\), and \(5\), the reconstructed spectra follow the \(\kappa^{-5/3}\) scaling throughout the inertial range before a premature dissipative roll-off. This roll-off sets in earlier for smaller \(C\); for \(C\ceq5\), the deviation begins around \(\kappa\approx 300\). The \(C\ceq2\) case departs from DNS well before the end of the inertial range, consistent with the much coarser truncation. In each case, the deviation begins where the energy approaches the cutoff value \(10^{-C}\). The energy $E(\kappa)$ below this threshold is not faithfully preserved by the truncation, leading to the differences at high wavenumbers observed in \cref{fig:energy_spectrum}.

\begin{figure}[b!]
    % \centering
    \hspace{-1em}
    \includegraphics[width=1.05\textwidth]{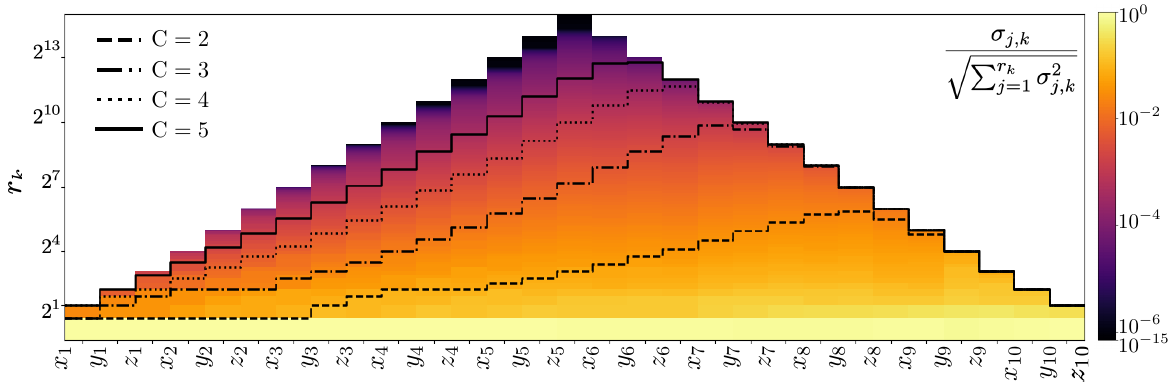}
    \caption{Schmidt spectrum for the \(u\) velocity using the interleaved ordering scheme. For each truncation level $C$, the maximum height of the corresponding contour line (i.e.\ the largest value of $\chi_k$ across all bonds) gives the maximum retained bond dimension $\chi_{\max}$ reported in \Cref{tab:velocity_compression}.}
    \label{fig:Schmidt spectrum}
\end{figure}

The distribution of correlations across length scales in the MPS representation is characterized by the Schmidt spectrum at each bond \citep{Vidal2004,gourianov2022quantum}. \Cref{fig:Schmidt spectrum} shows the Schmidt spectra for the $u$-velocity component using the interleaved ordering scheme.
For reference, the MPS encoding of DNS data without truncation retains all singular values produced by each SVD; in that case, the entire triangular envelope visible in \cref{fig:Schmidt spectrum} is completely populated. The black lines corresponding to different cutoff levels indicate the number of singular values retained at each bond. The characteristic pyramidal structure of the spectra follows directly from the successive reshaping steps in the MPS construction. At bond $k$, the reshaped matrix $M_k$ has dimensions $d^k\times d^{N-k}$, so the maximum possible rank is $\min(d^k,d^{N-k})$. Consequently, the maximum number of singular values grows as $d^k$ for $k$ below the center of the chain and decreases as $d^{N-k}$ beyond it. The peak depends on the parity of $N$: when $N$ is even, the maximum bond size is attained at the single bond $k=N/2$, with value $d^{N/2}$. When $N$ is odd, it is attained at the two adjacent bonds $k=(N-1)/2$ and $k=(N+1)/2$, both with value $d^{(N-1)/2}$. This algebraic constraint accounts for the pyramidal envelope of the  spectra. The truncation profiles in \cref{fig:Schmidt spectrum} exhibit a pronounced asymmetry across the MPS bonds. With interleaved ordering as employed here, the bonds near the left end of the chain correspond to the largest length scales, while the bonds toward the right end correspond to progressively finer spatial scales. This asymmetry in the truncated contours can be understood through the energy cascade of turbulence \citep{gourianov2022quantum}. The energy injected at the largest scales is transferred to progressively smaller scales, generating an increasingly wide range of fluctuations \citep{pope2000turbulent}. The large-scale motions are dominated by a small number of energetic structures with strong spatially localized correlations. Such structures admit a low-rank representation, so few retained singular values suffice to capture them. The fine scales, in contrast, are populated by a much larger number of weaker structures whose correlations extend widely across the chain. Reproduction of them therefore requires a larger number of retained singular values on the corresponding bonds, which produces the observed skew in \cref{fig:Schmidt spectrum}. 

The reduction in storage achieved by MPS relative to DNS is quantified by the compression ratios (CR) in \Cref{tab:velocity_compression}. As the truncation level is reduced, the CR decreases, reflecting the growth in the bond dimensions and storage.
\begin{table}
\centering
\begin{tabular*}{\textwidth}{
@{\extracolsep{\fill}}
l
@{\hspace{2em}}
c c c
@{\hspace{4em}}
c c c
@{\hspace{4em}}
c c c
}
\toprule
& \multicolumn{3}{c}{\hspace{-5em}\textbf{$\boldsymbol{u}$}}
& \multicolumn{3}{c}{\hspace{-5em}\textbf{$\boldsymbol{v}$}}
& \multicolumn{3}{c}{\hspace{-2em}\textbf{$\boldsymbol{w}$}} \\[0.5em]
\cmidrule(lr){2-4}\cmidrule(lr){5-7}\cmidrule(lr){8-10}
$\boldsymbol{C}$ 
& $\boldsymbol{\chi_{\max}}$ & \textbf{CR} & $\boldsymbol{\mathcal{I}}$
& $\boldsymbol{\chi_{\max}}$ & \textbf{CR} & $\boldsymbol{\mathcal{I}}$
& $\boldsymbol{\chi_{\max}}$ & \textbf{CR} & $\boldsymbol{\mathcal{I}}$ \\
\midrule
2 & 58   & 41721 & 0.1492 & 32   & 113575 & 0.1426 & 73   & 25817 & 0.1594 \\
3 & 945  & 223   & 0.0179 & 709  & 383    & 0.0192 & 1061 & 185   & 0.0174 \\
4 & 3304 & 20    & 0.0018 & 2952 & 27     & 0.0019 & 3385 & 18    & 0.0018 \\
5 & 7123 & 4     & 0.0002 & 6755 & 5     & 0.0002 & 7284 & 4     & 0.0002 \\
\bottomrule
\end{tabular*}
\caption{The maximum bond dimension ($\chi_{\max}$), compression ratio (CR) and infidelity ($\mathcal{I}$) for the three velocity components for different truncation levels $C\ceq2$ to $5$.}
\label{tab:velocity_compression}
\end{table}
\Cref{tab:velocity_compression} also reports the maximum bond dimension \(\chi_{\max}\) for each case, which increases systematically as the cutoff is decreased. The precision of the reconstructed field is assessed by infidelity $\mathcal{I}$. For the velocity component \(u\), the infidelity at \(C\ceq2\) is approximately \(0.15\). At this level, the approximation omits a substantial portion of the kinetic energy and consequently does not reliably reproduce higher-order statistics or the tails of the velocity distributions. As the cutoff is reduced by an order of magnitude (i.e.\ \(C\) increases by one), the infidelity decreases by roughly the same order, demonstrating progressively improved agreement with DNS. The same monotonic behavior is observed for the $v$ and $w$ components. The impact of compression  is summarized in Table~\ref{tab:cr_memory_dns}. The equivalent DNS resolutions illustrate the effective reduction in storage achieved by the MPS representation while preserving the large-scale flow structures. For example, at \(C\ceq4\), the average CR corresponds to approximately \(367^3\) equivalent grid points---a substantial reduction from the original \(1024^3\) DNS---while the reconstruction maintains infidelities on the order of \(10^{-3}\) (\Cref{tab:velocity_compression}).

\begin{table}
\centering
\begin{tabular}{l c c}
\toprule
\textbf{C} & $\textbf{CR}_{\text{Avg}}$ & \textbf{Equivalent DNS} \\
\midrule
2 & 60371 & $\sim 26^3$  \\
3 & 264   & $\sim 159^3$ \\
4 & 22    & $\sim 367^3$ \\
5 & 4     & $\sim 630^3$ \\
\bottomrule
\end{tabular}
\caption{Estimated DNS-equivalent resolution based on the average compression ratios ($\text{CRs}_{\text{Avg}}$) for $C\ceq2$ to $5$.}
\label{tab:cr_memory_dns}
\end{table}

\begin{figure}[t!]
    \centering

    \noindent\hspace*{-8em}
    \begin{minipage}{\plotwidth}
      \centering
      \raggedright(a)\\
      \includegraphics[width=\plotwidths]{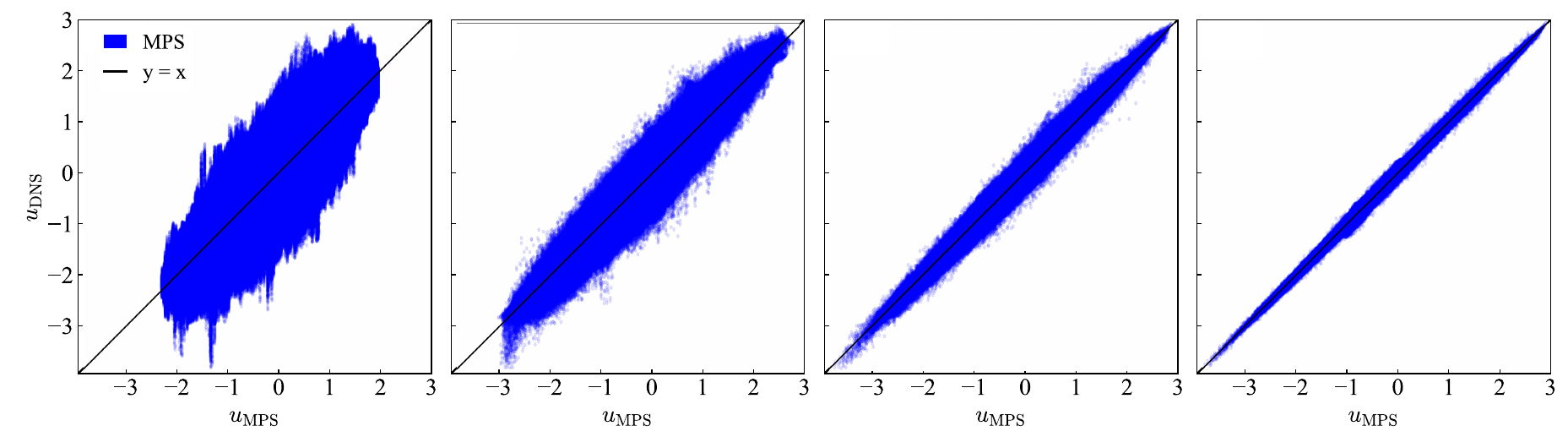}
    \end{minipage}
    \noindent\hspace*{-8em}
    \begin{minipage}{\plotwidth}
        \raggedright
        (b)\\
        \includegraphics[width=\plotwidths]{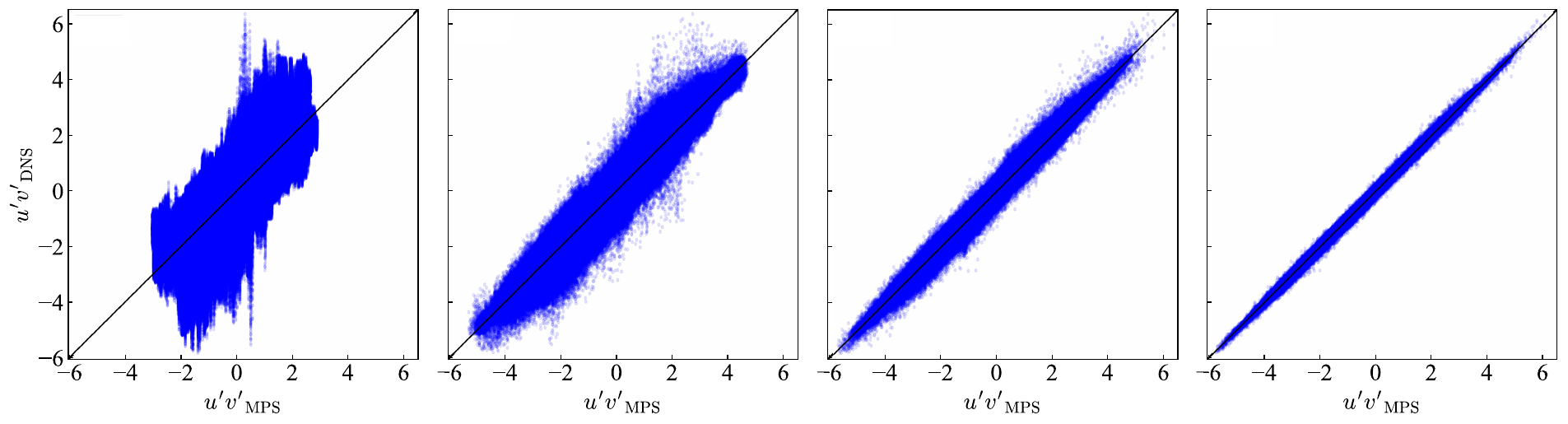}
    \end{minipage}

    \caption{Scatter plots of (a) velocity $u$ and (b) the pointwise Reynolds-stress contribution $u'v'$ for $C\ceq2$ to $5$ (left to right).}
    \label{fig:scatter_plots_grouped}
\end{figure}

Infidelity \(\mathcal{I}\) provides an aggregate measure of error between two fields. To complement it, pointwise comparisons between the MPS-reconstructed and DNS fields are conducted, to examine  whether the errors are concentrated or spread uniformly.
These comparisons are made in Fig.~\ref{fig:scatter_plots_grouped}, which shows scatter plots for the velocity component \(u\) (\cref{fig:scatter_plots_grouped}(a)) and the Reynolds stress component \(u'v'\) (\cref{fig:scatter_plots_grouped}(b)) for \(C\ceq2\) to \(5\). The \(45^{\circ}\) line indicates a perfect agreement. The velocity fluctuations are defined by the Reynolds decomposition, \(u=\bar{u}+u'\), where the overbar denotes the Reynolds-averaged values \citep{pope2000turbulent}. For homogeneous flow, the Reynolds averages are evaluated by a spatial average over the entire domain.
The plots shown in \cref{fig:scatter_plots_grouped} provide a visual evaluation of the way MPS preserves both the magnitude and spatial correspondence of the physical quantities. For $C\ceq2$, the reconstructed values for velocity and Reynolds stress show poor agreement with DNS. As \(C\) increases from \(C\ceq3\) to \(5\), the scatter decreases. In particular, \(C\ceq4\) and \(C\ceq5\) produce dense clustering, demonstrating high pointwise agreement with DNS.

For further assessments, several quantities defined in Appendix~\ref{app:turbulence_definitions} are considered. These quantities are of significant importance in Reynolds averaged turbulence modeling \citep{pope2000turbulent}. These include total kinetic energy, quantities related to dissipation, and length and time scales derived from the velocity field and its gradients. The relative errors for these quantities are provided in Table~\ref{tab:scalar_statistics_error}. The total kinetic energy shows a clear trend of convergence toward that of DNS as the cutoff is decreased. The relative error decreases from approximately 14.9\% in $C\ceq2$ to 1.8\% in $C\ceq3$, and continues to drop by an order of magnitude in $C\ceq4$ and $C\ceq5$. The root mean square (RMS) of each of the velocity components and the integral length scale exhibit a similar behavior. Statistics involving velocity gradients, such as the mean dissipation rate and the Taylor microscale, rely on fine-scale features and are therefore affected more strongly by truncation. The MPS truncation discards singular values associated with these  scales first. fine-scale information is encoded in correlations that extend more widely across the MPS chain, making it harder to retain at a fixed cutoff. The mean dissipation shows the largest relative error: approximately 39\% for $C\ceq2$, reduced to 2.6\% for $C\ceq5$. The Taylor microscale, the Taylor-scale Reynolds number, and the Kolmogorov length and time scales are all defined through $\bar{\varepsilon}$. Their errors therefore inherit the sensitivity of the mean dissipation to truncation, moderated by the exponent with which $\bar{\varepsilon}$ enters each definition. The Kolmogorov length scale depends on $\bar{\varepsilon}^{-1/4}$ and its error is accordingly about half that of the Kolmogorov time scale, which depends on $\bar{\varepsilon}^{-1/2}$. The errors of these derived quantities are consequently intermediate between those of the low-order moments and that of $\bar{\varepsilon}$ itself. The values reported are magnitudes of the relative deviation. The deviation in $\bar{\varepsilon}$ does not carry the same sign at every cutoff, so partial cancellation against the deviation in $u_{\text{rms}}$ can make an individual entry at $C\ceq2$ appear smaller than the corresponding entry at $C\ceq3$. Overall, the results demonstrate a systematic convergence of both low-order and gradient-based statistics, such as the total kinetic energy and the mean dissipation rate, as the cutoff is decreased. Due to its poor performance, case \(C\ceq2\) is excluded from subsequent considerations.

\begin{table}
\centering
\begin{tabular}{l c >{\centering\arraybackslash}p{1.5cm} >{\centering\arraybackslash}p{1.25cm} >{\centering\arraybackslash}p{1.25cm} >{\centering\arraybackslash}p{1.25cm}}
\toprule
\multirow{2}{*}{\textbf{Statistic}} & \multirow{2}{*}{\textbf{DNS value}} & \multicolumn{4}{c}{\textbf{Rel. error of MPS to DNS (\%)}} \\
\cmidrule(lr){3-6}
 & & $C\,$=$\,2$ & $C\,$=$\,3$ & $C\,$=$\,4$ & $C\,$=$\,5$ \\
\midrule
Total kinetic energy ($E_t$)        & 0.6919  & 14.92 & 1.83  & 0.18 & 0.02 \\
Mean dissipation ($\bar{\varepsilon}$)            & 0.0919  & 38.99 & 19.16 & 3.96 & 2.63 \\
RMS velocity ($u_{\text{rms}}$)                 & 0.6792  & 7.76  & 0.92  & 0.09 & 0.01 \\
Taylor microscale ($\lambda$)           & 0.1180  & 18.12 & 9.24  & 1.93 & 1.34 \\
Taylor-scale Reynolds number ($R_{\lambda}$) & 433.36  & 8.93 & 10.09  & 1.83 & 1.32 \\
Kolmogorov length scale ($\eta$)     & 0.0029  & 13.15 & 4.29  & 1.01 & 0.67 \\
Kolmogorov time scale  ($\tau_{\eta}$)      & 0.0449  & 28.03 & 8.39  & 2.04 & 1.33 \\
Integral length scale ($L$ )            & 0.6808  & 12.87 & 1.74  & 0.18 & 0.02 \\
Large eddy turnover time ($T_L$)    & 1.0024  & 22.36 & 2.68  & 0.27 & 0.02 \\
\bottomrule
\end{tabular}
\caption{Relative error (\%) of single-point turbulence statistics computed from MPS-reconstructed velocity fields using spectral derivatives for $C\ceq2$ to $5$. Reported values are magnitudes of the relative deviation from DNS.}
\label{tab:scalar_statistics_error}
\end{table}

\begin{figure}[t!]
    \centering
    \begin{minipage}{0.65\linewidth}
        \raggedright
        (a)\\
        \includegraphics[height=3cm]{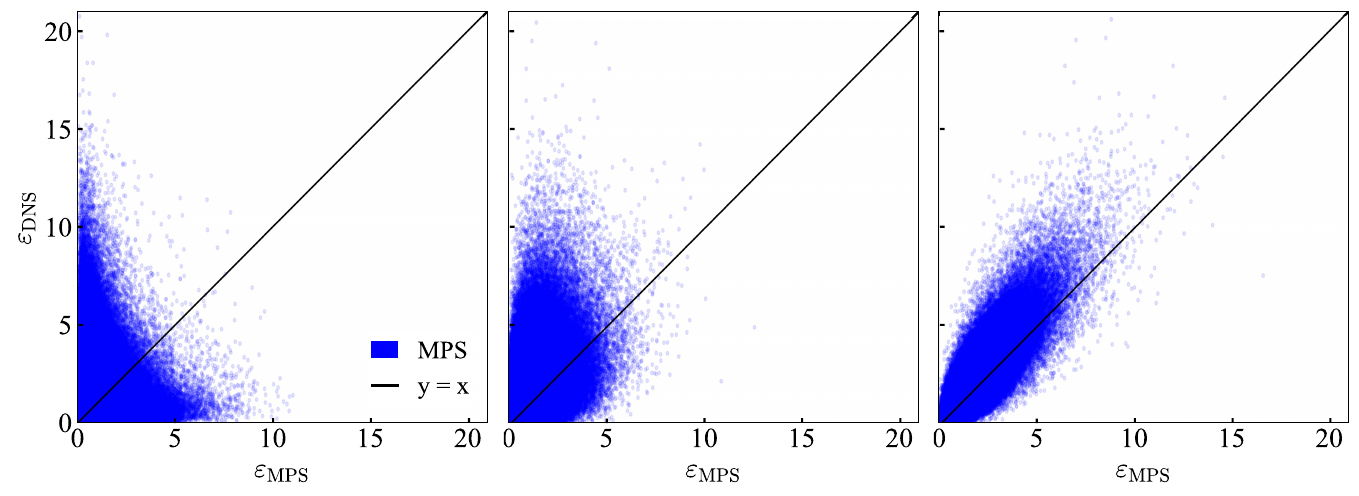}
    \end{minipage}
    \hspace{1.0em}
    \begin{minipage}{0.3\linewidth}
        \raggedright
        (b)\\
        \resizebox{!}{3cm}{\input{PDF_Dissipation.tex}}
    \end{minipage}
    \caption{(a) Pointwise scatter plot of the dissipation rate (\(\varepsilon\)) at \(C\ceq3\), \(4\), and \(5\) (left to right). (b) The corresponding PDF of \(\varepsilon\) at \(C\ceq5\) compared with the DNS distribution and a reference log-normal distribution.}
    \label{fig:scatter_dissipation}
\end{figure}

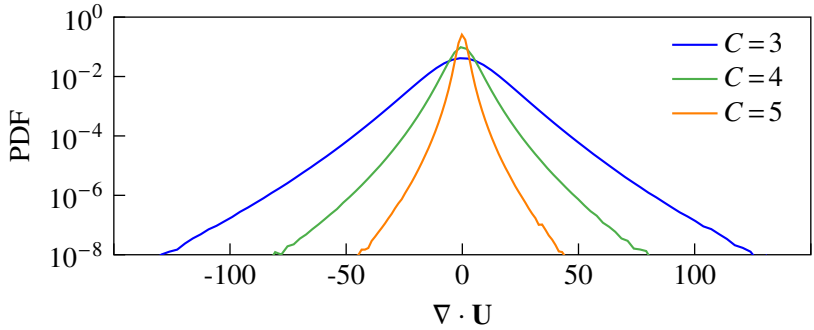
\begin{figure}[t!]
    \centering
    {\input{PDF_Div.tex}}
    \caption{Probability density functions of \(\nabla\cdot\mathbf{U}\) for MPS reconstructions at \(C\ceq3\), \(4\), and \(5\). The DNS divergence is not shown: its magnitude is bounded by \(\sim 10^{-4}\) (set by single-precision storage of the JHTDB data), so on the scale of the MPS distributions it collapses to a near-vertical spike at \(\nabla\cdot\mathbf{U}=0\).}
    \label{fig:div}
\end{figure}

The PDFs of the velocity component \(u\) and its spatial gradients are shown in \cref{fig:u_velocity_pdf_comparison,fig:pdf_velocity_gradients}, respectively.
\begin{figure}[b!]
    \centering
    \begin{minipage}[t]{0.25\textwidth}
        \centering
        \hspace{4em}
        \textbf{(a)}\\
        \resizebox{!}{3.5cm}{\input{C3/PDF_Velocity_u_1024Cubed_SameLimits.tex}}
    \end{minipage}
    \hspace{0.08\textwidth}
    \begin{minipage}[t]{0.25\textwidth}
        \centering
        \hspace{1.1em}
        \textbf{(b)}\\
        \resizebox{!}{3.5cm}{\input{C4/PDF_Velocity_u_1024Cubed_SameLimits.tex}}
    \end{minipage}
    \hspace{0.04\textwidth}
    \begin{minipage}[t]{0.25\textwidth}
        \centering
        \hspace{1em}
        \textbf{(c)}\\
        \resizebox{!}{3.5cm}{\input{C5/PDF_Velocity_u_1024Cubed_SameLimits.tex}}
    \end{minipage}

    \caption{Probability density function of the \(u\) velocity component from MPS reconstructions for (a) \(C\ceq3\), (b) \(C\ceq4\), and (c) \(C\ceq5\).}
    \label{fig:u_velocity_pdf_comparison}
\end{figure}
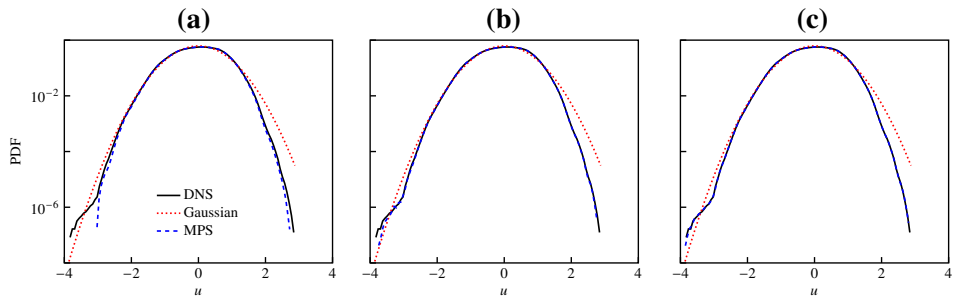
\begin{figure}[t!]
    \centering

    % First row: du/dx
    \begin{minipage}[t]{0.25\textwidth}
        \centering
        \hspace{4em}
        \textbf{(a)}\\
        \resizebox{!}{3.5cm}{\input{C3/PDF_grad_dux_dx_SameLimits.tex}}
    \end{minipage}
    \hspace{0.08\textwidth}
    \begin{minipage}[t]{0.25\textwidth}
        \centering
        \hspace{1.5em}
        \textbf{(b)}\\
        \resizebox{!}{3.5cm}{\input{C4/PDF_grad_dux_dx_SameLimits.tex}}
    \end{minipage}
    \hspace{0.04\textwidth}
    \begin{minipage}[t]{0.25\textwidth}
        \centering
        \hspace{1.5em}
        \textbf{(c)}\\
        \resizebox{!}{3.5cm}{\input{C5/PDF_grad_dux_dx_SameLimits.tex}}
    \end{minipage}

    \vspace{0.6cm}

    % Second row: du/dy
    \begin{minipage}[t]{0.25\textwidth}
        \centering
        \hspace{4em}
        \textbf{(d)}\\
        \resizebox{!}{3.5cm}{\input{C3/PDF_grad_dux_dy_SameLimits.tex}}
    \end{minipage}
    \hspace{0.08\textwidth}
    \begin{minipage}[t]{0.25\textwidth}
        \centering
        \hspace{1.5em}
        \textbf{(e)}\\
        \resizebox{!}{3.5cm}{\input{C4/PDF_grad_dux_dy_SameLimits.tex}}
    \end{minipage}
    \hspace{0.04\textwidth}
    \begin{minipage}[t]{0.25\textwidth}
        \centering
        \hspace{1.5em}
        \textbf{(f)}\\
        \resizebox{!}{3.5cm}{\input{C5/PDF_grad_dux_dy_SameLimits.tex}}
    \end{minipage}

    \caption{Probability density functions of velocity gradients. The top row shows $ \partial u / \partial x $, and the bottom row shows $ \partial u / \partial y $ for (a,d) $C\ceq3$, (b,e) $C\ceq4$, and (c,f) $C\ceq5$.}
    \label{fig:pdf_velocity_gradients}
\end{figure}
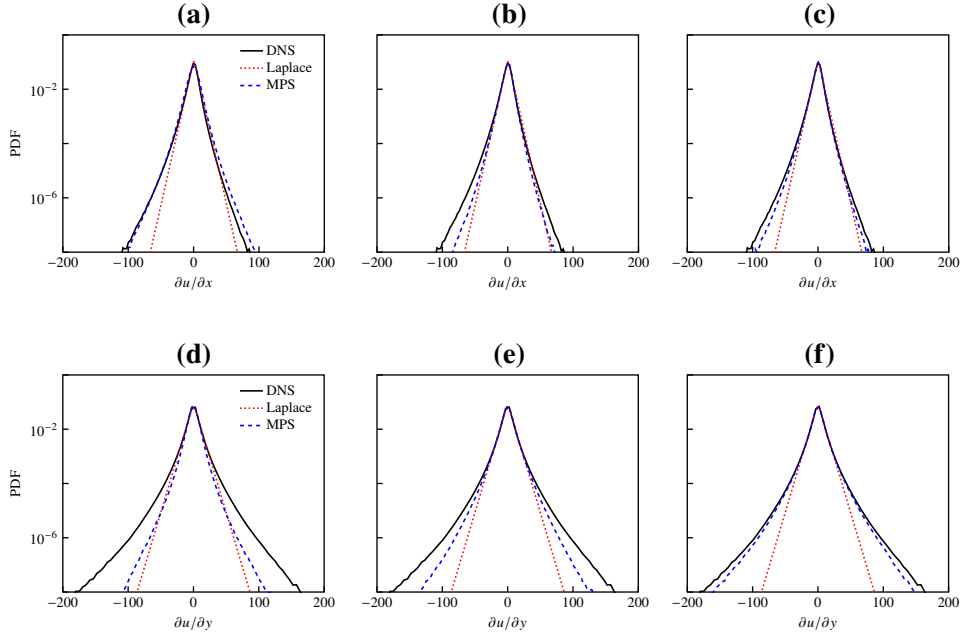
These figures demonstrate the ability of the MPS representation to preserve the statistical distributions of these quantities as the cutoff is decreased. In \cref{fig:u_velocity_pdf_comparison}, a Gaussian PDF with the same mean and variance as the DNS is included as a reference to illustrate how MPS preserves the near-Gaussian distribution. In \cref{fig:pdf_velocity_gradients}, the Laplace PDF with the same scale parameters as the DNS is shown for reference. The MPS captures the characteristic heavy-tailed shape of the longitudinal and transverse velocity-gradient PDFs. The improved agreement at lower cutoff is also evident at the pointwise level in \cref{fig:scatter_dissipation}(a), which shows the scatter plot of the dissipation against DNS. The lognormal distribution serves as a classical reference for turbulent dissipation, following the refined similarity hypothesis of \cite{kolmogorov_1962}. For the case with the best overall accuracy (\(C\ceq5\)), the reconstructed PDF closely follows the DNS curve (\cref{fig:scatter_dissipation}(b)).
\Cref{fig:div} shows the PDFs of the divergence of the velocity field. The DNS divergence remains close to zero, as required for incompressibility, with magnitudes on the order of \(10^{-4}\) set by the single-precision storage of the JHTDB data. Consequently, the divergence in DNS does not reach machine precision. MPS reconstructions do not strictly preserve incompressibility (\(\displaystyle \nabla \cdot \mathbf{U}=0\)), but the distribution becomes progressively narrower as the truncation error decreases. The extreme values of the divergence reduce from approximately \( \pm 130 \) for \(C\ceq3\) to about \( \pm 50 \) for \(C\ceq5\).

\begin{figure}[t!]
    \centering
    \begin{minipage}{0.27\textwidth}
    \centering
    \hspace{1em}
    \textbf{(a)}\\
    % \hspace{-2em}
    \resizebox{\linewidth}{3.1cm}
    {\input{PDF_Strain_rate_eigen_values_alpha.tex}}
    \end{minipage}
    \begin{minipage}{0.23\textwidth}
    \centering
    \textbf{(b)}\\
    \resizebox{\linewidth}{3.2cm}{\input{PDF_Strain_rate_eigen_values_beta.tex}}
    \end{minipage}
    \begin{minipage}{0.23\textwidth}
    \centering
    \hspace{0.5em}
    \textbf{(c)}\\
    \resizebox{\linewidth}{3.2cm}{\input{PDF_Strain_rate_eigen_values_gamma.tex}}
    \end{minipage}
    \begin{minipage}{0.23\textwidth}
    \centering
    \hspace{0.5em}
    \textbf{(d)}\\
    \resizebox{\linewidth}{3.2cm}{\input{PDF_Strain_rate_eigen_values_stilde.tex}}
\end{minipage}

\caption{Probability density functions of eigenvalues of the deviatoric strain rate tensor. Panels (a)--(c) show the PDFs of the three eigenvalues $\alpha, \beta$, and $\gamma$, and panel (d) shows the PDF of the normalized strain-rate invariant $s^*$. All panels are shown for cutoffs $C\ceq3$ to $5$.}
\label{fig:strainrate_eigenvalue_pdf_grid}
\end{figure}
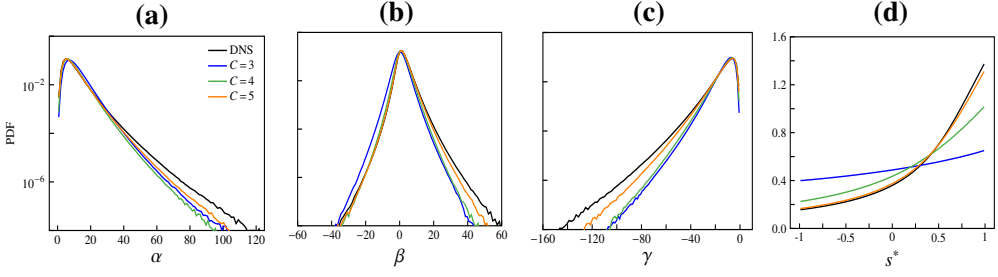

PDFs that are more sensitive to fine-scale features are associated with the three eigenvalues of the strain rate tensor. These eigenvalues serve as rotationally invariant measures of local deformation and characterize  stretching, compression, and flow symmetry \citep{Lund1994,Ryu_Livescu_2014}. \Cref{fig:strainrate_eigenvalue_pdf_grid} shows these PDFs. For the eigenvalues \(\alpha\), \(\beta\), and \(\gamma\) (\cref{fig:strainrate_eigenvalue_pdf_grid}(a)--(c)), the MPS reconstructions reproduce the central shape of the distributions, but underestimate the tails at \(C\ceq3\). These tail regions correspond to localized events of intense strain, concentrated in the small scale features of the flow that truncation captures less effectively at low \(C\). \Cref{fig:strainrate_eigenvalue_pdf_grid}(d) shows the PDF of \(s^*\), a normalized invariant constructed from the eigenvalues of the deviatoric strain rate tensor that describes the asymmetry and anisotropy of the local strain field. The deviatoric strain rate tensor \(S^*\) and the scalar quantity \(s^*\) are defined as
\begin{equation}
    S^* = S - \frac{1}{3} \operatorname{tr}(S) I, \qquad
    s^* = \frac{-3\sqrt{6}\, \alpha \beta \gamma}
{(\alpha^2 + \beta^2 + \gamma^2)^{3/2}},
\end{equation}
where \(S\) is the strain rate tensor, and \(\alpha\), \(\beta\), \(\gamma\) are the eigenvalues of \(S^*\). The quantity \(s^*\) takes values in \([-1, +1]\): \(s^*\!=\!-1\) corresponds to the axisymmetric extension (two extensional and one compressive), \(s^*\!=\!+1\) to the axisymmetric contraction (two compressive eigenvalues and one extensional), and \(s^*\!=\!0\) to a planar shear-like configuration. The disagreement between MPS and DNS is most pronounced in \(s^*\): the \(C\ceq3\) curve does not capture the rise of DNS toward \(s^*\!=\!+1\), indicating that the geometric preference of the local strain for axisymmetric contraction is not preserved at this cutoff. At \(C\ceq4\) and \(C\ceq5\), the agreement improves markedly across all four panels, and the \(C\ceq5\) distributions nearly overlap with DNS.

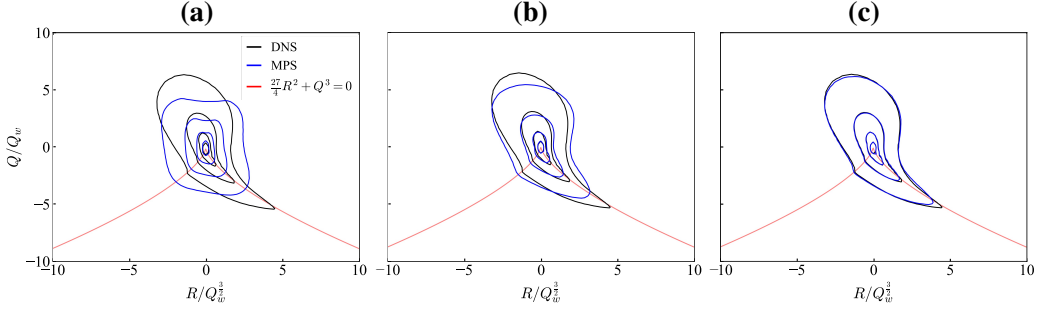
\begin{figure}[t!]
\centering
\begin{tikzpicture}
  \node[anchor=south west, inner sep=0] (img)
    {\includegraphics[width=1.02\textwidth]{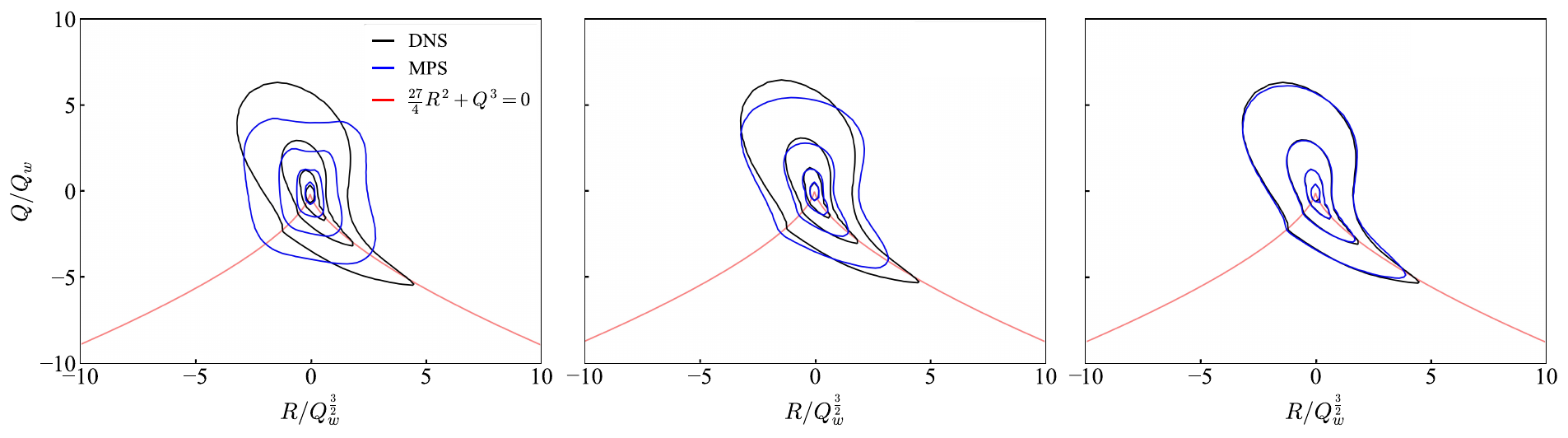}};
  \begin{scope}[x={(img.south east)}, y={(img.north west)}]
    \node at (0.19,1.02) {\textbf{(a)}};
    \node at (0.51,1.02) {\textbf{(b)}};
    \node at (0.83,1.02) {\textbf{(c)}};
  \end{scope}
\end{tikzpicture}
\caption{Joint PDF of the second (\(Q\)) and third (\(R\)) invariants of the velocity gradient tensor for cutoffs \(C\ceq3\) to \(5\) (left to right). Axes are normalized by \(Q_w = \langle S_{ij}S_{ij}\rangle\). The red curve shows the incompressible Vieillefosse relation, \(27R^2/4 + Q^3 = 0\), as a visual reference.}
\label{fig:joint_pdf_qr_contour}
\end{figure}

The flow structure is further examined through the invariants of the velocity gradient tensor, \( A = \nabla \mathbf{U} \), and its second and third invariants:
\begin{equation}\label{eq:NS}
\begin{aligned}
Q &= \frac{1}{2}\,\left[\mathrm{tr}(A)^2-\mathrm{tr}(A^2)\right],
\qquad
R = \frac{1}{3}\,\left[\mathrm{tr}(A)^3 - 3\,Q\,\mathrm{tr}(A) - \mathrm{tr}(A^3)\right],
\end{aligned}
\end{equation}

\noindent where \( \mathrm{tr}(\cdot) \) denotes the matrix trace operator. The variable \( Q \) is a measure of the balance between the rotation and deformation components of the local flow field, while \( R \) characterizes the asymmetry of the velocity gradient tensor and relates to vortex stretching and compression \citep{Ryu_Livescu_2014,Wang_Shi_Wang_Xiao_He_Chen_2012,Pirozzoli_2004,Perry1987}. The \((Q,R)\) plane is naturally partitioned by the Vieillefosse line \(27R^2/4 + Q^3 = 0\) \citep{vieillefosse_1982}, which separates topologically distinct regions of the local flow geometry. The characteristic teardrop distribution, with its preferential alignment along the lower-right branch of the Vieillefosse line, is a well-known signature of turbulent velocity gradient statistics. \Cref{fig:joint_pdf_qr_contour} shows the joint PDFs of \(Q\) and \(R\) invariants for DNS and MPS-reconstructed fields, computed directly from the velocity gradient tensor. The MPS reconstructions reproduce the characteristic teardrop shape and its alignment along the Vieillefosse line, confirming that the higher-order gradient statistics are captured along with the lower-order ones.

\begin{figure}[t!]
    \centering
    {\input{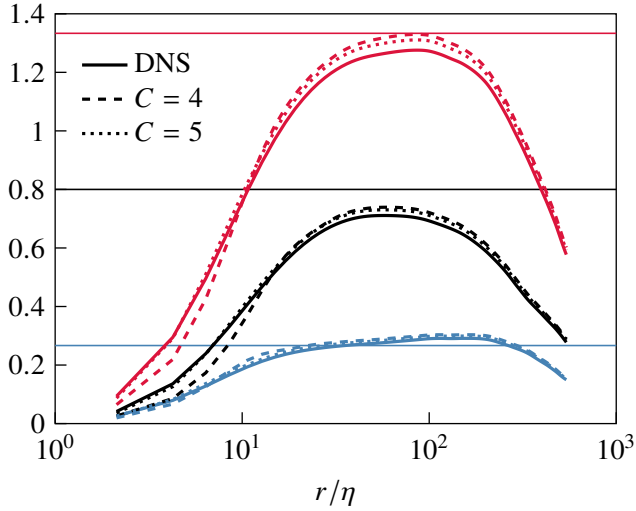}}
    \caption{Third-order velocity structure functions for DNS and MPS reconstructions at \(C\ceq4\) and \(C\ceq5\), plotted against \(r/\eta\). Horizontal lines mark the Kolmogorov theoretical values: bottom (blue), the 4/15-law for the transverse structure function; middle (black), the 4/5-law for the longitudinal one; top (red), the 4/3-law for the mixed combination.}
    \label{fig:sf}
\end{figure}

To further assess the performance of MPS in preserving small structures under compression, the longitudinal and transverse structure functions of the flow are considered. The classical scaling relations of \cite{Kolmogorov} are examined by evaluating the \(\tfrac{4}{5}\)-law for the third-order longitudinal structure function, \(-D_{LLL}/(\bar{\varepsilon} r)\); the \(\tfrac{4}{15}\)-law for the third-order transverse structure function, \(-D_{LTT}/(\bar{\varepsilon} r)\); and the \(\tfrac{4}{3}\)-law for the mixed combination, \(-(D_{LLL} + 2D_{LTT})/(\bar{\varepsilon} r)\). Here, \(r\) denotes the magnitude of the separation vector \(\mathbf{r}\) used to compute the velocity increments \(\delta u_L(\mathbf{r}) = [\mathbf{U}(\mathbf{x}+\mathbf{r}) - \mathbf{U}(\mathbf{x})] \cdot \mathbf{r}/r\) (longitudinal) and the corresponding transverse increment. The results shown in \cref{fig:sf} indicate that highly-truncated fields capture the Kolmogorov scaling within the inertial range to a degree very similar to that observed in the DNS \citep{Taylor2003}. The slopes and magnitudes of the structure functions for \(C\ceq4\) and \(C\ceq5\) closely follow those of the DNS.

\section{MPS representation of the scalar field}\label{sec:scalar}

\begin{figure}[b!]
    \centering
    \begin{tikzpicture}
      \node[anchor=south west, inner sep=0] (img) 
        {\includegraphics[width=1.02\textwidth]{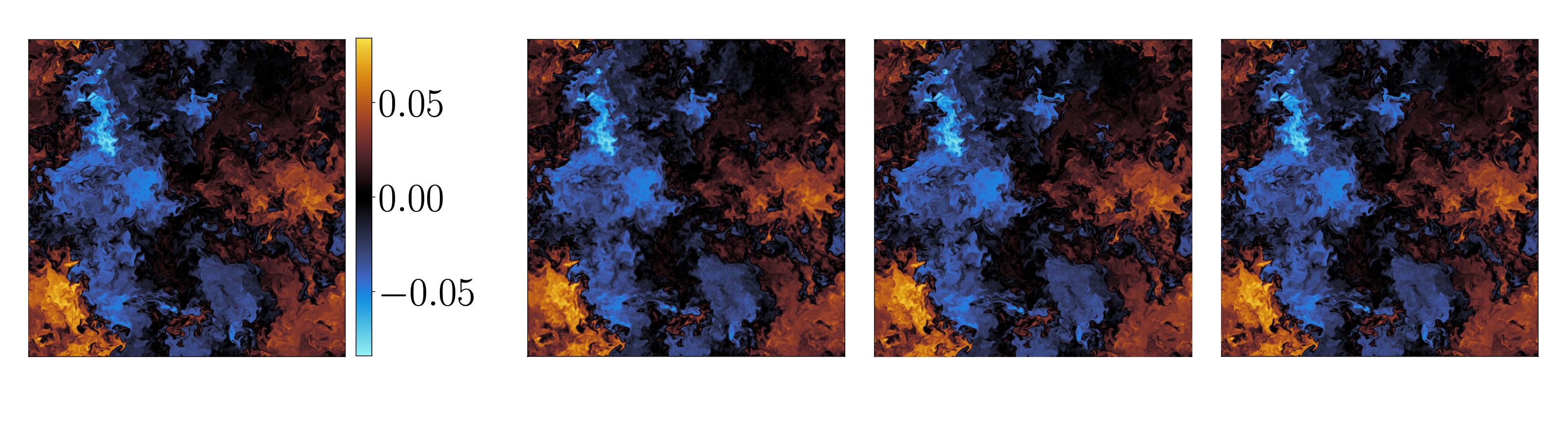}};
      \begin{scope}[x={(img.south east)}, y={(img.north west)}]
        \node at (0.12,0.97) {\textbf{(a)}};
        \node at (0.43,0.97) {\textbf{(b)}};
        \node at (0.65,0.97) {\textbf{(c)}};
        \node at (0.87,0.97) {\textbf{(d)}};
      \end{scope}
    \end{tikzpicture}
    \vspace{-2em}
    \caption{A slice of the passive scalar field (\(\phi\)) in the \(xy\)-plane at \(z=\pi\). From left to right: (a) the original DNS field, and (b)--(d) MPS reconstructions for \(C\ceq3\), \(4\), and \(5\), corresponding to CRs of approximately 57, 7, and 2 (see \Cref{tab:scalar_compression_phi1}).}
    \label{fig:scalar_slice}
\end{figure}
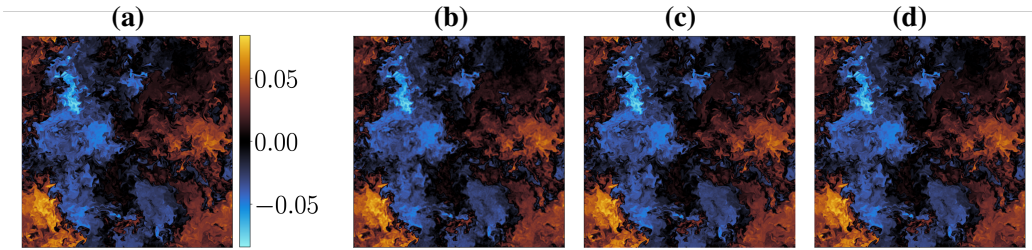

\begin{figure}[h]
    \centering
    \begin{minipage}{0.65\linewidth}
        \raggedright
        (a)\\
        \includegraphics[height=3cm]{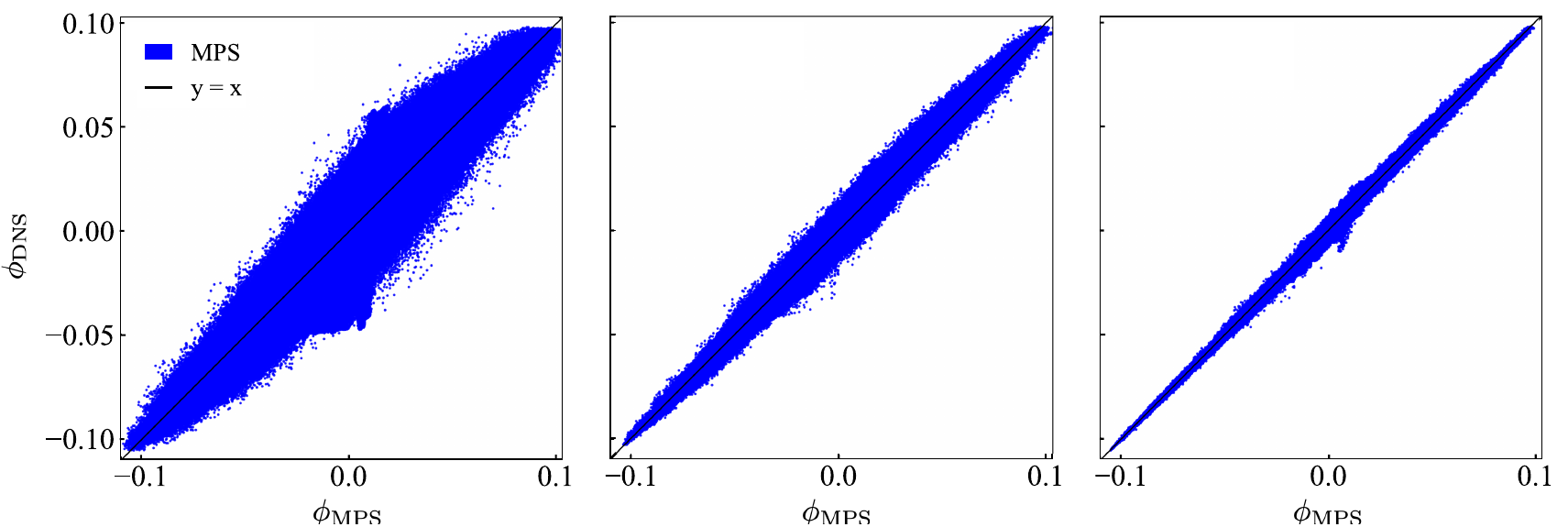}
    \end{minipage}
    \hspace{1.0em}
    \begin{minipage}{0.3\linewidth}
        \raggedright
        (b)\\
        \resizebox{!}{3cm}{\input{scalar_PDF}}
    \end{minipage}
    \caption{(a) Pointwise scatter plots of the passive scalar \(\phi\) (MPS against DNS) for \(C\ceq3\), \(4\), and \(5\) (left to right). (b) PDFs of \(\phi\) from DNS and from the MPS reconstructions at the same cutoffs, with a Gaussian reference of matching mean and variance.}
    \label{fig:scalar_scatter_pdf}
\end{figure}

\begin{figure}[h]
    \centering
    \begin{minipage}{0.65\linewidth}
        \raggedright
        (a)\\
        \includegraphics[height=3cm]{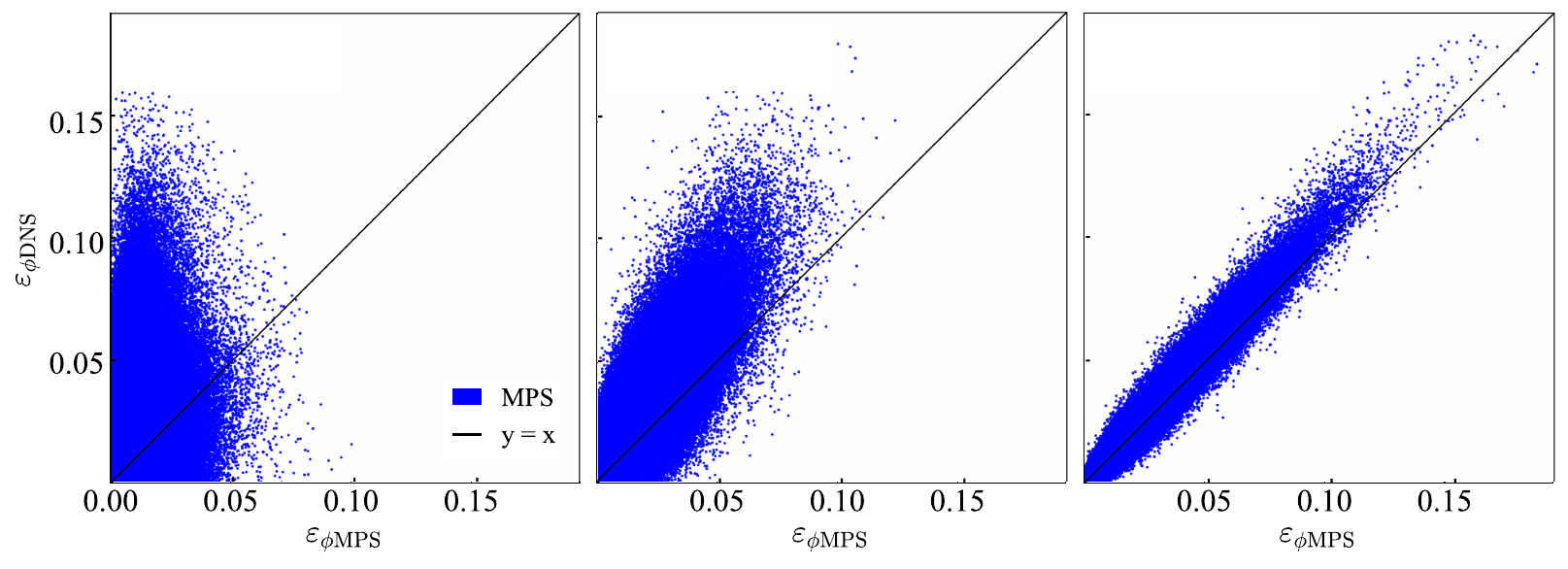}
    \end{minipage}
    \hspace{1.0em}
    \begin{minipage}{0.3\linewidth}
        \raggedright
        (b)\\
        \resizebox{!}{3cm}{\input{scalar_disp_PDF}}
    \end{minipage}
    \caption{(a) Scatter plots of the scalar dissipation \(\varepsilon_\phi\) (MPS against DNS) for \(C\ceq3\), \(4\), and \(5\) (left to right). (b) PDFs of \(\varepsilon_\phi\) from DNS and from the MPS reconstructions at the same cutoffs, with a log-normal reference.}
    \label{fig:scalar_disp_scatter_pdf}
\end{figure}

The comparison between the MPS and DNS scalar fields is made in \cref{fig:scalar_slice}, which presents contours of the conserved scalar in an \(x\)--\(y\) plane through the center of the domain (\(z=\pi\)).
Relative to the DNS field, the reconstructed field with $C\ceq3$ appears smoother and shows attenuated filaments and reduced contrast, indicating the loss of the finest scalar gradients. Reconstructions for $C\ceq4$ and $C\ceq5$ recover the small scale texture with only mild smoothing, producing structures that are almost indistinguishable from the DNS. The scalar field exhibits a finer-scale structure than the velocity field, and identical truncation levels \(C\) therefore yield lower CRs. Despite this, the achieved storage reduction remains substantial and infidelity decreases rapidly as the cutoff is decreased (\Cref{tab:scalar_compression_phi1}).
\begin{table}
    \centering
    \begin{tabular}{l c c c c c}
        \toprule
        $\boldsymbol{C}$ & $\boldsymbol{\chi_{\max}}$ & \textbf{CR} & $\boldsymbol{\mathcal{I}}$ & \textbf{Equivalent DNS} \\
        \midrule
        3 & 1891  & 57.1  & 0.0196 & $\sim 266^3$ \\
        4 & 5979  & 6.5   & 0.0018 & $\sim 549^3$ \\
        5 & 12071 & 1.7   & 0.0002 & $\sim 858^3$ \\
        \bottomrule
  \end{tabular}
  \caption{Compression metrics for scalar \(\phi\): maximum bond dimension ($\chi_{\max}$), compression ratio (CR) and infidelity ($\mathcal{I}$).}
  \label{tab:scalar_compression_phi1}
\end{table}
The values of the scalar energy and the scalar dissipation rate are given in \Cref{tab:scalar_statistics_error_scalar}. The scalar energy is recovered with high accuracy even at moderate truncation levels, with relative errors below $0.2\%$ for $C\ceq4$. The scalar dissipation rate \(\varepsilon_\phi\) (defined in Appendix~\ref{app:turbulence_definitions}) is directly related to the scalar gradients and shows a higher sensitivity to truncation than the scalar energy. Its mean value \(\bar{\varepsilon}_\phi\), reported in \Cref{tab:scalar_statistics_error_scalar}, has a relative error of about 10\% at \(C\ceq4\) (CR \(\approx 7\)) and below 2\% at \(C\ceq5\) (CR \(\approx 2\)). As the cutoff is reduced, the scatter of passive scalar tightens around the diagonal (\cref{fig:scalar_scatter_pdf}(a)), indicating improved pointwise correspondence between the MPS and DNS scalar values. The corresponding scalar PDFs (\cref{fig:scalar_scatter_pdf}(b)) show that the bulk statistics (mean, variance, and central shape) are preserved even for $C\ceq3$, while the tails are better captured for $C\ceq4$ and $C\ceq5$. A greater sensitivity to truncation is observed in the scalar-dissipation statistics, as expected. The PDFs of \(\varepsilon_\phi\) (\cref{fig:scalar_disp_scatter_pdf}(b)) exhibit the largest deviations in the tails for $C\ceq3$, where rare, large-amplitude dissipation events are under-represented. The agreement with DNS improves markedly for $C\ceq4$ and further for $C\ceq5$. A similar behavior is observed in the dissipation scatter plots (\cref{fig:scalar_disp_scatter_pdf}(a)), where the greatest spread and outliers occur for $C\ceq3$, while a stronger correlation with DNS is recovered at smaller truncation errors. Taken together, these results indicate that MPS reconstructions preserve the large-scale features and low-order statistics of the scalar field even at moderate truncation levels, whereas quantities that depend on fine-scale features (such as scalar dissipation) require smaller truncation errors.

\begin{table}
\centering
\begin{tabular}{l c c c c}
\toprule
\multirow{2}{*}{\textbf{Statistic}} &
\multirow{2}{*}{\textbf{DNS value}} &
\multicolumn{3}{c}{\hspace{0.5cm}}{\textbf{ Rel. error of MPS to DNS (\%)}} \\
\cmidrule(lr){3-5}
 & & {\hspace{0.5cm}} $C\,$=$\,3$ & \hspace{0.4cm} $C\,$=$\,4$ & $C\,$=$\,5$ \\
\midrule
Scalar energy ($E_\phi$)       & $2.6\times10^{-4}$ & {\hspace{0.5cm}} 1.96 & \hspace{0.4cm} 0.17 & 0.02 \\
Mean scalar dissipation ($\bar{\varepsilon}_{\phi}$)  & $5.0\times10^{-4}$ & {\hspace{0.5cm}} 34.40 & \hspace{0.4cm} 10.47 & 1.84 \\
\bottomrule
\end{tabular}
\caption{Relative error (\%) of scalar statistics computed from MPS-reconstructed scalar fields for $C\ceq3$ to $5$ with respect to DNS.}
\label{tab:scalar_statistics_error_scalar}
\end{table}

\FloatBarrier

\section{Conclusions}\label{sec:conclusions}

The capability of the MPS encoding is assessed as a reduced-order modeling of isotropic turbulent flow. {\it A priori} assessments are conducted using DNS data of the velocity field and a conserved passive scalar field. The MPS representation reduces storage by factors ranging from a few to the order \(10^4\), depending on the cutoff, while retaining infidelities as low as \(2\times 10^{-4}\). For sufficiently small cutoffs (\(C\ceq4\)), the reconstructed fields achieve infidelities of about \(0.2\%\) while compressing the velocity and scalar fields by factors of about \(22\) and \(7\) relative to the original DNS storage, respectively. At this level of truncation, key statistical quantities are also accurately reproduced: the total kinetic energy and the scalar energy are recovered with about $0.2\%$ relative error, while the mean dissipation and the mean scalar dissipation remain within approximately $10\%$ of the DNS generated values. Pointwise quantities involving the velocity gradients are inherently more difficult to reproduce because they depend on the fine-scale features. Consequently, statistics based on gradients, such as dissipation or strain-rate measures, exhibit greater sensitivity to truncation. For \(C\ceq4\) the reconstructed fields reproduce a wide range of statistics for both velocity and scalar fields, with mean-dissipation errors of about 4\% and 10\% (velocity and scalar, respectively), and with strain-rate eigenvalue distributions and other higher-order gradient-sensitive statistics in close agreement with DNS.

The CRs quantify how much storage is saved, but the results also reveal why the truncation preserves the flow physics so well. The cutoff adaptively selects the singular values at each bond, retaining more information at scales where inter-tensor correlations are stronger. This leads to a balanced preservation of both large-scale energy-containing structures and small-scale dissipative motions. Such adaptivity enables MPS to serve not only as a storage-efficient representation but also as a reduced-order modeling tool capable of maintaining statistical and spectral fidelity across length scales. The observed alignment of reconstructed statistics with Kolmogorov laws confirms that the essential physical features of turbulence are retained even at large CRs.

Several directions are suggested for future research. First, other TN architectures such as the multiscale entanglement renormalization ansatz \citep{Vidal2008MERA}, tree tensor networks \citep{Vidal2006Tree}, and  projected entangled pair states \citep{Verstraete2004} should  be considered. These architectures can potentially capture the fine-scale and multiscale correlations that limit MPS efficiency in three-dimensional isotropic turbulence. The time evolution of the Navier--Stokes equations using these TN ans\"atze could be investigated to assess how TNs perform under temporal dynamics. Second, comparisons between MPS-compressed  and ``filtered'' fields as typically done in large eddy simulation (LES) methods \citep{pope2000turbulent,moin2024fundamentals} could provide insight into how TN truncation parallels explicit subgrid-scale modeling \citep{Givi2006}.  Finally, the implementation of MPS and other TNs to more complex flows---chemically reacting, multiphase, wall-bounded, and anisotropic---would certainly assess the generality of the present findings.

\bibliographystyle{jfm}
\bibliography{jfm}

\clearpage
\appendix

\appendix
\renewcommand{\theHsection}{appendix.\Alph{section}}
\section{Definitions of turbulence statistics}
\label{app:turbulence_definitions}

The turbulence statistics considered in  this work are defined as follows \citep{pope2000turbulent}:

\begin{center}
\renewcommand{\arraystretch}{1.5}
\begin{tabular}{l l}
\toprule
\textbf{Quantity} & \textbf{Definition} \\ \\
\midrule
Total kinetic energy 
& $E_{\text{t}} = \dfrac{1}{2}\overline{u_i u_i}$ \\ \\

Strain-rate tensor
& $S_{ij}=\dfrac{1}{2}\left(\dfrac{\partial u_i}{\partial x_j}
+\dfrac{\partial u_j}{\partial x_i}\right)$ \\ \\ 

RMS velocity 
& $u_{\text{rms}} = \sqrt{\dfrac{2}{3}E_{\text{t}}}$ \\ \\

Dissipation 
& $\varepsilon = 2\nu S_{ij} S_{ij} \to \bar{\varepsilon} = 2\nu \overline{S_{ij} S_{ij}}$\\ \\

Taylor microscale 
& $\lambda = \sqrt{\dfrac{15\nu {u_{\text{rms}}}^{2}}{\bar{\varepsilon}}}$ \\ \\

Taylor-scale Reynolds number \quad {} 
& $R_\lambda = \dfrac{u_{\text{rms}} \lambda}{\nu}$ \\ \\

Kolmogorov length scale 
& $\eta = \nu^{3/4}\bar{\varepsilon}^{-1/4}$ \\ \\

Kolmogorov time scale 
& $\tau_\eta = \sqrt{\dfrac{\nu}{\bar{\varepsilon}}}$ \\ \\

Integral length scale 
& $L = \dfrac{\pi}{2u_{\text{rms}}^2} \displaystyle\int \dfrac{E(k)}{k} \, dk$ \\ \\

Large-eddy turnover time 
& $T_L = \dfrac{L}{u_{\text{rms}}}$ \\ \\

Scalar energy 
& $E_\phi = \tfrac{1}{2}\,\overline{(\phi-\bar\phi)^2}$ \\ \\

Scalar dissipation
& $\varepsilon_\phi = 2D\,\nabla\phi \cdot \nabla\phi$ \\ \\
\bottomrule
\end{tabular}
\end{center}
Here, $u_i$ denotes the velocity components, $\nu$ the kinematic viscosity, $E(k)$ the energy
spectrum, and $D$ the molecular diffusivity. For the hydrodynamic dataset
$\nu = 1.85 \times 10^{-4}$; for the scalar dataset $\nu = 5 \times 10^{-4}$ and, since
$Sc\ceq1 \Rightarrow D = \nu$. 
\end{document}

%% file: C3/PDF_Velocity_u_1024Cubed_SameLimits.tex
% This file was created by tikzplotlib v0.9.6.
\begin{tikzpicture}

\begin{axis}[
legend cell align={left},
legend style={fill opacity=0.8, text opacity=1, at={(0.5,0.09)}, anchor=south, draw=none},
log basis y={10},
tick pos=left,
x grid style={white!69.0196078431373!black},
xlabel={\(\displaystyle u\)},
xmin=-4, xmax=4,
xtick style={color=black},
y grid style={white!69.0196078431373!black},
ylabel={PDF},
ymin=1e-08, ymax=1,
ymode=log,
ytick style={color=black},
ytick={1e-10,1e-08,1e-06,0.0001,0.01,1,100},
yticklabels={\(\displaystyle 10^{-10}\),,\(\displaystyle 10^{-6}\),,\(\displaystyle 10^{-2}\),,\(\displaystyle 10^{2}\)}
]
\addplot [very thick, black]
table {%
-3.82805346488953 8.29281779528482e-08
-3.76067063331604 1.65856355905696e-07
-3.69328780174255 1.65856355905696e-07
-3.62590497016907 3.17891348819251e-07
-3.55852213859558 4.00819526772099e-07
-3.49113930702209 4.83747704724948e-07
-3.42375647544861 6.0813997165422e-07
-3.35637364387512 6.91068149607068e-07
-3.28899081230164 8.70745868504906e-07
-3.22160798072815 1.11953040236345e-06
-3.15422514915466 1.29920812126129e-06
-3.08684231758118 1.85206264094694e-06
-3.01945948600769 2.36345307165619e-06
-2.9520766544342 5.74968700473077e-06
-2.88469382286072 1.05318786000118e-05
-2.81731099128723 2.03865104134084e-05
-2.74992815971375 3.55900097047642e-05
-2.68254532814026 5.4497634278013e-05
-2.61516249656677 9.35844488197898e-05
-2.54777966499329 0.000155531797750566
-2.4803968334198 0.000260726191483756
-2.41301400184631 0.00043162734488158
-2.34563117027283 0.000701973205007872
-2.27824833869934 0.00108960715148546
-2.21086550712585 0.00159979512361438
-2.14348267555237 0.00222729882482059
-2.07609984397888 0.00311066359773732
-2.0087170124054 0.00434959675498988
-1.94133418083191 0.0060098188776059
-1.87395134925842 0.00836363845925655
-1.80656851768494 0.0113805793945442
-1.73918568611145 0.0154459536409636
-1.67180285453796 0.0210303924297604
-1.60442002296448 0.028030995713458
-1.53703719139099 0.0376147149908458
-1.4696543598175 0.0502142280303929
-1.40227152824402 0.0649747111737613
-1.33488869667053 0.0828873496465694
-1.26750586509705 0.10392139091616
-1.20012303352356 0.127701128872317
-1.13274020195007 0.15239314589277
-1.06535737037659 0.175925370880464
-0.9979745388031 0.203734810145158
-0.930591707229614 0.23467514748301
-0.863208875656128 0.267841733222095
-0.795826044082641 0.302967828878028
-0.728443212509155 0.338573871363862
-0.661060380935669 0.373496696234272
-0.593677549362183 0.405694335320791
-0.526294717788696 0.43598256693131
-0.45891188621521 0.460485936168284
-0.391529054641724 0.484513933446506
-0.324146223068237 0.503694004727056
-0.256763391494751 0.519113822170602
-0.189380559921265 0.533056205017102
-0.121997728347778 0.546188158423196
-0.0546148967742919 0.556947481372606
0.0127679347991945 0.56583529502334
0.0801507663726808 0.566962648317157
0.147533597946167 0.564841552845568
0.214916429519654 0.56039086664165
0.28229926109314 0.551128839587911
0.349682092666626 0.538539968997861
0.417064924240113 0.522481728976352
0.484447755813599 0.504747994226114
0.551830587387085 0.483805685342696
0.619213418960571 0.455751442096692
0.686596250534058 0.418027178982765
0.753979082107544 0.374295432801588
0.82136191368103 0.327010767049642
0.888744745254517 0.28141265311682
0.956127576828003 0.236429562194416
1.02351040840149 0.194828987258447
1.09089323997498 0.157134992929476
1.15827607154846 0.123890896986733
1.22565890312195 0.096158318321637
1.29304173469543 0.0725735721545741
1.36042456626892 0.0539289957617903
1.42780739784241 0.0391861348562376
1.49519022941589 0.0282266232852485
1.56257306098938 0.0200330705189736
1.62995589256287 0.0136901567932569
1.69733872413635 0.00893750145069032
1.76472155570984 0.00552612645833288
1.83210438728333 0.00319617637056972
1.89948721885681 0.00180898145250043
1.9668700504303 0.00106671897437049
2.03425288200378 0.000644103158159767
2.10163571357727 0.000413645751628809
2.16901854515076 0.000280186670576688
2.23640137672424 0.000171509293369483
2.30378420829773 9.79105354363287e-05
2.37116703987122 5.31707834307682e-05
2.4385498714447 2.36483520795537e-05
2.50593270301819 1.23010130630059e-05
2.57331553459167 7.17328739292141e-06
2.64069836616516 3.44151938504318e-06
2.70808119773865 1.39595766220629e-06
2.77546402931213 4.97569067717086e-07
2.84284686088562 1.24392266929273e-07
};
\addlegendentry{DNS}
\addplot [very thick, red, dotted]
table {%
-3.86174488067627 1.13221398160252e-08
-3.79368141444042 2.10971153085155e-08
-3.72561794820458 3.88786635039558e-08
-3.65755448196874 7.08587138863232e-08
-3.58949101573289 1.27722935468882e-07
-3.52142754949705 2.27686984366697e-07
-3.4533640832612 4.01422033271285e-07
-3.38530061702536 6.99935297492419e-07
-3.31723715078951 1.2070027846996e-06
-3.24917368455367 2.05850694438501e-06
-3.18111021831782 3.47208292048374e-06
-3.11304675208198 5.79190631144215e-06
-3.04498328584613 9.55535123822828e-06
-2.97691981961029 1.55906948255169e-05
-2.90885635337444 2.51581077853315e-05
-2.8407928871386 4.01498718916337e-05
-2.77272942090275 6.33700488651422e-05
-2.70466595466691 9.89185195238932e-05
-2.63660248843106 0.00015270908019641
-2.56853902219522 0.000233155573980608
-2.50047555595937 0.000352063037242453
-2.43241208972353 0.000525761460084459
-2.36434862348768 0.000776516613820324
-2.29628515725184 0.00113424387811753
-2.228221691016 0.00163853538066419
-2.16015822478015 0.00234098640032248
-2.09209475854431 0.00330777260194311
-2.02403129230846 0.00462238471333362
-1.95596782607262 0.00638837230913405
-1.88790435983677 0.00873188558574487
-1.81984089360093 0.0118037374825211
-1.75177742736508 0.0157806444684557
-1.68371396112924 0.0208652511446592
-1.61565049489339 0.027284511659507
-1.54758702865755 0.0352860009528181
-1.4795235624217 0.0451317720084593
-1.41146009618586 0.0570894708021805
-1.34339662995001 0.0714205740708776
-1.27533316371417 0.0883658264854737
-1.20726969747832 0.108128216217906
-1.13920623124248 0.130854125974172
-1.07114276500663 0.15661360667349
-1.00307929877079 0.18538101210487
-0.935015832534944 0.21701746911788
-0.866952366299099 0.25125680187332
-0.798888900063254 0.28769654631845
-0.73082543382741 0.325795556657264
-0.662761967591565 0.364879406745641
-0.59469850135572 0.404154330863887
-0.526635035119876 0.442729854417416
-0.458571568884031 0.479649579213016
-0.390508102648186 0.513928869799113
-0.322444636412341 0.544597507493511
-0.254381170176496 0.570744810846227
-0.186317703940651 0.591564333336821
-0.118254237704807 0.606395094305955
-0.0501907714689622 0.614756408262821
0.0178726947668828 0.616373753914636
0.0859361610027274 0.611193741125615
0.153999627238572 0.59938703757565
0.222063093474417 0.581339031339991
0.290126559710262 0.557628941981202
0.358190025946107 0.528998959450144
0.426253492181951 0.496315703843709
0.494316958417796 0.46052679461308
0.562380424653641 0.42261555518641
0.630443890889485 0.383556847265537
0.69850735712533 0.34427674639425
0.766570823361175 0.305618280183448
0.83463428959702 0.26831481483785
0.902697755832865 0.232971966465088
0.97076122206871 0.200058204160231
1.03882468830455 0.169903667758696
1.1068881545404 0.142706196139332
1.17495162077624 0.118543185556589
1.24301508701209 0.0973876855708481
1.31107855324793 0.0791270880938057
1.37914201948378 0.063582852906682
1.44720548571962 0.0505299100004922
1.51526895195547 0.0397146488251843
1.58333241819131 0.0308707098805193
1.65139588442716 0.0237321013107046
1.719459350663 0.0180434449384941
1.78752281689885 0.0135673931922153
1.85558628313469 0.0100894397603656
1.92364974937054 0.00742046942096296
1.99171321560638 0.00539745953958464
2.05977668184223 0.00388276493483397
2.12784014807807 0.00276239952045854
2.19590361431392 0.00194368359297648
2.26396708054976 0.00135256559191235
2.3320305467856 0.000930860901618777
2.40009401302145 0.000633585064509498
2.46815747925729 0.000426499730000141
2.53622094549314 0.000283939803687997
2.60428441172898 0.000186950866572067
2.67234787796483 0.000121736938330852
2.74041134420067 7.83990780076549e-05
2.80847481043652 4.99336396923617e-05
2.87653827667236 3.14535144849892e-05
};
\addlegendentry{Gaussian}
\addplot [very thick, blue, dashed]
table {%
-3.02418443686854 1.92556687011483e-07
-2.96614505600445 1.86138130777767e-06
-2.90810567514037 5.51995836099586e-06
-2.85006629427629 9.62783435057417e-06
-2.7920269134122 1.54045349609187e-05
-2.73398753254812 2.06035655102287e-05
-2.67594815168403 2.95574514562629e-05
-2.61790877081995 4.86526562515681e-05
-2.55986938995587 8.60086535317959e-05
-2.50183000909178 0.00015125327764752
-2.4437906282277 0.000264412424047935
-2.38575124736361 0.000406936465217602
-2.32771186649953 0.000603151729282303
-2.26967248563544 0.000930498097201825
-2.21163310477136 0.00142671667963043
-2.15359372390728 0.00201119041027261
-2.09555434304319 0.00265035024002606
-2.03751496217911 0.00339649135580497
-1.97947558131502 0.00448435640546693
-1.92143620045094 0.00599420943471456
-1.86339681958686 0.00799569177868367
-1.80535743872277 0.0104724040261971
-1.74731805785869 0.0137560732098039
-1.6892786769946 0.0180140952760795
-1.63123929613052 0.0233644432885995
-1.57319991526643 0.0303518440685286
-1.51516053440235 0.0394237672637006
-1.45712115353827 0.0506173763147086
-1.39908177267418 0.0640672362529926
-1.3410423918101 0.0795789931957958
-1.28300301094601 0.0969305009118684
-1.22496363008193 0.11631423585627
-1.16692424921784 0.136620108728379
-1.10888486835376 0.158198107353229
-1.05084548748968 0.179652067378863
-0.992806106625592 0.204629654312901
-0.934766725761508 0.232129846171524
-0.876727344897424 0.26055008992708
-0.81868796403334 0.289797236282221
-0.760648583169256 0.320044810904733
-0.702609202305172 0.352906037672005
-0.644569821441087 0.383889965112039
-0.586530440577003 0.413232635630331
-0.528491059712919 0.437445579960247
-0.470451678848835 0.460123735169947
-0.41241229798475 0.480645913387132
-0.354372917120666 0.49799701994344
-0.296333536256582 0.512514510432859
-0.238294155392498 0.526455357830245
-0.180254774528414 0.537425408567629
-0.12221539366433 0.548806615970958
-0.0641760128002455 0.558309256382193
-0.00613663193616132 0.565498103549518
0.0519027489279229 0.568444894809198
0.109942129792007 0.567568665604952
0.167981510656091 0.565814185351247
0.226020891520176 0.561134688789885
0.28406027238426 0.553529357554945
0.342099653248344 0.543710186043048
0.400139034112428 0.530419570484918
0.458178414976512 0.515504851596585
0.516217795840596 0.499538532501311
0.57425717670468 0.481343514171393
0.632296557568765 0.455226777923386
0.690335938432849 0.421510631558564
0.748375319296933 0.381182249025674
0.806414700161017 0.338685180758927
0.864454081025102 0.29656189711445
0.922493461889186 0.256353316933662
0.98053284275327 0.217618453310527
1.03857222361735 0.182716270078449
1.09661160448144 0.150169215592427
1.15465098534552 0.121645665174291
1.21269036620961 0.0976810066458862
1.27072974707369 0.0760978250368784
1.32876912793777 0.0586694872026867
1.38680850880186 0.0445756214246928
1.44484788966594 0.0333622432204849
1.50288727053003 0.0249177338971587
1.56092665139411 0.018326695011052
1.6189660322582 0.012974613988349
1.67700541312228 0.00888509506959912
1.73504479398636 0.005842378487006
1.79308417485045 0.00363360886946903
1.85112355571453 0.00215658675535686
1.90916293657862 0.00130956198197451
1.9672023174427 0.000795804694637292
2.02524169830679 0.000517271446875181
2.08328107917087 0.000351319675452451
2.14132046003495 0.000240278652609163
2.19935984089904 0.000152954195049455
2.25739922176312 9.51871889460099e-05
2.31543860262721 5.53600475158015e-05
2.37347798349129 3.2670451229615e-05
2.43151736435537 1.72659162686963e-05
2.48955674521946 9.30690653888836e-06
2.54759612608354 4.33252545775844e-06
2.60563550694763 1.92556687011483e-06
2.66367488781171 5.7767006103445e-07
2.72171426867579 1.60463905842903e-07
};
\addlegendentry{MPS}
\end{axis}

% \draw ({$(current bounding box.south west)!0.5!(current bounding box.south east)$}|-{$(current bounding box.south west)!-0.2!(current bounding box.north west)$}) node[
%   scale=0.6,
%   fill=white,
%   draw=black,
%   line width=0.4pt,
%   inner sep=3.3pt,
%   fill opacity=0.5,
%   anchor=base,
%   text=black,
%   rotate=0.0
% ]{DNS Normal Fit: $\mu = -0.0000$, $\sigma = 0.6470$};
\end{tikzpicture}

%% file: C4/PDF_Velocity_u_1024Cubed_SameLimits.tex
% This file was created by tikzplotlib v0.9.6.
\begin{tikzpicture}

\begin{axis}[
legend cell align={left},
legend style={fill opacity=0.8, draw opacity=1, text opacity=1, at={(0.5,0.09)}, anchor=south, draw=white!80!black},
log basis y={10},
tick pos=left,
x grid style={white!69.0196078431373!black},
xlabel={\(\displaystyle u\)},
xmin=-4, xmax=4,
xtick style={color=black},
y grid style={white!69.0196078431373!black},
ymin=1e-08, ymax=1,
ymode=log,
ytick style={color=black},
ytick={1e-10,1e-08,1e-06,0.0001,0.01,1,100},
yticklabels={,,,,,,}
]
\addplot [very thick, black]
table {%
-3.82805346488953 8.29281779528482e-08
-3.76067063331604 1.65856355905696e-07
-3.69328780174255 1.65856355905696e-07
-3.62590497016907 3.17891348819251e-07
-3.55852213859558 4.00819526772099e-07
-3.49113930702209 4.83747704724948e-07
-3.42375647544861 6.0813997165422e-07
-3.35637364387512 6.91068149607068e-07
-3.28899081230164 8.70745868504906e-07
-3.22160798072815 1.11953040236345e-06
-3.15422514915466 1.29920812126129e-06
-3.08684231758118 1.85206264094694e-06
-3.01945948600769 2.36345307165619e-06
-2.9520766544342 5.74968700473077e-06
-2.88469382286072 1.05318786000118e-05
-2.81731099128723 2.03865104134084e-05
-2.74992815971375 3.55900097047642e-05
-2.68254532814026 5.4497634278013e-05
-2.61516249656677 9.35844488197898e-05
-2.54777966499329 0.000155531797750566
-2.4803968334198 0.000260726191483756
-2.41301400184631 0.00043162734488158
-2.34563117027283 0.000701973205007872
-2.27824833869934 0.00108960715148546
-2.21086550712585 0.00159979512361438
-2.14348267555237 0.00222729882482059
-2.07609984397888 0.00311066359773732
-2.0087170124054 0.00434959675498988
-1.94133418083191 0.0060098188776059
-1.87395134925842 0.00836363845925655
-1.80656851768494 0.0113805793945442
-1.73918568611145 0.0154459536409636
-1.67180285453796 0.0210303924297604
-1.60442002296448 0.028030995713458
-1.53703719139099 0.0376147149908458
-1.4696543598175 0.0502142280303929
-1.40227152824402 0.0649747111737613
-1.33488869667053 0.0828873496465694
-1.26750586509705 0.10392139091616
-1.20012303352356 0.127701128872317
-1.13274020195007 0.15239314589277
-1.06535737037659 0.175925370880464
-0.9979745388031 0.203734810145158
-0.930591707229614 0.23467514748301
-0.863208875656128 0.267841733222095
-0.795826044082641 0.302967828878028
-0.728443212509155 0.338573871363862
-0.661060380935669 0.373496696234272
-0.593677549362183 0.405694335320791
-0.526294717788696 0.43598256693131
-0.45891188621521 0.460485936168284
-0.391529054641724 0.484513933446506
-0.324146223068237 0.503694004727056
-0.256763391494751 0.519113822170602
-0.189380559921265 0.533056205017102
-0.121997728347778 0.546188158423196
-0.0546148967742919 0.556947481372606
0.0127679347991945 0.56583529502334
0.0801507663726808 0.566962648317157
0.147533597946167 0.564841552845568
0.214916429519654 0.56039086664165
0.28229926109314 0.551128839587911
0.349682092666626 0.538539968997861
0.417064924240113 0.522481728976352
0.484447755813599 0.504747994226114
0.551830587387085 0.483805685342696
0.619213418960571 0.455751442096692
0.686596250534058 0.418027178982765
0.753979082107544 0.374295432801588
0.82136191368103 0.327010767049642
0.888744745254517 0.28141265311682
0.956127576828003 0.236429562194416
1.02351040840149 0.194828987258447
1.09089323997498 0.157134992929476
1.15827607154846 0.123890896986733
1.22565890312195 0.096158318321637
1.29304173469543 0.0725735721545741
1.36042456626892 0.0539289957617903
1.42780739784241 0.0391861348562376
1.49519022941589 0.0282266232852485
1.56257306098938 0.0200330705189736
1.62995589256287 0.0136901567932569
1.69733872413635 0.00893750145069032
1.76472155570984 0.00552612645833288
1.83210438728333 0.00319617637056972
1.89948721885681 0.00180898145250043
1.9668700504303 0.00106671897437049
2.03425288200378 0.000644103158159767
2.10163571357727 0.000413645751628809
2.16901854515076 0.000280186670576688
2.23640137672424 0.000171509293369483
2.30378420829773 9.79105354363287e-05
2.37116703987122 5.31707834307682e-05
2.4385498714447 2.36483520795537e-05
2.50593270301819 1.23010130630059e-05
2.57331553459167 7.17328739292141e-06
2.64069836616516 3.44151938504318e-06
2.70808119773865 1.39595766220629e-06
2.77546402931213 4.97569067717086e-07
2.84284686088562 1.24392266929273e-07
};
% \addlegendentry{DNS}
\addplot [very thick, red, dotted]
table {%
-3.86174488067627 1.13221398160252e-08
-3.79368141444042 2.10971153085155e-08
-3.72561794820458 3.88786635039558e-08
-3.65755448196874 7.08587138863232e-08
-3.58949101573289 1.27722935468882e-07
-3.52142754949705 2.27686984366697e-07
-3.4533640832612 4.01422033271285e-07
-3.38530061702536 6.99935297492419e-07
-3.31723715078951 1.2070027846996e-06
-3.24917368455367 2.05850694438501e-06
-3.18111021831782 3.47208292048374e-06
-3.11304675208198 5.79190631144215e-06
-3.04498328584613 9.55535123822828e-06
-2.97691981961029 1.55906948255169e-05
-2.90885635337444 2.51581077853315e-05
-2.8407928871386 4.01498718916337e-05
-2.77272942090275 6.33700488651422e-05
-2.70466595466691 9.89185195238932e-05
-2.63660248843106 0.00015270908019641
-2.56853902219522 0.000233155573980608
-2.50047555595937 0.000352063037242453
-2.43241208972353 0.000525761460084459
-2.36434862348768 0.000776516613820324
-2.29628515725184 0.00113424387811753
-2.228221691016 0.00163853538066419
-2.16015822478015 0.00234098640032248
-2.09209475854431 0.00330777260194311
-2.02403129230846 0.00462238471333362
-1.95596782607262 0.00638837230913405
-1.88790435983677 0.00873188558574487
-1.81984089360093 0.0118037374825211
-1.75177742736508 0.0157806444684557
-1.68371396112924 0.0208652511446592
-1.61565049489339 0.027284511659507
-1.54758702865755 0.0352860009528181
-1.4795235624217 0.0451317720084593
-1.41146009618586 0.0570894708021805
-1.34339662995001 0.0714205740708776
-1.27533316371417 0.0883658264854737
-1.20726969747832 0.108128216217906
-1.13920623124248 0.130854125974172
-1.07114276500663 0.15661360667349
-1.00307929877079 0.18538101210487
-0.935015832534944 0.21701746911788
-0.866952366299099 0.25125680187332
-0.798888900063254 0.28769654631845
-0.73082543382741 0.325795556657264
-0.662761967591565 0.364879406745641
-0.59469850135572 0.404154330863887
-0.526635035119876 0.442729854417416
-0.458571568884031 0.479649579213016
-0.390508102648186 0.513928869799113
-0.322444636412341 0.544597507493511
-0.254381170176496 0.570744810846227
-0.186317703940651 0.591564333336821
-0.118254237704807 0.606395094305955
-0.0501907714689622 0.614756408262821
0.0178726947668828 0.616373753914636
0.0859361610027274 0.611193741125615
0.153999627238572 0.59938703757565
0.222063093474417 0.581339031339991
0.290126559710262 0.557628941981202
0.358190025946107 0.528998959450144
0.426253492181951 0.496315703843709
0.494316958417796 0.46052679461308
0.562380424653641 0.42261555518641
0.630443890889485 0.383556847265537
0.69850735712533 0.34427674639425
0.766570823361175 0.305618280183448
0.83463428959702 0.26831481483785
0.902697755832865 0.232971966465088
0.97076122206871 0.200058204160231
1.03882468830455 0.169903667758696
1.1068881545404 0.142706196139332
1.17495162077624 0.118543185556589
1.24301508701209 0.0973876855708481
1.31107855324793 0.0791270880938057
1.37914201948378 0.063582852906682
1.44720548571962 0.0505299100004922
1.51526895195547 0.0397146488251843
1.58333241819131 0.0308707098805193
1.65139588442716 0.0237321013107046
1.719459350663 0.0180434449384941
1.78752281689885 0.0135673931922153
1.85558628313469 0.0100894397603656
1.92364974937054 0.00742046942096296
1.99171321560638 0.00539745953958464
2.05977668184223 0.00388276493483397
2.12784014807807 0.00276239952045854
2.19590361431392 0.00194368359297648
2.26396708054976 0.00135256559191235
2.3320305467856 0.000930860901618777
2.40009401302145 0.000633585064509498
2.46815747925729 0.000426499730000141
2.53622094549314 0.000283939803687997
2.60428441172898 0.000186950866572067
2.67234787796483 0.000121736938330852
2.74041134420067 7.83990780076549e-05
2.80847481043652 4.99336396923617e-05
2.87653827667236 3.14535144849892e-05
};
% \addlegendentry{DNS (Normal Fit)}
\addplot [very thick, blue, dashed]
table {%
-3.74360834481144 4.24661728486555e-08
-3.67781556113386 8.4932345697311e-08
-3.61202277745627 1.98175473293726e-07
-3.54622999377869 3.1141860089014e-07
-3.48043721010111 4.10506337537003e-07
-3.41464442642353 4.95438683234314e-07
-3.34885164274594 7.07769547477592e-07
-3.28305885906836 1.07580971216594e-06
-3.21726607539078 9.62566584569525e-07
-3.1514732917132 1.45800526780384e-06
-3.08568050803561 1.5287822225516e-06
-3.01988772435803 2.59043654376799e-06
-2.95409494068045 4.48725893100796e-06
-2.88830215700286 1.00220167922826e-05
-2.82250937332528 1.66892059295217e-05
-2.7567165896477 2.78578093887178e-05
-2.69092380597011 4.47876569643823e-05
-2.62513102229253 7.97373172188249e-05
-2.55933823861495 0.000138453878877567
-2.49354545493737 0.000231030135687634
-2.42775267125978 0.00037688728403182
-2.3619598875822 0.000610677720954611
-2.29616710390462 0.000962679827697128
-2.23037432022703 0.00142675016458723
-2.16458153654945 0.00199435303088236
-2.09878875287187 0.00274191338231914
-2.03299596919429 0.0038202144432922
-1.9672031855167 0.00525837385298477
-1.90141040183912 0.00723231481011787
-1.83561761816154 0.00990703255159949
-1.76982483448395 0.0133439614741507
-1.70403205080637 0.018046042618209
-1.63823926712879 0.0240889507037908
-1.57244648345121 0.0320686276898721
-1.50665369977362 0.0427875710001651
-1.44086091609604 0.0559916347455613
-1.37506813241846 0.0715051222135951
-1.30927534874087 0.0902692678146748
-1.24348256506329 0.111786721887788
-1.17768978138571 0.135831578548771
-1.11189699770813 0.159017443620762
-1.04610421403054 0.183209233239802
-0.98031143035296 0.211462657494778
-0.914518646675377 0.242543323828717
-0.848725862997794 0.275748119817764
-0.782933079320211 0.310163116800652
-0.717140295642628 0.344274197539853
-0.651347511965045 0.378767728631731
-0.585554728287463 0.409725979144736
-0.51976194460988 0.438571480936959
-0.453969160932297 0.462697433832692
-0.388176377254714 0.485608217192361
-0.322383593577131 0.504347378528415
-0.256590809899548 0.519569420652188
-0.190798026221965 0.533284394923126
-0.125005242544382 0.545545582232764
-0.0592124588667995 0.556684487681748
0.00658032481078341 0.565063799665117
0.0723731084883663 0.567025099858132
0.138165892165949 0.565453044605448
0.203958675843532 0.562095980398631
0.269751459521115 0.553567980188735
0.335544243198698 0.541595024949605
0.401337026876281 0.526833117964045
0.467129810553863 0.510018608513837
0.532922594231446 0.49049461568197
0.598715377909029 0.465595141447805
0.664508161586612 0.431741632202316
0.730300945264195 0.390644823563344
0.796093728941778 0.344728330792294
0.861886512619361 0.299212364808171
0.927679296296944 0.255153229194395
0.993472079974527 0.212703773862454
1.05926486365211 0.174133900683442
1.12505764732969 0.139251138806628
1.19085043100728 0.109523119165654
1.25664321468486 0.084381432036946
1.32243599836244 0.0635272288067729
1.38822878204002 0.047096740957637
1.45402156571761 0.0342505113400539
1.51981434939519 0.0248035431497062
1.58560713307277 0.0175260443316771
1.65139991675036 0.0118919014705457
1.71719270042794 0.00768408411227277
1.78298548410552 0.00467862566125495
1.8487782677831 0.00269702663761809
1.91457105146069 0.00155978252873113
1.98036383513827 0.000930816042669674
2.04615661881585 0.000573137624156408
2.11194940249344 0.000375330191027364
2.17774218617102 0.000260020376352319
2.2435349698486 0.000152708357563764
2.30932775352618 8.87967674265392e-05
2.37512053720377 4.70808302982091e-05
2.44091332088135 2.0567783049699e-05
2.50670610455893 1.00220167922828e-05
2.57249888823652 6.76627687388573e-06
2.6382916719141 3.19911835459874e-06
2.70408445559168 1.08996510311548e-06
2.76987723926926 3.11418600890145e-07
};
% \addlegendentry{Comp cutoff 0.0001}
\end{axis}

% \draw ({$(current bounding box.south west)!0.5!(current bounding box.south east)$}|-{$(current bounding box.south west)!-0.2!(current bounding box.north west)$}) node[
%   scale=0.6,
%   fill=white,
%   draw=black,
%   line width=0.4pt,
%   inner sep=3.3pt,
%   fill opacity=0.5,
%   anchor=base,
%   text=black,
%   rotate=0.0
% ]{DNS Normal Fit: $\mu = -0.0000$, $\sigma = 0.6470$};
\end{tikzpicture}

%% file: C5/PDF_Velocity_u_1024Cubed_SameLimits.tex
% This file was created by tikzplotlib v0.9.6.
\begin{tikzpicture}

\begin{axis}[
legend cell align={left},
legend style={fill opacity=0.8, draw opacity=1, text opacity=1, at={(0.5,0.09)}, anchor=south, draw=white!80!black},
log basis y={10},
tick pos=left,
x grid style={white!69.0196078431373!black},
xlabel={\(\displaystyle u\)},
xmin=-4, xmax=4,
xtick style={color=black},
y grid style={white!69.0196078431373!black},
ymin=1e-08, ymax=1,
ymode=log,
ytick style={color=black},
ytick={1e-10,1e-08,1e-06,0.0001,0.01,1,100},
yticklabels={,,,,,,}
]
\addplot [very thick, black]
table {%
-3.82805346488953 8.29281779528482e-08
-3.76067063331604 1.65856355905696e-07
-3.69328780174255 1.65856355905696e-07
-3.62590497016907 3.17891348819251e-07
-3.55852213859558 4.00819526772099e-07
-3.49113930702209 4.83747704724948e-07
-3.42375647544861 6.0813997165422e-07
-3.35637364387512 6.91068149607068e-07
-3.28899081230164 8.70745868504906e-07
-3.22160798072815 1.11953040236345e-06
-3.15422514915466 1.29920812126129e-06
-3.08684231758118 1.85206264094694e-06
-3.01945948600769 2.36345307165619e-06
-2.9520766544342 5.74968700473077e-06
-2.88469382286072 1.05318786000118e-05
-2.81731099128723 2.03865104134084e-05
-2.74992815971375 3.55900097047642e-05
-2.68254532814026 5.4497634278013e-05
-2.61516249656677 9.35844488197898e-05
-2.54777966499329 0.000155531797750566
-2.4803968334198 0.000260726191483756
-2.41301400184631 0.00043162734488158
-2.34563117027283 0.000701973205007872
-2.27824833869934 0.00108960715148546
-2.21086550712585 0.00159979512361438
-2.14348267555237 0.00222729882482059
-2.07609984397888 0.00311066359773732
-2.0087170124054 0.00434959675498988
-1.94133418083191 0.0060098188776059
-1.87395134925842 0.00836363845925655
-1.80656851768494 0.0113805793945442
-1.73918568611145 0.0154459536409636
-1.67180285453796 0.0210303924297604
-1.60442002296448 0.028030995713458
-1.53703719139099 0.0376147149908458
-1.4696543598175 0.0502142280303929
-1.40227152824402 0.0649747111737613
-1.33488869667053 0.0828873496465694
-1.26750586509705 0.10392139091616
-1.20012303352356 0.127701128872317
-1.13274020195007 0.15239314589277
-1.06535737037659 0.175925370880464
-0.9979745388031 0.203734810145158
-0.930591707229614 0.23467514748301
-0.863208875656128 0.267841733222095
-0.795826044082641 0.302967828878028
-0.728443212509155 0.338573871363862
-0.661060380935669 0.373496696234272
-0.593677549362183 0.405694335320791
-0.526294717788696 0.43598256693131
-0.45891188621521 0.460485936168284
-0.391529054641724 0.484513933446506
-0.324146223068237 0.503694004727056
-0.256763391494751 0.519113822170602
-0.189380559921265 0.533056205017102
-0.121997728347778 0.546188158423196
-0.0546148967742919 0.556947481372606
0.0127679347991945 0.56583529502334
0.0801507663726808 0.566962648317157
0.147533597946167 0.564841552845568
0.214916429519654 0.56039086664165
0.28229926109314 0.551128839587911
0.349682092666626 0.538539968997861
0.417064924240113 0.522481728976352
0.484447755813599 0.504747994226114
0.551830587387085 0.483805685342696
0.619213418960571 0.455751442096692
0.686596250534058 0.418027178982765
0.753979082107544 0.374295432801588
0.82136191368103 0.327010767049642
0.888744745254517 0.28141265311682
0.956127576828003 0.236429562194416
1.02351040840149 0.194828987258447
1.09089323997498 0.157134992929476
1.15827607154846 0.123890896986733
1.22565890312195 0.096158318321637
1.29304173469543 0.0725735721545741
1.36042456626892 0.0539289957617903
1.42780739784241 0.0391861348562376
1.49519022941589 0.0282266232852485
1.56257306098938 0.0200330705189736
1.62995589256287 0.0136901567932569
1.69733872413635 0.00893750145069032
1.76472155570984 0.00552612645833288
1.83210438728333 0.00319617637056972
1.89948721885681 0.00180898145250043
1.9668700504303 0.00106671897437049
2.03425288200378 0.000644103158159767
2.10163571357727 0.000413645751628809
2.16901854515076 0.000280186670576688
2.23640137672424 0.000171509293369483
2.30378420829773 9.79105354363287e-05
2.37116703987122 5.31707834307682e-05
2.4385498714447 2.36483520795537e-05
2.50593270301819 1.23010130630059e-05
2.57331553459167 7.17328739292141e-06
2.64069836616516 3.44151938504318e-06
2.70808119773865 1.39595766220629e-06
2.77546402931213 4.97569067717086e-07
2.84284686088562 1.24392266929273e-07
};
% \addlegendentry{DNS}
\addplot [very thick, red, dotted]
table {%
-3.86174488067627 1.13221398160252e-08
-3.79368141444042 2.10971153085155e-08
-3.72561794820458 3.88786635039558e-08
-3.65755448196874 7.08587138863232e-08
-3.58949101573289 1.27722935468882e-07
-3.52142754949705 2.27686984366697e-07
-3.4533640832612 4.01422033271285e-07
-3.38530061702536 6.99935297492419e-07
-3.31723715078951 1.2070027846996e-06
-3.24917368455367 2.05850694438501e-06
-3.18111021831782 3.47208292048374e-06
-3.11304675208198 5.79190631144215e-06
-3.04498328584613 9.55535123822828e-06
-2.97691981961029 1.55906948255169e-05
-2.90885635337444 2.51581077853315e-05
-2.8407928871386 4.01498718916337e-05
-2.77272942090275 6.33700488651422e-05
-2.70466595466691 9.89185195238932e-05
-2.63660248843106 0.00015270908019641
-2.56853902219522 0.000233155573980608
-2.50047555595937 0.000352063037242453
-2.43241208972353 0.000525761460084459
-2.36434862348768 0.000776516613820324
-2.29628515725184 0.00113424387811753
-2.228221691016 0.00163853538066419
-2.16015822478015 0.00234098640032248
-2.09209475854431 0.00330777260194311
-2.02403129230846 0.00462238471333362
-1.95596782607262 0.00638837230913405
-1.88790435983677 0.00873188558574487
-1.81984089360093 0.0118037374825211
-1.75177742736508 0.0157806444684557
-1.68371396112924 0.0208652511446592
-1.61565049489339 0.027284511659507
-1.54758702865755 0.0352860009528181
-1.4795235624217 0.0451317720084593
-1.41146009618586 0.0570894708021805
-1.34339662995001 0.0714205740708776
-1.27533316371417 0.0883658264854737
-1.20726969747832 0.108128216217906
-1.13920623124248 0.130854125974172
-1.07114276500663 0.15661360667349
-1.00307929877079 0.18538101210487
-0.935015832534944 0.21701746911788
-0.866952366299099 0.25125680187332
-0.798888900063254 0.28769654631845
-0.73082543382741 0.325795556657264
-0.662761967591565 0.364879406745641
-0.59469850135572 0.404154330863887
-0.526635035119876 0.442729854417416
-0.458571568884031 0.479649579213016
-0.390508102648186 0.513928869799113
-0.322444636412341 0.544597507493511
-0.254381170176496 0.570744810846227
-0.186317703940651 0.591564333336821
-0.118254237704807 0.606395094305955
-0.0501907714689622 0.614756408262821
0.0178726947668828 0.616373753914636
0.0859361610027274 0.611193741125615
0.153999627238572 0.59938703757565
0.222063093474417 0.581339031339991
0.290126559710262 0.557628941981202
0.358190025946107 0.528998959450144
0.426253492181951 0.496315703843709
0.494316958417796 0.46052679461308
0.562380424653641 0.42261555518641
0.630443890889485 0.383556847265537
0.69850735712533 0.34427674639425
0.766570823361175 0.305618280183448
0.83463428959702 0.26831481483785
0.902697755832865 0.232971966465088
0.97076122206871 0.200058204160231
1.03882468830455 0.169903667758696
1.1068881545404 0.142706196139332
1.17495162077624 0.118543185556589
1.24301508701209 0.0973876855708481
1.31107855324793 0.0791270880938057
1.37914201948378 0.063582852906682
1.44720548571962 0.0505299100004922
1.51526895195547 0.0397146488251843
1.58333241819131 0.0308707098805193
1.65139588442716 0.0237321013107046
1.719459350663 0.0180434449384941
1.78752281689885 0.0135673931922153
1.85558628313469 0.0100894397603656
1.92364974937054 0.00742046942096296
1.99171321560638 0.00539745953958464
2.05977668184223 0.00388276493483397
2.12784014807807 0.00276239952045854
2.19590361431392 0.00194368359297648
2.26396708054976 0.00135256559191235
2.3320305467856 0.000930860901618777
2.40009401302145 0.000633585064509498
2.46815747925729 0.000426499730000141
2.53622094549314 0.000283939803687997
2.60428441172898 0.000186950866572067
2.67234787796483 0.000121736938330852
2.74041134420067 7.83990780076549e-05
2.80847481043652 4.99336396923617e-05
2.87653827667236 3.14535144849892e-05
};
% \addlegendentry{DNS (Normal Fit)}
\addplot [very thick, blue, dashed]
table {%
-3.85182321949955 4.15637876355713e-08
-3.78460201847851 9.6982171149667e-08
-3.71738081745747 1.80109746420809e-07
-3.65015961643644 2.21673534056382e-07
-3.5829384154154 3.87928684598665e-07
-3.51571721439436 4.57201663991284e-07
-3.44849601337333 5.81893026898002e-07
-3.38127481235229 6.09602218655046e-07
-3.31405361133125 8.45130348589955e-07
-3.24683241031021 1.06680388264633e-06
-3.17961120928918 1.23305903318862e-06
-3.11239000826814 1.5101509507591e-06
-3.0451688072471 2.10589857353561e-06
-2.97794760622607 3.54677654490211e-06
-2.91072640520503 8.28504833535721e-06
-2.84350520418399 1.65423874789574e-05
-2.77628400316296 2.73074084765705e-05
-2.70906280214192 4.52906739268945e-05
-2.64184160112088 7.55075475379545e-05
-2.57462040009984 0.000127365299911269
-2.50739919907881 0.000212307827242498
-2.44017799805777 0.000351615788801057
-2.37295679703673 0.000574979583554619
-2.3057355960157 0.000915774932974542
-2.23851439499466 0.00137840759855021
-2.17129319397362 0.00194826098162977
-2.10407199295259 0.00269244674464881
-2.03685079193155 0.00378499246643745
-1.96962959091051 0.00523756371672537
-1.90240838988948 0.00726999136831717
-1.83518718886844 0.010029729885148
-1.7679659878474 0.0134999320786294
-1.70074478682636 0.0184770571020303
-1.63352358580533 0.0247075658507738
-1.56630238478429 0.0331054044511779
-1.49908118376325 0.0443826991313995
-1.43185998274222 0.0581335933597723
-1.36463878172118 0.0745677764449184
-1.29741758070014 0.0941826422334731
-1.23019637967911 0.116745128713566
-1.16297517865807 0.141396320213711
-1.09575397763703 0.165135519085554
-1.02853277661599 0.190542714319001
-0.961311575594958 0.220179759237951
-0.894090374573921 0.252389630320731
-0.826869173552884 0.286659333445755
-0.759647972531847 0.322392511388019
-0.69242677151081 0.357416237039135
-0.625205570489773 0.390826331762871
-0.557984369468736 0.422397755358379
-0.490763168447699 0.449132137749416
-0.423541967426663 0.473446856534051
-0.356320766405626 0.495149582212311
-0.289099565384589 0.511983387260974
-0.221878364363552 0.526576488518206
-0.154657163342515 0.539986310305245
-0.0874359623214778 0.551990527922017
-0.0202147613004409 0.562048645874124
0.047006439720596 0.567559699313488
0.114227640741633 0.565710581819965
0.18144884176267 0.563230678430689
0.248670042783707 0.556188955384073
0.315891243804744 0.545435696938354
0.383112444825781 0.530878853158535
0.450333645846818 0.514138123284981
0.517554846867855 0.49491046574204
0.584776047888892 0.471394256590867
0.651997248909928 0.438490381091332
0.719218449930966 0.397560455071799
0.786439650952003 0.351455865200829
0.85366085197304 0.304994838141692
0.920882052994076 0.259702945010364
0.988103254015113 0.216168867585713
1.05532445503615 0.176565214320439
1.12254565605719 0.140774456677702
1.18976685707822 0.110377861248906
1.25698805809926 0.0845683008419555
1.3242092591203 0.0633722792987638
1.39143046014133 0.046699464654206
1.45865166116237 0.0336766571484412
1.52587286218341 0.0242234308523945
1.59309406320445 0.016919287905237
1.66031526422548 0.0113724341901192
1.72753646524652 0.00723722524906446
1.79475766626756 0.00437343871718596
1.86197886728859 0.00246035455449176
1.92920006830963 0.00142141226415716
1.99642126933067 0.000847679594231598
2.0636424703517 0.000528982325237916
2.13086367137274 0.000346129368833159
2.19808487239378 0.000230803712740327
2.26530607341482 0.000132837865283288
2.33252727443585 7.62834049071518e-05
2.39974847545689 3.72827175091075e-05
2.46696967647793 1.72766810605191e-05
2.53419087749896 8.89465055401237e-06
2.60141207852 5.37558320086722e-06
2.66863327954104 2.70164619631213e-06
2.73585448056207 9.42112519739616e-07
2.80307568158311 2.77091917570477e-07
};
% \addlegendentry{Comp cutoff 1e-05}
\end{axis}

% \draw ({$(current bounding box.south west)!0.5!(current bounding box.south east)$}|-{$(current bounding box.south west)!-0.2!(current bounding box.north west)$}) node[
%   scale=0.6,
%   fill=white,
%   draw=black,
%   line width=0.4pt,
%   inner sep=3.3pt,
%   fill opacity=0.5,
%   anchor=base,
%   text=black,
%   rotate=0.0
% ]{DNS Normal Fit: $\mu = -0.0000$, $\sigma = 0.6470$};
\end{tikzpicture}

%% file: C3/PDF_grad_dux_dx_SameLimits.tex
% This file was created by tikzplotlib v0.9.6.
\begin{tikzpicture}

\begin{axis}[
legend cell align={left},
legend style={fill opacity=0.8, text opacity=1, draw=none},
log basis y={10},
tick pos=left,
x grid style={white!69.0196078431373!black},
xlabel={\(\displaystyle \partial u / \partial x\)},
xmin=-200, xmax=200,
xtick style={color=black},
y grid style={white!69.0196078431373!black},
ylabel={PDF},
ymin=1e-08, ymax=1,
ymode=log,
ytick style={color=black},
ytick={1e-10,1e-08,1e-06,0.0001,0.01,1,100},
yticklabels={\(\displaystyle 10^{-10}\),,\(\displaystyle 10^{-6}\),,\(\displaystyle 10^{-2}\),,\(\displaystyle 10^{2}\)}
]
\addplot [very thick, black]
table {%
-162.087262675767 3.2796837341557e-10
-159.247590774125 0
-156.407918872483 0
-153.568246970841 0
-150.728575069199 3.27968373415573e-10
-147.888903167557 0
-145.049231265914 6.55936746831146e-10
-142.209559364272 3.2796837341557e-10
-139.36988746263 3.2796837341557e-10
-136.530215560988 0
-133.690543659346 3.2796837341557e-10
-130.850871757704 6.5593674683114e-10
-128.011199856061 3.2796837341557e-10
-125.171527954419 1.31187349366228e-09
-122.331856052777 3.27968373415572e-09
-119.492184151135 5.24749397464912e-09
-116.652512249493 6.5593674683114e-09
-113.812840347851 6.55936746831143e-09
-110.973168446209 6.23139909489583e-09
-108.133496544566 1.44306084302851e-08
-105.293824642924 1.31187349366228e-08
-102.454152741282 1.37746716834539e-08
-99.6144808396399 2.3613722885921e-08
-96.7748089379977 2.65654382466612e-08
-93.9351370363555 3.87002680630375e-08
-91.0954651347134 5.4114781613569e-08
-88.2557932330712 7.96963147399835e-08
-85.4161213314291 1.02982069252489e-07
-82.5764494297869 1.37418748461124e-07
-79.7367775281447 1.78742763511487e-07
-76.8971056265026 2.63358603852703e-07
-74.0574337248604 3.28952278535817e-07
-71.2177618232183 4.93920370363848e-07
-68.3780899215761 6.9890060374858e-07
-65.5384180199339 9.2651065489899e-07
-62.6987461182918 1.33614315329503e-06
-59.8590742166496 1.89237751460784e-06
-57.0194023150075 2.76838104000083e-06
-54.1797304133653 4.08156640715677e-06
-51.3400585117231 5.93491568532818e-06
-48.500386610081 8.89614212889733e-06
-45.6607147084388 1.35342708657403e-05
-42.8210428067967 2.06718465763834e-05
-39.9813709051545 3.22219087829595e-05
-37.1416990035123 5.051598465208e-05
-34.3020271018702 8.07271193376286e-05
-31.462355200228 0.000131042715313552
-28.6226832985858 0.000216078683141114
-25.7830113969437 0.000364289886866228
-22.9433394953015 0.000628792772373774
-20.1036675936594 0.00110749049068554
-17.2639956920172 0.00200225347906952
-14.4243237903751 0.00372829920935267
-11.5846518887329 0.00713308779142123
-8.74497998709073 0.0140824752493984
-5.90530808544857 0.0283876984188114
-3.0656361838064 0.0559935658230039
-0.225964282164242 0.0907783168824396
2.61370761947791 0.0803380681665387
5.45337952112006 0.0391803065160595
8.29305142276222 0.0159174878158121
11.1327233244044 0.00653152393731354
13.9723952260465 0.00282124560617493
16.8120671276887 0.00129004129498075
19.6517390293309 0.000618188570956128
22.491410930973 0.000310768400040164
25.3310828326152 0.000162273503672049
28.1707547342574 8.78745340994742e-05
31.0104266358995 4.84415846902265e-05
33.8500985375417 2.75099871620983e-05
36.6897704391838 1.59359832642625e-05
39.529442340826 9.3546419149323e-06
42.3691142424681 5.6472874218427e-06
45.2087861441103 3.44268401574324e-06
48.0484580457525 2.07505589860031e-06
50.8881299473946 1.36828405388976e-06
53.7278018490368 8.39271067570444e-07
56.567473750679 5.34916417040795e-07
59.4071456523211 3.60437242383711e-07
62.2468175539633 2.33185513498473e-07
65.0864894556054 1.4660186291676e-07
67.9261613572476 9.74066069044243e-08
70.7658332588897 5.73944653477247e-08
73.6055051605319 4.03401099301151e-08
76.4451770621741 2.98451219808169e-08
79.2848489638162 1.93501340315186e-08
82.1245208654584 1.08229563227138e-08
84.9641927671006 1.31187349366228e-08
87.8038646687427 7.21530421514254e-09
90.6435365703849 3.93562048098684e-09
93.483208472027 4.59155722781803e-09
96.3228803736692 1.9678102404934e-09
99.1625522753114 2.62374698732459e-09
102.002224176954 1.9678102404934e-09
104.841896078596 3.27968373415573e-10
107.681567980238 0
110.52123988188 0
113.360911783522 3.27968373415573e-10
116.200583685164 3.27968373415567e-10
119.040255586806 3.27968373415573e-10
};
\addlegendentry{DNS}
\addplot [very thick, red, dotted, dotted]
table {%
-163.507098626588 4.36474597384841e-19
-160.638743170384 8.81809200908925e-19
-157.77038771418 1.78151826352917e-18
-154.902032257976 3.59919960011372e-18
-152.033676801772 7.27145942124472e-18
-149.165321345567 1.46905223353376e-17
-146.296965889363 2.96792478624751e-17
-143.428610433159 5.99609553408023e-17
-140.560254976955 1.21139059252489e-16
-137.691899520751 2.44737122568696e-16
-134.823544064546 4.94442168635006e-16
-131.955188608342 9.98921028238636e-16
-129.086833152138 2.01811917339504e-15
-126.218477695934 4.07720418620694e-15
-123.35012223973 8.23717161759967e-15
-120.481766783525 1.66415497382564e-14
-117.613411327321 3.36209066106069e-14
-114.745055871117 6.79242846428305e-14
-111.876700414913 1.3722736562924e-13
-109.008344958709 2.77240312158793e-13
-106.139989502504 5.60108330677794e-13
-103.2716340463 1.13158631099426e-12
-100.403278590096 2.28614271399975e-12
-97.5349231338918 4.61869188235577e-12
-94.6665676776876 9.3311386789222e-12
-91.7982122214834 1.88516903190493e-11
-88.9298567652792 3.8086051457804e-11
-86.061501309075 7.69452123972537e-11
-83.1931458528708 1.55452337121845e-10
-80.3247903966666 3.1406020418635e-10
-77.4564349404624 6.34495522420244e-10
-74.5880794842582 1.28187068149666e-09
-71.719724028054 2.58976208029467e-09
-68.8513685718498 5.23209379022659e-09
-65.9830131156456 1.05703939516378e-08
-63.1146576594414 2.13553565307899e-08
-60.2463022032372 4.31442058492521e-08
-57.377946747033 8.7164196752176e-08
-54.5095912908288 1.76097741189128e-07
-51.6412358346246 3.55770093770054e-07
-48.7728803784204 7.18761971428216e-07
-45.9045249222162 1.4521141057609e-06
-43.036169466012 2.93370470332455e-06
-40.1678140098078 5.92696073412133e-06
-37.2994585536036 1.19742329567142e-05
-34.4311030973994 2.419153109556e-05
-31.5627476411952 4.88741265401303e-05
-28.694392184991 9.87403499028251e-05
-25.8260367287868 0.000199485032043016
-22.9576812725826 0.000403019414538909
-20.0893258163784 0.000814219727825288
-17.2209703601742 0.00164496731736451
-14.35261490397 0.00332332585753564
-11.4842594477658 0.00671411196975021
-8.61590399156157 0.0135645138258488
-5.74754853535737 0.027404373975385
-2.87919307915317 0.0553650298583971
-0.0108376229489693 0.111853915509053
2.85751783325523 0.066499929593055
5.72587328945946 0.0329158847121698
8.59422874566366 0.0162925806540706
11.4625842018679 0.00806444021452208
14.3309396580721 0.00399170624681563
17.1992951142763 0.00197579972533917
20.0676505704805 0.000977973907214384
22.9360060266847 0.000484073841556985
25.8043614828889 0.000239605047078592
28.672716939093 0.000118598803853722
31.5410723952972 5.87035892900865e-05
34.4094278515014 2.90568815499148e-05
37.2777833077057 1.43824658017691e-05
40.1461387639099 7.1189787583956e-06
43.0144942201141 3.5237252958567e-06
45.8828496763183 1.7441602766432e-06
48.7512051325225 8.63317885249763e-07
51.6195605887267 4.27321835597905e-07
54.4879160449309 2.11514152896223e-07
57.3562715011351 1.04694478841245e-07
60.2246269573393 5.18212788589037e-08
63.0929824135435 2.56503014513722e-08
65.9613378697477 1.26962896137254e-08
68.8296933259519 6.28436161895353e-09
71.6980487821561 3.11060964733208e-09
74.5664042383603 1.53967784872422e-09
77.4347596945645 7.62103943156372e-10
80.3031151507687 3.77223339710801e-10
83.1714706069729 1.86716588072256e-10
86.0398260631771 9.24202735919589e-11
88.9081815193813 4.57458389690962e-11
91.7765369755855 2.26431030947366e-11
94.6448924317897 1.12077978962247e-11
97.5132478879939 5.54759359426392e-12
100.381603344198 2.74592698512924e-12
103.249958800402 1.35916859797684e-12
106.118314256606 6.72756154016736e-13
108.986669712811 3.32998307524982e-13
111.855025169015 1.6482624819177e-13
114.723380625219 8.15850755966267e-14
117.591736081423 4.03826734705701e-14
120.460091537628 1.99884636338821e-14
};
\addlegendentry{Laplace}
\addplot [very thick, blue, dashed]
table {%
-130.526592077004 3.63032322453592e-10
-127.961193579368 0
-125.395795081733 0
-122.830396584098 7.26064644907184e-10
-120.264998086463 7.26064644907184e-10
-117.699599588828 7.26064644907184e-10
-115.134201091192 1.08909696736078e-09
-112.568802593557 2.54122625717516e-09
-110.003404095922 2.17819393472155e-09
-107.438005598287 4.7194201918967e-09
-104.872607100651 3.99335554698951e-09
-102.307208603016 9.80187270624699e-09
-99.7418101053808 1.45212928981437e-08
-97.1764116077456 1.52473575430509e-08
-94.6110131101103 2.65013595391124e-08
-92.0456146124751 3.30359413432769e-08
-89.4802161148399 5.00984604985957e-08
-86.9148176172046 5.84482039150287e-08
-84.3494191195694 8.16822725520582e-08
-81.7840206219341 1.18348537119871e-07
-79.2186221242989 1.57192995622405e-07
-76.6532236266637 2.18908490439516e-07
-74.0878251290284 2.99864698346667e-07
-71.5224266313932 4.17850203144085e-07
-68.9570281337579 5.64878293737792e-07
-66.3916296361227 8.12103305328686e-07
-63.8262311384875 1.177313821717e-06
-61.2608326408522 1.65433829342102e-06
-58.695434143217 2.28056904965347e-06
-56.1300356455817 3.17290249824441e-06
-53.5646371479465 4.54988409731087e-06
-50.9992386503113 6.56761774550794e-06
-48.433840152676 9.65520764797574e-06
-45.8684416550408 1.41292179898938e-05
-43.3030431574055 2.10326406336714e-05
-40.7376446597703 3.18205091277023e-05
-38.1722461621351 4.86837235379941e-05
-35.6068476644998 7.48848553464372e-05
-33.0414491668646 0.000118192433221216
-30.4760506692293 0.000188725257086081
-27.9106521715941 0.000307666625988517
-25.3452536739589 0.000512678965189155
-22.7798551763236 0.000870054881026597
-20.2144566786884 0.00150511530940576
-17.6490581810531 0.00265084677119859
-15.0836596834179 0.00473289522564156
-12.5182611857826 0.00851170520464277
-9.95286268814741 0.0152355067159167
-7.38746419051216 0.0265024087025241
-4.82206569287693 0.0429360872622446
-2.25666719524169 0.0606658561959097
0.308731302393554 0.069078286218622
2.8741298000288 0.0606506462306959
5.43952829766404 0.042065336811288
8.00492679529928 0.0247697578025681
10.5703252929345 0.0133637599816125
13.1357237905698 0.00696602212229979
15.701122288205 0.00362024653636158
18.2665207858402 0.00190701096804268
20.8319192834755 0.00103306329142239
23.3973177811107 0.000572954673815414
25.962716278746 0.000324976007123105
28.5281147763812 0.000189936695946107
31.0935132740164 0.000113108165545253
33.6589117716517 6.88853831855691e-05
36.2243102692869 4.22638599477247e-05
38.7897087669222 2.6604460718689e-05
41.3551072645574 1.67688260064539e-05
43.9205057621927 1.09134776775999e-05
46.4859042598279 7.10889893828632e-06
49.0513027574631 4.85483124817189e-06
51.6167012550984 3.18052617701592e-06
54.1820997527336 2.14007554086393e-06
56.7474982503688 1.46483542110024e-06
59.3128967480041 9.93619466555482e-07
61.8782952456393 7.23523418650009e-07
64.4436937432746 5.13690736271833e-07
67.0090922409098 3.60854128518871e-07
69.5744907385451 2.34155847982567e-07
72.1398892361803 1.89502872320777e-07
74.7052877338155 1.24157054279129e-07
77.2706862314508 8.16822725520582e-08
79.836084729086 6.4982785719193e-08
82.4014832267213 4.46529756617918e-08
84.9668817243565 3.19468443759161e-08
87.5322802219917 2.25080039921227e-08
90.097678719627 1.66994868328652e-08
92.6630772172622 1.27061312858757e-08
95.2284757148975 9.4388403837934e-09
97.7938742125327 8.34974341643262e-09
100.359272710168 7.98671109397912e-09
102.924671207803 2.17819393472155e-09
105.490069705438 2.17819393472155e-09
108.055468203074 3.63032322453592e-09
110.620866700709 3.63032322453592e-10
113.186265198344 1.45212928981437e-09
115.751663695979 1.45212928981437e-09
118.317062193615 3.63032322453592e-10
120.88246069125 3.63032322453592e-10
123.447859188885 1.08909696736079e-09
};
% \addlegendentry{Comp cutoff 0.001}
\addlegendentry{MPS}
\end{axis}

% \draw ({$(current bounding box.south west)!0.5!(current bounding box.south east)$}|-{$(current bounding box.south west)!-0.2!(current bounding box.north west)$}) node[
%   scale=0.6,
%   fill=white,
%   draw=black,
%   line width=0.4pt,
%   inner sep=3.3pt,
%   fill opacity=0.5,
%   anchor=base,
%   text=black,
%   rotate=0.0
% ]{DNS Laplace Fit: $\mu = 0.3629$, $b = 4.0787$};
\end{tikzpicture}

%% file: C4/PDF_grad_dux_dx_SameLimits.tex
% This file was created by tikzplotlib v0.9.6.
\begin{tikzpicture}

\begin{axis}[
legend cell align={left},
legend style={fill opacity=0.8, draw opacity=1, text opacity=1, draw=white!80!black},
log basis y={10},
tick pos=left,
x grid style={white!69.0196078431373!black},
xlabel={\(\displaystyle \partial u / \partial x\)},
xmin=-200, xmax=200,
xtick style={color=black},
y grid style={white!69.0196078431373!black},
ymin=1e-08, ymax=1,
ymode=log,
ytick style={color=black},
ytick={1e-10,1e-08,1e-06,0.0001,0.01,1,100},
yticklabels={,,,,,,}
]
\addplot [very thick, black]
table {%
-162.087262675767 3.2796837341557e-10
-159.247590774125 0
-156.407918872483 0
-153.568246970841 0
-150.728575069199 3.27968373415573e-10
-147.888903167557 0
-145.049231265914 6.55936746831146e-10
-142.209559364272 3.2796837341557e-10
-139.36988746263 3.2796837341557e-10
-136.530215560988 0
-133.690543659346 3.2796837341557e-10
-130.850871757704 6.5593674683114e-10
-128.011199856061 3.2796837341557e-10
-125.171527954419 1.31187349366228e-09
-122.331856052777 3.27968373415572e-09
-119.492184151135 5.24749397464912e-09
-116.652512249493 6.5593674683114e-09
-113.812840347851 6.55936746831143e-09
-110.973168446209 6.23139909489583e-09
-108.133496544566 1.44306084302851e-08
-105.293824642924 1.31187349366228e-08
-102.454152741282 1.37746716834539e-08
-99.6144808396399 2.3613722885921e-08
-96.7748089379977 2.65654382466612e-08
-93.9351370363555 3.87002680630375e-08
-91.0954651347134 5.4114781613569e-08
-88.2557932330712 7.96963147399835e-08
-85.4161213314291 1.02982069252489e-07
-82.5764494297869 1.37418748461124e-07
-79.7367775281447 1.78742763511487e-07
-76.8971056265026 2.63358603852703e-07
-74.0574337248604 3.28952278535817e-07
-71.2177618232183 4.93920370363848e-07
-68.3780899215761 6.9890060374858e-07
-65.5384180199339 9.2651065489899e-07
-62.6987461182918 1.33614315329503e-06
-59.8590742166496 1.89237751460784e-06
-57.0194023150075 2.76838104000083e-06
-54.1797304133653 4.08156640715677e-06
-51.3400585117231 5.93491568532818e-06
-48.500386610081 8.89614212889733e-06
-45.6607147084388 1.35342708657403e-05
-42.8210428067967 2.06718465763834e-05
-39.9813709051545 3.22219087829595e-05
-37.1416990035123 5.051598465208e-05
-34.3020271018702 8.07271193376286e-05
-31.462355200228 0.000131042715313552
-28.6226832985858 0.000216078683141114
-25.7830113969437 0.000364289886866228
-22.9433394953015 0.000628792772373774
-20.1036675936594 0.00110749049068554
-17.2639956920172 0.00200225347906952
-14.4243237903751 0.00372829920935267
-11.5846518887329 0.00713308779142123
-8.74497998709073 0.0140824752493984
-5.90530808544857 0.0283876984188114
-3.0656361838064 0.0559935658230039
-0.225964282164242 0.0907783168824396
2.61370761947791 0.0803380681665387
5.45337952112006 0.0391803065160595
8.29305142276222 0.0159174878158121
11.1327233244044 0.00653152393731354
13.9723952260465 0.00282124560617493
16.8120671276887 0.00129004129498075
19.6517390293309 0.000618188570956128
22.491410930973 0.000310768400040164
25.3310828326152 0.000162273503672049
28.1707547342574 8.78745340994742e-05
31.0104266358995 4.84415846902265e-05
33.8500985375417 2.75099871620983e-05
36.6897704391838 1.59359832642625e-05
39.529442340826 9.3546419149323e-06
42.3691142424681 5.6472874218427e-06
45.2087861441103 3.44268401574324e-06
48.0484580457525 2.07505589860031e-06
50.8881299473946 1.36828405388976e-06
53.7278018490368 8.39271067570444e-07
56.567473750679 5.34916417040795e-07
59.4071456523211 3.60437242383711e-07
62.2468175539633 2.33185513498473e-07
65.0864894556054 1.4660186291676e-07
67.9261613572476 9.74066069044243e-08
70.7658332588897 5.73944653477247e-08
73.6055051605319 4.03401099301151e-08
76.4451770621741 2.98451219808169e-08
79.2848489638162 1.93501340315186e-08
82.1245208654584 1.08229563227138e-08
84.9641927671006 1.31187349366228e-08
87.8038646687427 7.21530421514254e-09
90.6435365703849 3.93562048098684e-09
93.483208472027 4.59155722781803e-09
96.3228803736692 1.9678102404934e-09
99.1625522753114 2.62374698732459e-09
102.002224176954 1.9678102404934e-09
104.841896078596 3.27968373415573e-10
107.681567980238 0
110.52123988188 0
113.360911783522 3.27968373415573e-10
116.200583685164 3.27968373415567e-10
119.040255586806 3.27968373415573e-10
};
% \addlegendentry{DNS}
\addplot [very thick, red, dotted]
table {%
-163.507098626588 4.36474597384841e-19
-160.638743170384 8.81809200908925e-19
-157.77038771418 1.78151826352917e-18
-154.902032257976 3.59919960011372e-18
-152.033676801772 7.27145942124472e-18
-149.165321345567 1.46905223353376e-17
-146.296965889363 2.96792478624751e-17
-143.428610433159 5.99609553408023e-17
-140.560254976955 1.21139059252489e-16
-137.691899520751 2.44737122568696e-16
-134.823544064546 4.94442168635006e-16
-131.955188608342 9.98921028238636e-16
-129.086833152138 2.01811917339504e-15
-126.218477695934 4.07720418620694e-15
-123.35012223973 8.23717161759967e-15
-120.481766783525 1.66415497382564e-14
-117.613411327321 3.36209066106069e-14
-114.745055871117 6.79242846428305e-14
-111.876700414913 1.3722736562924e-13
-109.008344958709 2.77240312158793e-13
-106.139989502504 5.60108330677794e-13
-103.2716340463 1.13158631099426e-12
-100.403278590096 2.28614271399975e-12
-97.5349231338918 4.61869188235577e-12
-94.6665676776876 9.3311386789222e-12
-91.7982122214834 1.88516903190493e-11
-88.9298567652792 3.8086051457804e-11
-86.061501309075 7.69452123972537e-11
-83.1931458528708 1.55452337121845e-10
-80.3247903966666 3.1406020418635e-10
-77.4564349404624 6.34495522420244e-10
-74.5880794842582 1.28187068149666e-09
-71.719724028054 2.58976208029467e-09
-68.8513685718498 5.23209379022659e-09
-65.9830131156456 1.05703939516378e-08
-63.1146576594414 2.13553565307899e-08
-60.2463022032372 4.31442058492521e-08
-57.377946747033 8.7164196752176e-08
-54.5095912908288 1.76097741189128e-07
-51.6412358346246 3.55770093770054e-07
-48.7728803784204 7.18761971428216e-07
-45.9045249222162 1.4521141057609e-06
-43.036169466012 2.93370470332455e-06
-40.1678140098078 5.92696073412133e-06
-37.2994585536036 1.19742329567142e-05
-34.4311030973994 2.419153109556e-05
-31.5627476411952 4.88741265401303e-05
-28.694392184991 9.87403499028251e-05
-25.8260367287868 0.000199485032043016
-22.9576812725826 0.000403019414538909
-20.0893258163784 0.000814219727825288
-17.2209703601742 0.00164496731736451
-14.35261490397 0.00332332585753564
-11.4842594477658 0.00671411196975021
-8.61590399156157 0.0135645138258488
-5.74754853535737 0.027404373975385
-2.87919307915317 0.0553650298583971
-0.0108376229489693 0.111853915509053
2.85751783325523 0.066499929593055
5.72587328945946 0.0329158847121698
8.59422874566366 0.0162925806540706
11.4625842018679 0.00806444021452208
14.3309396580721 0.00399170624681563
17.1992951142763 0.00197579972533917
20.0676505704805 0.000977973907214384
22.9360060266847 0.000484073841556985
25.8043614828889 0.000239605047078592
28.672716939093 0.000118598803853722
31.5410723952972 5.87035892900865e-05
34.4094278515014 2.90568815499148e-05
37.2777833077057 1.43824658017691e-05
40.1461387639099 7.1189787583956e-06
43.0144942201141 3.5237252958567e-06
45.8828496763183 1.7441602766432e-06
48.7512051325225 8.63317885249763e-07
51.6195605887267 4.27321835597905e-07
54.4879160449309 2.11514152896223e-07
57.3562715011351 1.04694478841245e-07
60.2246269573393 5.18212788589037e-08
63.0929824135435 2.56503014513722e-08
65.9613378697477 1.26962896137254e-08
68.8296933259519 6.28436161895353e-09
71.6980487821561 3.11060964733208e-09
74.5664042383603 1.53967784872422e-09
77.4347596945645 7.62103943156372e-10
80.3031151507687 3.77223339710801e-10
83.1714706069729 1.86716588072256e-10
86.0398260631771 9.24202735919589e-11
88.9081815193813 4.57458389690962e-11
91.7765369755855 2.26431030947366e-11
94.6448924317897 1.12077978962247e-11
97.5132478879939 5.54759359426392e-12
100.381603344198 2.74592698512924e-12
103.249958800402 1.35916859797684e-12
106.118314256606 6.72756154016736e-13
108.986669712811 3.32998307524982e-13
111.855025169015 1.6482624819177e-13
114.723380625219 8.15850755966267e-14
117.591736081423 4.03826734705701e-14
120.460091537628 1.99884636338821e-14
};
% \addlegendentry{DNS (Laplace Fit)}
\addplot [very thick, blue, dashed]
table {%
-141.361213384101 3.97552677234611e-10
-139.018573960581 0
-136.675934537062 0
-134.333295113543 0
-131.990655690023 3.97552677234611e-10
-129.648016266504 3.97552677234611e-10
-127.305376842984 0
-124.962737419465 3.97552677234611e-10
-122.620097995946 7.95105354469218e-10
-120.277458572426 0
-117.934819148907 0
-115.592179725387 0
-113.249540301868 7.95105354469222e-10
-110.906900878349 7.95105354469213e-10
-108.564261454829 1.59021070893844e-09
-106.22162203131 7.95105354469218e-10
-103.87898260779 1.59021070893844e-09
-101.536343184271 1.19265803170383e-09
-99.1937037607516 3.18042141787685e-09
-96.8510643372322 6.36084283575378e-09
-94.5084249137128 4.77063212681531e-09
-92.1657854901934 5.16818480404991e-09
-89.823146066674 5.96329015851917e-09
-87.4805066431546 8.34860622192673e-09
-85.1378672196352 8.74615889916145e-09
-82.7952277961158 1.35167910259767e-08
-80.4525883725964 1.86849758300266e-08
-78.109948949077 2.94188981153611e-08
-75.7673095255576 3.41895302421764e-08
-73.4246701020382 4.88989792998572e-08
-71.0820306785188 7.1161929224995e-08
-68.7393912549994 1.00580827340356e-07
-66.39675183148 1.33577699550829e-07
-64.0541124079606 1.77308494046636e-07
-61.7114729844412 2.3535118492289e-07
-59.3688335609218 3.82445675499694e-07
-57.0261941374024 5.16818480404991e-07
-54.683554713883 7.84371432183883e-07
-52.3409152903636 1.21690874501514e-06
-49.9982758668442 1.74923177983229e-06
-47.6556364433248 2.64571306699632e-06
-45.3129970198054 4.15402792442443e-06
-42.970357596286 6.37754004819759e-06
-40.6277181727666 9.84578960439232e-06
-38.2850787492472 1.57884070236953e-05
-35.9424393257278 2.53304663826805e-05
-33.5997999022084 4.0907772934764e-05
-31.257160478689 6.71108674439743e-05
-28.9145210551696 0.000111416523111063
-26.5718816316502 0.000187628563994969
-24.2292422081308 0.000320093911154897
-21.8866027846114 0.000553569044993913
-19.543963361092 0.000971310639836527
-17.2013239375726 0.00172046089258081
-14.8586845140533 0.00308324562921705
-12.5160450905339 0.00558432465915592
-10.1734056670144 0.0101880363433808
-7.83076624349505 0.0186350440088712
-5.48812681997565 0.0336826151965142
-3.14548739645626 0.0577227472400103
-0.80284797293686 0.084321600271222
1.53979145058254 0.088056573484312
3.88243087410194 0.0608079714889012
6.22507029762134 0.0318650254264894
8.56770972114073 0.0150935004484437
10.9103491446601 0.00705730225176266
13.2529885681795 0.00335450973523793
15.5956279916989 0.00163618926627132
17.9382674152183 0.000819856189048495
20.2809068387377 0.000421441617609639
22.6235462622571 0.000222317022847073
24.9661856857765 0.000119826749997961
27.3088251092959 6.59690961549561e-05
29.6514645328153 3.67553352210487e-05
31.9941039563347 2.11569583770715e-05
34.3367433798541 1.22283227990593e-05
36.6793828033735 7.18457198298389e-06
39.0220222268929 4.29833954626056e-06
41.3646616504123 2.63020851258419e-06
43.7073010739317 1.62797821327573e-06
46.0499404974511 9.75991822610959e-07
48.3925799209705 6.40059810347724e-07
50.7352193444899 4.01130651329718e-07
53.0778587680093 2.6119210894314e-07
55.4204981915287 1.67767229793004e-07
57.7631376150481 9.97857219858874e-08
60.1057770385675 6.08255596168955e-08
62.4484164620869 4.81038739453874e-08
64.7910558856063 3.22017668560035e-08
67.1336953091257 1.70947651210881e-08
69.4763347326451 1.74923177983229e-08
71.8189741561645 9.93881693086528e-09
74.1616135796839 7.55350086745752e-09
76.5042530032033 5.16818480404995e-09
78.8468924267227 2.38531606340764e-09
81.1895318502421 3.5779740951115e-09
83.5321712737615 1.59021070893844e-09
85.8748106972809 7.95105354469213e-10
88.2174501208003 1.59021070893844e-09
90.5600895443197 3.97552677234606e-10
};
% \addlegendentry{Comp cutoff 0.0001}
\end{axis}

% \draw ({$(current bounding box.south west)!0.5!(current bounding box.south east)$}|-{$(current bounding box.south west)!-0.2!(current bounding box.north west)$}) node[
%   scale=0.6,
%   fill=white,
%   draw=black,
%   line width=0.4pt,
%   inner sep=3.3pt,
%   fill opacity=0.5,
%   anchor=base,
%   text=black,
%   rotate=0.0
% ]{DNS Laplace Fit: $\mu = 0.3629$, $b = 4.0787$};
\end{tikzpicture}

%% file: C5/PDF_grad_dux_dx_SameLimits.tex
% This file was created by tikzplotlib v0.9.6.
\begin{tikzpicture}

\begin{axis}[
legend cell align={left},
legend style={fill opacity=0.8, draw opacity=1, text opacity=1, draw=white!80!black},
log basis y={10},
tick pos=left,
x grid style={white!69.0196078431373!black},
xlabel={\(\displaystyle \partial u / \partial x\)},
xmin=-200, xmax=200,
xtick style={color=black},
y grid style={white!69.0196078431373!black},
ymin=1e-08, ymax=1,
ymode=log,
ytick style={color=black},
ytick={1e-10,1e-08,1e-06,0.0001,0.01,1,100},
yticklabels={,,,,,,}
]
\addplot [very thick, black]
table {%
-162.087262675767 3.2796837341557e-10
-159.247590774125 0
-156.407918872483 0
-153.568246970841 0
-150.728575069199 3.27968373415573e-10
-147.888903167557 0
-145.049231265914 6.55936746831146e-10
-142.209559364272 3.2796837341557e-10
-139.36988746263 3.2796837341557e-10
-136.530215560988 0
-133.690543659346 3.2796837341557e-10
-130.850871757704 6.5593674683114e-10
-128.011199856061 3.2796837341557e-10
-125.171527954419 1.31187349366228e-09
-122.331856052777 3.27968373415572e-09
-119.492184151135 5.24749397464912e-09
-116.652512249493 6.5593674683114e-09
-113.812840347851 6.55936746831143e-09
-110.973168446209 6.23139909489583e-09
-108.133496544566 1.44306084302851e-08
-105.293824642924 1.31187349366228e-08
-102.454152741282 1.37746716834539e-08
-99.6144808396399 2.3613722885921e-08
-96.7748089379977 2.65654382466612e-08
-93.9351370363555 3.87002680630375e-08
-91.0954651347134 5.4114781613569e-08
-88.2557932330712 7.96963147399835e-08
-85.4161213314291 1.02982069252489e-07
-82.5764494297869 1.37418748461124e-07
-79.7367775281447 1.78742763511487e-07
-76.8971056265026 2.63358603852703e-07
-74.0574337248604 3.28952278535817e-07
-71.2177618232183 4.93920370363848e-07
-68.3780899215761 6.9890060374858e-07
-65.5384180199339 9.2651065489899e-07
-62.6987461182918 1.33614315329503e-06
-59.8590742166496 1.89237751460784e-06
-57.0194023150075 2.76838104000083e-06
-54.1797304133653 4.08156640715677e-06
-51.3400585117231 5.93491568532818e-06
-48.500386610081 8.89614212889733e-06
-45.6607147084388 1.35342708657403e-05
-42.8210428067967 2.06718465763834e-05
-39.9813709051545 3.22219087829595e-05
-37.1416990035123 5.051598465208e-05
-34.3020271018702 8.07271193376286e-05
-31.462355200228 0.000131042715313552
-28.6226832985858 0.000216078683141114
-25.7830113969437 0.000364289886866228
-22.9433394953015 0.000628792772373774
-20.1036675936594 0.00110749049068554
-17.2639956920172 0.00200225347906952
-14.4243237903751 0.00372829920935267
-11.5846518887329 0.00713308779142123
-8.74497998709073 0.0140824752493984
-5.90530808544857 0.0283876984188114
-3.0656361838064 0.0559935658230039
-0.225964282164242 0.0907783168824396
2.61370761947791 0.0803380681665387
5.45337952112006 0.0391803065160595
8.29305142276222 0.0159174878158121
11.1327233244044 0.00653152393731354
13.9723952260465 0.00282124560617493
16.8120671276887 0.00129004129498075
19.6517390293309 0.000618188570956128
22.491410930973 0.000310768400040164
25.3310828326152 0.000162273503672049
28.1707547342574 8.78745340994742e-05
31.0104266358995 4.84415846902265e-05
33.8500985375417 2.75099871620983e-05
36.6897704391838 1.59359832642625e-05
39.529442340826 9.3546419149323e-06
42.3691142424681 5.6472874218427e-06
45.2087861441103 3.44268401574324e-06
48.0484580457525 2.07505589860031e-06
50.8881299473946 1.36828405388976e-06
53.7278018490368 8.39271067570444e-07
56.567473750679 5.34916417040795e-07
59.4071456523211 3.60437242383711e-07
62.2468175539633 2.33185513498473e-07
65.0864894556054 1.4660186291676e-07
67.9261613572476 9.74066069044243e-08
70.7658332588897 5.73944653477247e-08
73.6055051605319 4.03401099301151e-08
76.4451770621741 2.98451219808169e-08
79.2848489638162 1.93501340315186e-08
82.1245208654584 1.08229563227138e-08
84.9641927671006 1.31187349366228e-08
87.8038646687427 7.21530421514254e-09
90.6435365703849 3.93562048098684e-09
93.483208472027 4.59155722781803e-09
96.3228803736692 1.9678102404934e-09
99.1625522753114 2.62374698732459e-09
102.002224176954 1.9678102404934e-09
104.841896078596 3.27968373415573e-10
107.681567980238 0
110.52123988188 0
113.360911783522 3.27968373415573e-10
116.200583685164 3.27968373415567e-10
119.040255586806 3.27968373415573e-10
};
% \addlegendentry{DNS}
\addplot [very thick, red, dotted]
table {%
-163.507098626588 4.36474597384841e-19
-160.638743170384 8.81809200908925e-19
-157.77038771418 1.78151826352917e-18
-154.902032257976 3.59919960011372e-18
-152.033676801772 7.27145942124472e-18
-149.165321345567 1.46905223353376e-17
-146.296965889363 2.96792478624751e-17
-143.428610433159 5.99609553408023e-17
-140.560254976955 1.21139059252489e-16
-137.691899520751 2.44737122568696e-16
-134.823544064546 4.94442168635006e-16
-131.955188608342 9.98921028238636e-16
-129.086833152138 2.01811917339504e-15
-126.218477695934 4.07720418620694e-15
-123.35012223973 8.23717161759967e-15
-120.481766783525 1.66415497382564e-14
-117.613411327321 3.36209066106069e-14
-114.745055871117 6.79242846428305e-14
-111.876700414913 1.3722736562924e-13
-109.008344958709 2.77240312158793e-13
-106.139989502504 5.60108330677794e-13
-103.2716340463 1.13158631099426e-12
-100.403278590096 2.28614271399975e-12
-97.5349231338918 4.61869188235577e-12
-94.6665676776876 9.3311386789222e-12
-91.7982122214834 1.88516903190493e-11
-88.9298567652792 3.8086051457804e-11
-86.061501309075 7.69452123972537e-11
-83.1931458528708 1.55452337121845e-10
-80.3247903966666 3.1406020418635e-10
-77.4564349404624 6.34495522420244e-10
-74.5880794842582 1.28187068149666e-09
-71.719724028054 2.58976208029467e-09
-68.8513685718498 5.23209379022659e-09
-65.9830131156456 1.05703939516378e-08
-63.1146576594414 2.13553565307899e-08
-60.2463022032372 4.31442058492521e-08
-57.377946747033 8.7164196752176e-08
-54.5095912908288 1.76097741189128e-07
-51.6412358346246 3.55770093770054e-07
-48.7728803784204 7.18761971428216e-07
-45.9045249222162 1.4521141057609e-06
-43.036169466012 2.93370470332455e-06
-40.1678140098078 5.92696073412133e-06
-37.2994585536036 1.19742329567142e-05
-34.4311030973994 2.419153109556e-05
-31.5627476411952 4.88741265401303e-05
-28.694392184991 9.87403499028251e-05
-25.8260367287868 0.000199485032043016
-22.9576812725826 0.000403019414538909
-20.0893258163784 0.000814219727825288
-17.2209703601742 0.00164496731736451
-14.35261490397 0.00332332585753564
-11.4842594477658 0.00671411196975021
-8.61590399156157 0.0135645138258488
-5.74754853535737 0.027404373975385
-2.87919307915317 0.0553650298583971
-0.0108376229489693 0.111853915509053
2.85751783325523 0.066499929593055
5.72587328945946 0.0329158847121698
8.59422874566366 0.0162925806540706
11.4625842018679 0.00806444021452208
14.3309396580721 0.00399170624681563
17.1992951142763 0.00197579972533917
20.0676505704805 0.000977973907214384
22.9360060266847 0.000484073841556985
25.8043614828889 0.000239605047078592
28.672716939093 0.000118598803853722
31.5410723952972 5.87035892900865e-05
34.4094278515014 2.90568815499148e-05
37.2777833077057 1.43824658017691e-05
40.1461387639099 7.1189787583956e-06
43.0144942201141 3.5237252958567e-06
45.8828496763183 1.7441602766432e-06
48.7512051325225 8.63317885249763e-07
51.6195605887267 4.27321835597905e-07
54.4879160449309 2.11514152896223e-07
57.3562715011351 1.04694478841245e-07
60.2246269573393 5.18212788589037e-08
63.0929824135435 2.56503014513722e-08
65.9613378697477 1.26962896137254e-08
68.8296933259519 6.28436161895353e-09
71.6980487821561 3.11060964733208e-09
74.5664042383603 1.53967784872422e-09
77.4347596945645 7.62103943156372e-10
80.3031151507687 3.77223339710801e-10
83.1714706069729 1.86716588072256e-10
86.0398260631771 9.24202735919589e-11
88.9081815193813 4.57458389690962e-11
91.7765369755855 2.26431030947366e-11
94.6448924317897 1.12077978962247e-11
97.5132478879939 5.54759359426392e-12
100.381603344198 2.74592698512924e-12
103.249958800402 1.35916859797684e-12
106.118314256606 6.72756154016736e-13
108.986669712811 3.32998307524982e-13
111.855025169015 1.6482624819177e-13
114.723380625219 8.15850755966267e-14
117.591736081423 4.03826734705701e-14
120.460091537628 1.99884636338821e-14
};
% \addlegendentry{DNS (Laplace Fit)}
\addplot [very thick, blue, dashed]
table {%
-138.367523252482 3.94530342078222e-10
-136.00693782904 0
-133.646352405598 0
-131.285766982156 0
-128.925181558714 3.94530342078222e-10
-126.564596135272 1.18359102623466e-09
-124.204010711829 3.9453034207822e-10
-121.843425288387 0
-119.482839864945 1.18359102623466e-09
-117.122254441503 1.97265171039111e-09
-114.761669018061 1.57812136831288e-09
-112.401083594619 1.9726517103911e-09
-110.040498171177 2.36718205246932e-09
-107.679912747735 5.12889444701686e-09
-105.319327324293 1.97265171039111e-09
-102.95874190085 6.7070158153297e-09
-100.598156477408 6.31248547325156e-09
-98.2375710539662 4.73436410493864e-09
-95.8769856305241 1.2624970946503e-08
-93.516400207082 9.8632585519555e-09
-91.1558147836398 1.2624970946503e-08
-88.7952293601977 1.81483957355982e-08
-86.4346439367556 2.99843059979445e-08
-84.0740585133135 3.66913218132747e-08
-81.7134730898714 3.90585038657438e-08
-79.3528876664293 6.74646884953756e-08
-76.9923022429872 8.67966752572084e-08
-74.631716819545 1.09284904755667e-07
-72.2711313961029 1.59390258199602e-07
-69.9105459726608 2.06339368906909e-07
-67.5499605492187 2.79722012533458e-07
-65.1893751257766 4.29643542523182e-07
-62.8287897023345 5.88639270380704e-07
-60.4682042788924 8.04447367497491e-07
-58.1076188554502 1.15557937194711e-06
-55.7470334320081 1.62901578244098e-06
-53.386448008566 2.28472521097497e-06
-51.0258625851239 3.31997282858822e-06
-48.6652771616818 4.80814127890727e-06
-46.3046917382397 7.12719062964305e-06
-43.9441063147975 1.0587221729669e-05
-41.5835208913554 1.59571742156958e-05
-39.2229354679133 2.40553040171932e-05
-36.8623500444712 3.68227004171865e-05
-34.5017646210291 5.63835147774247e-05
-32.141179197587 8.81069105232502e-05
-29.7805937741449 0.000139328390304923
-27.4200083507027 0.000220732624846608
-25.0594229272606 0.000356494460859143
-22.6988375038185 0.000585402148923953
-20.3382520803764 0.000970777019883905
-17.9776666569343 0.00163225803279217
-15.6170812334922 0.00278519602409906
-13.25649581005 0.00481575768716109
-10.8959103866079 0.00845054955298376
-8.53532496316582 0.01503597703484
-6.1747395397237 0.0269478485911292
-3.8141541162816 0.0476578548770763
-1.45356869283948 0.0773009181681252
0.907016730602635 0.0942825271083499
3.26760215404475 0.0707980098635573
5.62818757748687 0.0373620293127628
7.98877300092899 0.0175924145033435
10.3493584243711 0.00824562891518696
12.7099438478132 0.00396789659535168
15.0705292712553 0.0019772929653353
17.4311146946974 0.00101700847989898
19.7917001181396 0.000538744201609095
22.1522855415817 0.000293183387805169
24.5128709650238 0.000163380538079379
26.8734563884659 9.37818349637039e-05
29.234041811908 5.43899529589037e-05
31.5946272353501 3.22165586734231e-05
33.9552126587922 1.93884046007501e-05
36.3157980822344 1.18299923072154e-05
38.6763835056765 7.20333498566419e-06
41.0369689291186 4.44832960693196e-06
43.3975543525607 2.85797779801461e-06
45.7581397760028 1.80734349706034e-06
48.1187251994449 1.20095036128609e-06
50.4793106228871 7.30670193528868e-07
52.8398960463292 4.98686352386867e-07
55.2004814697713 3.3061642666155e-07
57.5610668932134 2.22515112932117e-07
59.9216523166555 1.42425453490237e-07
62.2822377400976 9.46872820987734e-08
64.6428231635397 6.50975064429059e-08
67.0034085869819 4.57655196810738e-08
69.363994010424 3.15624273662574e-08
71.7245794338661 1.89374564197547e-08
74.0851648573082 9.46872820987734e-09
76.4457502807503 1.2624970946503e-08
78.8063357041924 9.46872820987734e-09
81.1669211276346 6.7070158153297e-09
83.5275065510767 1.97265171039111e-09
85.8880919745188 2.76171239454756e-09
88.2486773979609 1.57812136831287e-09
90.609262821403 1.97265171039111e-09
92.9698482448451 1.18359102623465e-09
95.3304336682872 7.89060684156445e-10
};
% \addlegendentry{Comp cutoff 1e-05}
\end{axis}

% \draw ({$(current bounding box.south west)!0.5!(current bounding box.south east)$}|-{$(current bounding box.south west)!-0.2!(current bounding box.north west)$}) node[
%   scale=0.6,
%   fill=white,
%   draw=black,
%   line width=0.4pt,
%   inner sep=3.3pt,
%   fill opacity=0.5,
%   anchor=base,
%   text=black,
%   rotate=0.0
% ]{DNS Laplace Fit: $\mu = 0.3629$, $b = 4.0787$};
\end{tikzpicture}

%% file: C3/PDF_grad_dux_dy_SameLimits.tex
% This file was created by tikzplotlib v0.9.6.
\begin{tikzpicture}

\begin{axis}[
legend cell align={left},
legend style={fill opacity=0.8, text opacity=1, draw=none},
log basis y={10},
tick pos=left,
x grid style={white!69.0196078431373!black},
xlabel={\(\displaystyle \partial u / \partial y\)},
xmin=-200, xmax=200,
xtick style={color=black},
y grid style={white!69.0196078431373!black},
ylabel={PDF},
ymin=1e-08, ymax=1,
ymode=log,
ytick style={color=black},
ytick={1e-10,1e-08,1e-06,0.0001,0.01,1,100},
yticklabels={\(\displaystyle 10^{-10}\),,\(\displaystyle 10^{-6}\),,\(\displaystyle 10^{-2}\),,\(\displaystyle 10^{2}\)}
]
\addplot [very thick, black]
table {%
-287.377180683043 1.73681402825852e-10
-282.014933721293 3.47362805651701e-10
-276.652686759543 0
-271.290439797792 1.7368140282585e-10
-265.928192836042 1.73681402825852e-10
-260.565945874291 5.21044208477551e-10
-255.203698912541 3.47362805651704e-10
-249.84145195079 5.21044208477554e-10
-244.47920498904 5.21044208477554e-10
-239.116958027289 3.47362805651703e-10
-233.754711065539 1.21576981978096e-09
-228.392464103788 1.38945122260682e-09
-223.030217142038 2.25785823673605e-09
-217.667970180288 1.04208841695511e-09
-212.305723218537 2.25785823673607e-09
-206.943476256787 2.43153963956192e-09
-201.581229295036 2.77890244521362e-09
-196.218982333286 3.99467226499458e-09
-190.856735371535 6.25253050173068e-09
-185.494488409785 5.21044208477551e-09
-180.132241448034 1.0594565572377e-08
-174.769994486284 1.05945655723769e-08
-169.407747524533 1.52839634486749e-08
-164.045500562783 1.87575915051919e-08
-158.683253601033 2.74416616464845e-08
-153.321006639282 3.0394245494524e-08
-147.958759677532 4.86307927912384e-08
-142.596512715781 6.00937653777449e-08
-137.234265754031 8.14565779253243e-08
-131.87201879228 1.24529565826135e-07
-126.50977183053 1.51623864666968e-07
-121.147524868779 2.16407027921011e-07
-115.785277907029 2.93347889372863e-07
-110.423030945278 4.0901970365488e-07
-105.060783983528 5.80269566841169e-07
-99.6985370217776 8.06576434723253e-07
-94.3362900600271 1.18850183953731e-06
-88.9740430982767 1.81010758025102e-06
-83.6117961365262 2.68928284135548e-06
-78.2495491747758 4.1970110992867e-06
-72.8873022130253 6.56150971735784e-06
-67.5250552512748 1.03189331860923e-05
-62.1628082895244 1.67177034290023e-05
-56.8005613277739 2.71076723088505e-05
-51.4383143660235 4.56386095833546e-05
-46.076067404273 7.91016317844085e-05
-40.7138204425226 0.000142037345956592
-35.3515734807721 0.000265393346543834
-29.9893265190217 0.000523133769434955
-24.6270795572712 0.00109872870131906
-19.2648325955208 0.00249008520891489
-13.9025856337703 0.00628523748878517
-8.54033867201986 0.0182516645260164
-3.1780917102694 0.0560703561352009
2.18415525148106 0.0653943930940704
7.54640221323149 0.0227952898641087
12.9086491749819 0.00758304207188145
18.2708961367324 0.00290341082190863
23.6331430984829 0.00124692027410338
28.9953900602333 0.000583465304653168
34.3576370219838 0.000292260470109769
39.7198839837342 0.000154905748454766
45.0821309454847 8.54693130562123e-05
50.4443779072352 4.90358178226285e-05
55.8066248689856 2.87135305593784e-05
61.168871830736 1.74146868985424e-05
66.5311187924865 1.05744185296492e-05
71.893365754237 6.60736160770383e-06
77.2556127159874 4.21924231884843e-06
82.6178596777379 2.70612993742957e-06
87.9801066394883 1.80819708481995e-06
93.3423536012388 1.21055937769618e-06
98.7046005629892 8.18386770115416e-07
104.06684752474 5.72106540908357e-07
109.42909448649 3.9008843074686e-07
114.791341448241 2.57569520390739e-07
120.153588409991 1.84970694009531e-07
125.515835371742 1.33387317370254e-07
130.878082333492 8.54512501903184e-08
136.240329295242 6.40884376427395e-08
141.602576256993 4.23782622895079e-08
146.964823218743 2.86574314662653e-08
152.327070180494 1.80628658938886e-08
157.689317142244 1.8583910102366e-08
163.051564103995 1.0594565572377e-08
168.413811065745 6.77357471020816e-09
173.776058027496 4.86307927912386e-09
179.138304989246 5.03676068194971e-09
184.500551950996 4.34203507064626e-09
189.862798912747 2.95258384803949e-09
195.225045874497 1.91049543108435e-09
200.587292836248 1.04208841695511e-09
205.949539797998 1.21576981978095e-09
211.311786759749 3.47362805651704e-10
216.674033721499 8.68407014129252e-10
222.03628068325 1.73681402825852e-10
227.398527645 1.73681402825852e-10
232.760774606751 3.47362805651704e-10
238.123021568501 1.73681402825849e-10
243.485268530251 3.47362805651708e-10
};
\addlegendentry{DNS}
\addplot [very thick, red, dotted]
table {%
-290.058304163919 3.69518190535496e-25
-284.641893091443 1.01059981672955e-24
-279.225482018968 2.76390179355914e-24
-273.809070946493 7.55902880445887e-24
-268.392659874018 2.06732802879581e-23
-262.976248801543 5.65396069945361e-23
-257.559837729067 1.54630862377397e-22
-252.143426656592 4.22901835909255e-22
-246.727015584117 1.15659940108802e-21
-241.310604511642 3.16319784169534e-21
-235.894193439167 8.65106844798082e-21
-230.477782366691 2.36599128594303e-20
-225.061371294216 6.47077849264374e-20
-219.644960221741 1.76970112060967e-19
-214.228549149266 4.83997722970666e-19
-208.812138076791 1.32369128952174e-18
-203.395727004315 3.62017949010458e-18
-197.97931593184 9.90087314490759e-18
-192.562904859365 2.70780190041683e-17
-187.14649378689 7.40560052087154e-17
-181.730082714415 2.02536673994838e-16
-176.313671641939 5.53920025759957e-16
-170.897260569464 1.51492265023436e-15
-165.480849496989 4.14318047635927e-15
-160.064438424514 1.13312349360076e-14
-154.648027352039 3.09899329531075e-14
-149.231616279563 8.47547464916015e-14
-143.815205207088 2.31796792323662e-13
-138.398794134613 6.33943881088275e-13
-132.982383062138 1.73378087047946e-12
-127.565971989663 4.74173849849328e-12
-122.149560917187 1.29682385882338e-11
-116.733149844712 3.5466994254279e-11
-111.316738772237 9.69991161771481e-11
-105.900327699762 2.65284068666539e-10
-100.483916627286 7.25528642547082e-10
-95.0675055548113 1.98425715423524e-09
-89.6510944823361 5.42676914905975e-09
-84.2346834098608 1.48417372891052e-08
-78.8182723373857 4.05908487551899e-08
-73.4018612649105 1.11012408491839e-07
-67.9854501924353 3.03609193133295e-07
-62.5690391199601 8.30344493983548e-07
-57.1526280474848 2.27091930772365e-06
-51.7362169750096 6.21076497713756e-06
-46.3198059025344 1.69858970638214e-05
-40.9033948300592 4.64549375358448e-05
-35.486983757584 0.000127050176587714
-30.0705726851088 0.000347471081163637
-24.6541616126336 0.000950302907778114
-19.2377505401584 0.00259899504012606
-13.8213394676832 0.00710802330847658
-8.404928395208 0.0194398198433635
-2.98851732273278 0.0531662009453129
2.42789374974245 0.0593225052409706
7.84430482221762 0.021690826014213
13.2607158946929 0.00793108671435393
18.6771269671681 0.00289994195838299
24.0935380396432 0.00106034187556797
29.5099491121185 0.000387705998677958
34.9263601845937 0.000141761770306729
40.3427712570689 5.18341206713974e-05
45.7591823295441 1.89527547516062e-05
51.1755934020193 6.92993163618478e-06
56.5920044744945 2.53387716517168e-06
62.0084155469697 9.2649304859712e-07
67.4248266194449 3.38765185975634e-07
72.8412376919201 1.23866931762602e-07
78.2576487643953 4.52910081066737e-08
83.6740598368705 1.65603150585028e-08
89.0904709093458 6.05515413105732e-09
94.5068819818209 2.21402137708941e-09
99.9232930542962 8.09540195362936e-10
105.339704126771 2.96002258465002e-10
110.756115199247 1.08230990280973e-10
116.172526271722 3.9573844192763e-11
121.588937344197 1.44698772516766e-11
127.005348416672 5.29080133480892e-12
132.421759489147 1.9345415498374e-12
137.838170561623 7.07350507271032e-13
143.254581634098 2.58637370791356e-13
148.670992706573 9.45688012975925e-14
154.087403779048 3.45783679732741e-14
159.503814851523 1.26433190998432e-14
164.920225923999 4.62293414148449e-15
170.336636996474 1.69034095459694e-15
175.753048068949 6.18060403921346e-16
181.169459141424 2.25989118855912e-16
186.585870213899 8.26312145499778e-17
192.002281286375 3.02134795364102e-17
197.41869235885 1.10473306082772e-17
202.835103431325 4.03937300308303e-18
208.2515145038 1.4769662316262e-18
213.667925576275 5.40041547957857e-19
219.084336648751 1.97462113402282e-19
224.500747721226 7.22005304531465e-20
229.917158793701 2.63995786730779e-20
235.333569866176 9.6528065616956e-21
240.749980938652 3.52947581745054e-21
246.166392011127 1.29052617664601e-21
};
\addlegendentry{Laplace}
\addplot [very thick, blue, dashed]
table {%
-145.638599178791 6.00779374356035e-10
-142.538217859563 6.00779374356035e-10
-139.437836540335 0
-136.337455221107 6.00779374356035e-10
-133.237073901879 6.00779374356035e-10
-130.136692582651 1.20155874871207e-09
-127.036311263423 1.80233812306811e-09
-123.935929944195 6.00779374356035e-10
-120.835548624967 1.20155874871208e-09
-117.735167305739 1.8023381230681e-09
-114.634785986511 3.00389687178017e-09
-111.534404667283 4.20545562049226e-09
-108.434023348055 8.41091124098453e-09
-105.333642028827 1.32171462358328e-08
-102.233260709599 1.86241606050371e-08
-99.1328793903709 2.76358512203776e-08
-96.0324980711428 4.14537768305666e-08
-92.9321167519148 5.28685849433313e-08
-89.8317354326868 7.99036567893526e-08
-86.7313541134587 1.14148081127647e-07
-83.6309727942307 1.48392505465941e-07
-80.5305914750027 2.24090706634801e-07
-77.4302101557747 3.37638008388092e-07
-74.3298288365466 4.52386868890096e-07
-71.2294475173186 6.01980933104747e-07
-68.1290661980906 8.47699697216369e-07
-65.0286848788625 1.25082265740926e-06
-61.9283035596345 1.90266827858556e-06
-58.8279222404065 2.79182175263251e-06
-55.7275409211785 4.11113325871835e-06
-52.6271596019504 5.95492515861705e-06
-49.5267782827224 8.87230980048992e-06
-46.4263969634944 1.3408194076878e-05
-43.3260156442663 2.06433800822478e-05
-40.2256343250383 3.22324142135756e-05
-37.1252530058103 5.18634810500337e-05
-34.0248716865823 8.52956516741981e-05
-30.9244903673542 0.000144364880540258
-27.8241090481262 0.000252115262110388
-24.7237277288982 0.000456362826789583
-21.6233464096701 0.000852126840425999
-18.5229650904421 0.00166296031211438
-15.4225837712141 0.00337100280809074
-12.3222024519861 0.00710819956940178
-9.22182113275802 0.0153971015311452
-6.12143981352999 0.032966204339076
-3.02105849430197 0.0612443920446931
0.0793228249260522 0.0782765236548383
3.17970414415409 0.0603115417840209
6.28008546338212 0.0320742857779586
9.38046678261016 0.014851590855333
12.4808481018382 0.00677497248318027
15.5812294210662 0.00318617543668761
18.6816107402942 0.00156727237882257
21.7819920595223 0.00080324082195527
24.8823733787503 0.000427187478422418
27.9827546979783 0.000236243872598651
31.0831360172063 0.000134080739210031
34.1835173364344 7.87081058343841e-05
37.2838986556624 4.75018227922086e-05
40.3842799748904 2.94075495953536e-05
43.4846612941185 1.87353047892931e-05
46.5850426133465 1.21621776544636e-05
49.6854239325745 7.85699265782823e-06
52.7858052518026 5.21236185191296e-06
55.8861865710306 3.67496743293587e-06
58.9865678902586 2.5406959741517e-06
62.0869492094866 1.68458536569432e-06
65.1873305287147 1.12886444441499e-06
68.2877118479427 8.74734769062387e-07
71.3880931671707 5.8636066937149e-07
74.4884744863988 4.58394662633659e-07
77.5888558056268 3.03994363424154e-07
80.6892371248548 2.09672001650256e-07
83.7896184440828 1.8023381230681e-07
86.8899997633109 1.23760551117343e-07
89.9903810825389 9.85278173943906e-08
93.0907624017669 6.36826136817397e-08
96.191143720995 4.80623499484828e-08
99.291525040223 3.3643644963938e-08
102.391906359451 2.46319543485974e-08
105.492287678679 2.10272781024614e-08
108.592668997907 1.38179256101888e-08
111.693050317135 1.02132493640527e-08
114.793431636363 9.61246998969647e-09
117.893812955591 9.61246998969665e-09
120.994194274819 5.40701436920426e-09
124.094575594047 2.40311749742416e-09
127.194956913275 1.20155874871208e-09
130.295338232503 4.80623499484824e-09
133.395719551731 1.20155874871208e-09
136.496100870959 3.00389687178015e-09
139.596482190187 2.40311749742416e-09
142.696863509415 1.80233812306812e-09
145.797244828643 2.40311749742412e-09
148.897626147871 6.0077937435604e-10
151.998007467099 2.40311749742412e-09
155.098388786328 1.20155874871208e-09
158.198770105556 6.0077937435604e-10
161.299151424784 6.00779374356035e-10
};
% \addlegendentry{Comp cutoff 0.001}
\addlegendentry{MPS}
\end{axis}

% \draw ({$(current bounding box.south west)!0.5!(current bounding box.south east)$}|-{$(current bounding box.south west)!-0.2!(current bounding box.north west)$}) node[
%   scale=0.6,
%   fill=white,
%   draw=black,
%   line width=0.4pt,
%   inner sep=3.3pt,
%   fill opacity=0.5,
%   anchor=base,
%   text=black,
%   rotate=0.0
% ]{DNS Laplace Fit: $\mu = 0.0146$, $b = 5.3836$};
\end{tikzpicture}

%% file: C4/PDF_grad_dux_dy_SameLimits.tex
% This file was created by tikzplotlib v0.9.6.
\begin{tikzpicture}

\begin{axis}[
legend cell align={left},
legend style={fill opacity=0.8, draw opacity=1, text opacity=1, draw=white!80!black},
log basis y={10},
tick pos=left,
x grid style={white!69.0196078431373!black},
xlabel={\(\displaystyle \partial u / \partial y\)},
xmin=-200, xmax=200,
xtick style={color=black},
y grid style={white!69.0196078431373!black},
ymin=1e-08, ymax=1,
ymode=log,
ytick style={color=black},
ytick={1e-10,1e-08,1e-06,0.0001,0.01,1,100},
% yticklabels={\(\displaystyle 10^{-10}\),,\(\displaystyle 10^{-6}\),,\(\displaystyle 10^{-2}\),,\(\displaystyle 10^{2}\)}
% ]
yticklabels={,,,,,,}
]
\addplot [very thick, black]
table {%
-287.377180683043 1.73681402825852e-10
-282.014933721293 3.47362805651701e-10
-276.652686759543 0
-271.290439797792 1.7368140282585e-10
-265.928192836042 1.73681402825852e-10
-260.565945874291 5.21044208477551e-10
-255.203698912541 3.47362805651704e-10
-249.84145195079 5.21044208477554e-10
-244.47920498904 5.21044208477554e-10
-239.116958027289 3.47362805651703e-10
-233.754711065539 1.21576981978096e-09
-228.392464103788 1.38945122260682e-09
-223.030217142038 2.25785823673605e-09
-217.667970180288 1.04208841695511e-09
-212.305723218537 2.25785823673607e-09
-206.943476256787 2.43153963956192e-09
-201.581229295036 2.77890244521362e-09
-196.218982333286 3.99467226499458e-09
-190.856735371535 6.25253050173068e-09
-185.494488409785 5.21044208477551e-09
-180.132241448034 1.0594565572377e-08
-174.769994486284 1.05945655723769e-08
-169.407747524533 1.52839634486749e-08
-164.045500562783 1.87575915051919e-08
-158.683253601033 2.74416616464845e-08
-153.321006639282 3.0394245494524e-08
-147.958759677532 4.86307927912384e-08
-142.596512715781 6.00937653777449e-08
-137.234265754031 8.14565779253243e-08
-131.87201879228 1.24529565826135e-07
-126.50977183053 1.51623864666968e-07
-121.147524868779 2.16407027921011e-07
-115.785277907029 2.93347889372863e-07
-110.423030945278 4.0901970365488e-07
-105.060783983528 5.80269566841169e-07
-99.6985370217776 8.06576434723253e-07
-94.3362900600271 1.18850183953731e-06
-88.9740430982767 1.81010758025102e-06
-83.6117961365262 2.68928284135548e-06
-78.2495491747758 4.1970110992867e-06
-72.8873022130253 6.56150971735784e-06
-67.5250552512748 1.03189331860923e-05
-62.1628082895244 1.67177034290023e-05
-56.8005613277739 2.71076723088505e-05
-51.4383143660235 4.56386095833546e-05
-46.076067404273 7.91016317844085e-05
-40.7138204425226 0.000142037345956592
-35.3515734807721 0.000265393346543834
-29.9893265190217 0.000523133769434955
-24.6270795572712 0.00109872870131906
-19.2648325955208 0.00249008520891489
-13.9025856337703 0.00628523748878517
-8.54033867201986 0.0182516645260164
-3.1780917102694 0.0560703561352009
2.18415525148106 0.0653943930940704
7.54640221323149 0.0227952898641087
12.9086491749819 0.00758304207188145
18.2708961367324 0.00290341082190863
23.6331430984829 0.00124692027410338
28.9953900602333 0.000583465304653168
34.3576370219838 0.000292260470109769
39.7198839837342 0.000154905748454766
45.0821309454847 8.54693130562123e-05
50.4443779072352 4.90358178226285e-05
55.8066248689856 2.87135305593784e-05
61.168871830736 1.74146868985424e-05
66.5311187924865 1.05744185296492e-05
71.893365754237 6.60736160770383e-06
77.2556127159874 4.21924231884843e-06
82.6178596777379 2.70612993742957e-06
87.9801066394883 1.80819708481995e-06
93.3423536012388 1.21055937769618e-06
98.7046005629892 8.18386770115416e-07
104.06684752474 5.72106540908357e-07
109.42909448649 3.9008843074686e-07
114.791341448241 2.57569520390739e-07
120.153588409991 1.84970694009531e-07
125.515835371742 1.33387317370254e-07
130.878082333492 8.54512501903184e-08
136.240329295242 6.40884376427395e-08
141.602576256993 4.23782622895079e-08
146.964823218743 2.86574314662653e-08
152.327070180494 1.80628658938886e-08
157.689317142244 1.8583910102366e-08
163.051564103995 1.0594565572377e-08
168.413811065745 6.77357471020816e-09
173.776058027496 4.86307927912386e-09
179.138304989246 5.03676068194971e-09
184.500551950996 4.34203507064626e-09
189.862798912747 2.95258384803949e-09
195.225045874497 1.91049543108435e-09
200.587292836248 1.04208841695511e-09
205.949539797998 1.21576981978095e-09
211.311786759749 3.47362805651704e-10
216.674033721499 8.68407014129252e-10
222.03628068325 1.73681402825852e-10
227.398527645 1.73681402825852e-10
232.760774606751 3.47362805651704e-10
238.123021568501 1.73681402825849e-10
243.485268530251 3.47362805651708e-10
};
% \addlegendentry{DNS}
\addplot [very thick, red, dotted]
table {%
-290.058304163919 3.69518190535496e-25
-284.641893091443 1.01059981672955e-24
-279.225482018968 2.76390179355914e-24
-273.809070946493 7.55902880445887e-24
-268.392659874018 2.06732802879581e-23
-262.976248801543 5.65396069945361e-23
-257.559837729067 1.54630862377397e-22
-252.143426656592 4.22901835909255e-22
-246.727015584117 1.15659940108802e-21
-241.310604511642 3.16319784169534e-21
-235.894193439167 8.65106844798082e-21
-230.477782366691 2.36599128594303e-20
-225.061371294216 6.47077849264374e-20
-219.644960221741 1.76970112060967e-19
-214.228549149266 4.83997722970666e-19
-208.812138076791 1.32369128952174e-18
-203.395727004315 3.62017949010458e-18
-197.97931593184 9.90087314490759e-18
-192.562904859365 2.70780190041683e-17
-187.14649378689 7.40560052087154e-17
-181.730082714415 2.02536673994838e-16
-176.313671641939 5.53920025759957e-16
-170.897260569464 1.51492265023436e-15
-165.480849496989 4.14318047635927e-15
-160.064438424514 1.13312349360076e-14
-154.648027352039 3.09899329531075e-14
-149.231616279563 8.47547464916015e-14
-143.815205207088 2.31796792323662e-13
-138.398794134613 6.33943881088275e-13
-132.982383062138 1.73378087047946e-12
-127.565971989663 4.74173849849328e-12
-122.149560917187 1.29682385882338e-11
-116.733149844712 3.5466994254279e-11
-111.316738772237 9.69991161771481e-11
-105.900327699762 2.65284068666539e-10
-100.483916627286 7.25528642547082e-10
-95.0675055548113 1.98425715423524e-09
-89.6510944823361 5.42676914905975e-09
-84.2346834098608 1.48417372891052e-08
-78.8182723373857 4.05908487551899e-08
-73.4018612649105 1.11012408491839e-07
-67.9854501924353 3.03609193133295e-07
-62.5690391199601 8.30344493983548e-07
-57.1526280474848 2.27091930772365e-06
-51.7362169750096 6.21076497713756e-06
-46.3198059025344 1.69858970638214e-05
-40.9033948300592 4.64549375358448e-05
-35.486983757584 0.000127050176587714
-30.0705726851088 0.000347471081163637
-24.6541616126336 0.000950302907778114
-19.2377505401584 0.00259899504012606
-13.8213394676832 0.00710802330847658
-8.404928395208 0.0194398198433635
-2.98851732273278 0.0531662009453129
2.42789374974245 0.0593225052409706
7.84430482221762 0.021690826014213
13.2607158946929 0.00793108671435393
18.6771269671681 0.00289994195838299
24.0935380396432 0.00106034187556797
29.5099491121185 0.000387705998677958
34.9263601845937 0.000141761770306729
40.3427712570689 5.18341206713974e-05
45.7591823295441 1.89527547516062e-05
51.1755934020193 6.92993163618478e-06
56.5920044744945 2.53387716517168e-06
62.0084155469697 9.2649304859712e-07
67.4248266194449 3.38765185975634e-07
72.8412376919201 1.23866931762602e-07
78.2576487643953 4.52910081066737e-08
83.6740598368705 1.65603150585028e-08
89.0904709093458 6.05515413105732e-09
94.5068819818209 2.21402137708941e-09
99.9232930542962 8.09540195362936e-10
105.339704126771 2.96002258465002e-10
110.756115199247 1.08230990280973e-10
116.172526271722 3.9573844192763e-11
121.588937344197 1.44698772516766e-11
127.005348416672 5.29080133480892e-12
132.421759489147 1.9345415498374e-12
137.838170561623 7.07350507271032e-13
143.254581634098 2.58637370791356e-13
148.670992706573 9.45688012975925e-14
154.087403779048 3.45783679732741e-14
159.503814851523 1.26433190998432e-14
164.920225923999 4.62293414148449e-15
170.336636996474 1.69034095459694e-15
175.753048068949 6.18060403921346e-16
181.169459141424 2.25989118855912e-16
186.585870213899 8.26312145499778e-17
192.002281286375 3.02134795364102e-17
197.41869235885 1.10473306082772e-17
202.835103431325 4.03937300308303e-18
208.2515145038 1.4769662316262e-18
213.667925576275 5.40041547957857e-19
219.084336648751 1.97462113402282e-19
224.500747721226 7.22005304531465e-20
229.917158793701 2.63995786730779e-20
235.333569866176 9.6528065616956e-21
240.749980938652 3.52947581745054e-21
246.166392011127 1.29052617664601e-21
};
% \addlegendentry{DNS (Laplace Fit)}
\addplot [very thick, blue, dashed]
table {%
-175.256142928376 2.7570159501075e-10
-171.878133546243 5.51403190021504e-10
-168.500124164109 8.27104785032249e-10
-165.122114781976 8.27104785032249e-10
-161.744105399842 5.51403190021499e-10
-158.366096017709 1.92991116507526e-09
-154.988086635575 1.6542095700645e-09
-151.610077253442 3.58412073513975e-09
-148.232067871308 2.48131435509675e-09
-144.854058489175 5.51403190021504e-09
-141.476049107041 4.41122552017199e-09
-138.098039724908 6.06543509023649e-09
-134.720030342774 1.15794669904516e-08
-131.342020960641 1.26822733704945e-08
-127.964011578507 1.70934988906665e-08
-124.586002196374 2.56402483359997e-08
-121.20799281424 2.78458610960858e-08
-117.829983432107 3.61169089464084e-08
-114.451974049973 5.15561982670102e-08
-111.07396466784 6.53412780175479e-08
-107.695955285706 8.98787199735044e-08
-104.317945903573 1.06972218864171e-07
-100.939936521439 1.45570442165676e-07
-97.561927139306 1.84995770252214e-07
-94.1839177571725 2.80388522125932e-07
-90.8059083750391 3.5979058148903e-07
-87.4278989929056 4.72552533848425e-07
-84.0498896107721 6.82085746056597e-07
-80.6718802286386 9.85081798973409e-07
-77.2938708465051 1.37768087026872e-06
-73.9158614643717 1.97292061389692e-06
-70.5378520822382 2.70876817098063e-06
-67.1598427001047 4.02800030310707e-06
-63.7818333179712 5.69268653378196e-06
-60.4038239358378 8.28951985718824e-06
-57.0258145537043 1.24013334451785e-05
-53.6478051715708 1.84626330114899e-05
-50.2697957894373 2.72081633068258e-05
-46.8917864073038 4.17406700814375e-05
-43.5137770251704 6.45097620069958e-05
-40.1357676430369 0.000100553610030726
-36.7577582609034 0.000159859504431893
-33.3797488787699 0.00025749260746667
-30.0017394966365 0.000424517044949706
-26.623730114503 0.00071427721569324
-23.2457207323695 0.00123038646602944
-19.867711350236 0.00217695550919579
-16.4897019681026 0.00398311305072179
-13.1116925859691 0.00757134339383673
-9.7336832038356 0.0150140132691612
-6.35567382170213 0.0305921770672911
-2.97766443956866 0.0573060964658776
0.400344942564828 0.0722038094284091
3.77835432469831 0.05085112424626
7.15636370683178 0.0260867321812337
10.5343730889653 0.0127847378682802
13.9123824710987 0.00646390991392467
17.2903918532322 0.00340770562562905
20.6684012353657 0.00186741595462301
24.0464106174992 0.0010536201126867
27.4244199996326 0.000613051996577068
30.8024293817661 0.000365169519607688
34.1804387638996 0.000223274179703507
37.5584481460331 0.000138640131171891
40.9364575281666 8.73681812493365e-05
44.3144669103 5.58182692242818e-05
47.6924762924335 3.60554274907208e-05
51.070485674567 2.35156918448469e-05
54.4484950567005 1.54525229971625e-05
57.8265044388339 1.03399126192832e-05
61.2045138209674 6.899156713549e-06
64.5825232031009 4.66321677801182e-06
67.9605325852344 3.12066635392668e-06
71.3385419673678 2.15957059371922e-06
74.7165513495013 1.46176985674699e-06
78.0945607316348 1.04049781957057e-06
81.4725701137683 7.07174591202573e-07
84.8505794959018 5.10323652364898e-07
88.2285888780353 3.57033565538921e-07
91.6065982601687 2.47028629129636e-07
94.9846076423022 1.92164011722492e-07
98.3626170244357 1.27374136894966e-07
101.740626406569 8.49160912633109e-08
105.118635788703 6.86496971576767e-08
108.496645170836 4.65935695568167e-08
111.87465455297 4.13552392516124e-08
115.252663935103 2.17804260058496e-08
118.630673317236 1.92991116507525e-08
122.00868269937 1.2130870180473e-08
125.386692081503 1.2130870180473e-08
128.764701463637 1.2130870180473e-08
132.14271084577 8.54674944533324e-09
135.520720227904 2.48131435509675e-09
138.898729610037 4.13552392516124e-09
142.276738992171 4.41122552017207e-09
145.654748374304 1.92991116507525e-09
149.032757756438 1.37850797505375e-09
152.410767138571 1.6542095700645e-09
155.788776520705 5.51403190021499e-10
159.166785902838 1.102806380043e-09
};
% \addlegendentry{Comp cutoff 0.0001}
\end{axis}

% \draw ({$(current bounding box.south west)!0.5!(current bounding box.south east)$}|-{$(current bounding box.south west)!-0.2!(current bounding box.north west)$}) node[
%   scale=0.6,
%   fill=white,
%   draw=black,
%   line width=0.4pt,
%   inner sep=3.3pt,
%   fill opacity=0.5,
%   anchor=base,
%   text=black,
%   rotate=0.0
% ]{DNS Laplace Fit: $\mu = 0.0146$, $b = 5.3836$};
\end{tikzpicture}

%% file: C5/PDF_grad_dux_dy_SameLimits.tex
% This file was created by tikzplotlib v0.9.6.
\begin{tikzpicture}

\begin{axis}[
legend cell align={left},
legend style={fill opacity=0.8, draw opacity=1, text opacity=1, draw=white!80!black},
log basis y={10},
tick pos=left,
% x grid = true,
% title={PDF of \(\displaystyle \partial u / \partial y\)},
x grid style={white!69.0196078431373!black},
xlabel={\(\displaystyle \partial u / \partial y\)},
xmin=-200, xmax=200,
xtick style={color=black},
y grid style={white!69.0196078431373!black},
ymin=1e-08, ymax=1,
ymode=log,
ytick style={color=black},
ytick={1e-10,1e-08,1e-06,0.0001,0.01,1,100},
% yticklabels={\(\displaystyle 10^{-10}\),,\(\displaystyle 10^{-6}\),,\(\displaystyle 10^{-2}\),,\(\displaystyle 10^{2}\)}
% ]
yticklabels={,,,,,,}
]
\addplot [very thick, black]
table {%
-287.377180683043 1.73681402825852e-10
-282.014933721293 3.47362805651701e-10
-276.652686759543 0
-271.290439797792 1.7368140282585e-10
-265.928192836042 1.73681402825852e-10
-260.565945874291 5.21044208477551e-10
-255.203698912541 3.47362805651704e-10
-249.84145195079 5.21044208477554e-10
-244.47920498904 5.21044208477554e-10
-239.116958027289 3.47362805651703e-10
-233.754711065539 1.21576981978096e-09
-228.392464103788 1.38945122260682e-09
-223.030217142038 2.25785823673605e-09
-217.667970180288 1.04208841695511e-09
-212.305723218537 2.25785823673607e-09
-206.943476256787 2.43153963956192e-09
-201.581229295036 2.77890244521362e-09
-196.218982333286 3.99467226499458e-09
-190.856735371535 6.25253050173068e-09
-185.494488409785 5.21044208477551e-09
-180.132241448034 1.0594565572377e-08
-174.769994486284 1.05945655723769e-08
-169.407747524533 1.52839634486749e-08
-164.045500562783 1.87575915051919e-08
-158.683253601033 2.74416616464845e-08
-153.321006639282 3.0394245494524e-08
-147.958759677532 4.86307927912384e-08
-142.596512715781 6.00937653777449e-08
-137.234265754031 8.14565779253243e-08
-131.87201879228 1.24529565826135e-07
-126.50977183053 1.51623864666968e-07
-121.147524868779 2.16407027921011e-07
-115.785277907029 2.93347889372863e-07
-110.423030945278 4.0901970365488e-07
-105.060783983528 5.80269566841169e-07
-99.6985370217776 8.06576434723253e-07
-94.3362900600271 1.18850183953731e-06
-88.9740430982767 1.81010758025102e-06
-83.6117961365262 2.68928284135548e-06
-78.2495491747758 4.1970110992867e-06
-72.8873022130253 6.56150971735784e-06
-67.5250552512748 1.03189331860923e-05
-62.1628082895244 1.67177034290023e-05
-56.8005613277739 2.71076723088505e-05
-51.4383143660235 4.56386095833546e-05
-46.076067404273 7.91016317844085e-05
-40.7138204425226 0.000142037345956592
-35.3515734807721 0.000265393346543834
-29.9893265190217 0.000523133769434955
-24.6270795572712 0.00109872870131906
-19.2648325955208 0.00249008520891489
-13.9025856337703 0.00628523748878517
-8.54033867201986 0.0182516645260164
-3.1780917102694 0.0560703561352009
2.18415525148106 0.0653943930940704
7.54640221323149 0.0227952898641087
12.9086491749819 0.00758304207188145
18.2708961367324 0.00290341082190863
23.6331430984829 0.00124692027410338
28.9953900602333 0.000583465304653168
34.3576370219838 0.000292260470109769
39.7198839837342 0.000154905748454766
45.0821309454847 8.54693130562123e-05
50.4443779072352 4.90358178226285e-05
55.8066248689856 2.87135305593784e-05
61.168871830736 1.74146868985424e-05
66.5311187924865 1.05744185296492e-05
71.893365754237 6.60736160770383e-06
77.2556127159874 4.21924231884843e-06
82.6178596777379 2.70612993742957e-06
87.9801066394883 1.80819708481995e-06
93.3423536012388 1.21055937769618e-06
98.7046005629892 8.18386770115416e-07
104.06684752474 5.72106540908357e-07
109.42909448649 3.9008843074686e-07
114.791341448241 2.57569520390739e-07
120.153588409991 1.84970694009531e-07
125.515835371742 1.33387317370254e-07
130.878082333492 8.54512501903184e-08
136.240329295242 6.40884376427395e-08
141.602576256993 4.23782622895079e-08
146.964823218743 2.86574314662653e-08
152.327070180494 1.80628658938886e-08
157.689317142244 1.8583910102366e-08
163.051564103995 1.0594565572377e-08
168.413811065745 6.77357471020816e-09
173.776058027496 4.86307927912386e-09
179.138304989246 5.03676068194971e-09
184.500551950996 4.34203507064626e-09
189.862798912747 2.95258384803949e-09
195.225045874497 1.91049543108435e-09
200.587292836248 1.04208841695511e-09
205.949539797998 1.21576981978095e-09
211.311786759749 3.47362805651704e-10
216.674033721499 8.68407014129252e-10
222.03628068325 1.73681402825852e-10
227.398527645 1.73681402825852e-10
232.760774606751 3.47362805651704e-10
238.123021568501 1.73681402825849e-10
243.485268530251 3.47362805651708e-10
};
% \addlegendentry{DNS}
\addplot [very thick, red, dotted]
table {%
-290.058304163919 3.69518190535496e-25
-284.641893091443 1.01059981672955e-24
-279.225482018968 2.76390179355914e-24
-273.809070946493 7.55902880445887e-24
-268.392659874018 2.06732802879581e-23
-262.976248801543 5.65396069945361e-23
-257.559837729067 1.54630862377397e-22
-252.143426656592 4.22901835909255e-22
-246.727015584117 1.15659940108802e-21
-241.310604511642 3.16319784169534e-21
-235.894193439167 8.65106844798082e-21
-230.477782366691 2.36599128594303e-20
-225.061371294216 6.47077849264374e-20
-219.644960221741 1.76970112060967e-19
-214.228549149266 4.83997722970666e-19
-208.812138076791 1.32369128952174e-18
-203.395727004315 3.62017949010458e-18
-197.97931593184 9.90087314490759e-18
-192.562904859365 2.70780190041683e-17
-187.14649378689 7.40560052087154e-17
-181.730082714415 2.02536673994838e-16
-176.313671641939 5.53920025759957e-16
-170.897260569464 1.51492265023436e-15
-165.480849496989 4.14318047635927e-15
-160.064438424514 1.13312349360076e-14
-154.648027352039 3.09899329531075e-14
-149.231616279563 8.47547464916015e-14
-143.815205207088 2.31796792323662e-13
-138.398794134613 6.33943881088275e-13
-132.982383062138 1.73378087047946e-12
-127.565971989663 4.74173849849328e-12
-122.149560917187 1.29682385882338e-11
-116.733149844712 3.5466994254279e-11
-111.316738772237 9.69991161771481e-11
-105.900327699762 2.65284068666539e-10
-100.483916627286 7.25528642547082e-10
-95.0675055548113 1.98425715423524e-09
-89.6510944823361 5.42676914905975e-09
-84.2346834098608 1.48417372891052e-08
-78.8182723373857 4.05908487551899e-08
-73.4018612649105 1.11012408491839e-07
-67.9854501924353 3.03609193133295e-07
-62.5690391199601 8.30344493983548e-07
-57.1526280474848 2.27091930772365e-06
-51.7362169750096 6.21076497713756e-06
-46.3198059025344 1.69858970638214e-05
-40.9033948300592 4.64549375358448e-05
-35.486983757584 0.000127050176587714
-30.0705726851088 0.000347471081163637
-24.6541616126336 0.000950302907778114
-19.2377505401584 0.00259899504012606
-13.8213394676832 0.00710802330847658
-8.404928395208 0.0194398198433635
-2.98851732273278 0.0531662009453129
2.42789374974245 0.0593225052409706
7.84430482221762 0.021690826014213
13.2607158946929 0.00793108671435393
18.6771269671681 0.00289994195838299
24.0935380396432 0.00106034187556797
29.5099491121185 0.000387705998677958
34.9263601845937 0.000141761770306729
40.3427712570689 5.18341206713974e-05
45.7591823295441 1.89527547516062e-05
51.1755934020193 6.92993163618478e-06
56.5920044744945 2.53387716517168e-06
62.0084155469697 9.2649304859712e-07
67.4248266194449 3.38765185975634e-07
72.8412376919201 1.23866931762602e-07
78.2576487643953 4.52910081066737e-08
83.6740598368705 1.65603150585028e-08
89.0904709093458 6.05515413105732e-09
94.5068819818209 2.21402137708941e-09
99.9232930542962 8.09540195362936e-10
105.339704126771 2.96002258465002e-10
110.756115199247 1.08230990280973e-10
116.172526271722 3.9573844192763e-11
121.588937344197 1.44698772516766e-11
127.005348416672 5.29080133480892e-12
132.421759489147 1.9345415498374e-12
137.838170561623 7.07350507271032e-13
143.254581634098 2.58637370791356e-13
148.670992706573 9.45688012975925e-14
154.087403779048 3.45783679732741e-14
159.503814851523 1.26433190998432e-14
164.920225923999 4.62293414148449e-15
170.336636996474 1.69034095459694e-15
175.753048068949 6.18060403921346e-16
181.169459141424 2.25989118855912e-16
186.585870213899 8.26312145499778e-17
192.002281286375 3.02134795364102e-17
197.41869235885 1.10473306082772e-17
202.835103431325 4.03937300308303e-18
208.2515145038 1.4769662316262e-18
213.667925576275 5.40041547957857e-19
219.084336648751 1.97462113402282e-19
224.500747721226 7.22005304531465e-20
229.917158793701 2.63995786730779e-20
235.333569866176 9.6528065616956e-21
240.749980938652 3.52947581745054e-21
246.166392011127 1.29052617664601e-21
};
% \addlegendentry{DNS (Laplace Fit)}
% \addlegendentry{Laplace Fit}
\addplot [very thick, blue, dashed]
table {%
-230.466461467867 2.12610856877636e-10
-226.086052168489 2.12610856877636e-10
-221.705642869112 8.50443427510546e-10
-217.325233569734 6.37832570632909e-10
-212.944824270356 6.37832570632909e-10
-208.564414970978 1.06305428438818e-09
-204.1840056716 1.06305428438819e-09
-199.803596372223 6.37832570632909e-10
-195.423187072845 1.27566514126582e-09
-191.042777773467 1.48827599814345e-09
-186.662368474089 2.338719425654e-09
-182.281959174711 2.97655199628691e-09
-177.901549875334 4.03960628067509e-09
-173.521140575956 5.10266056506327e-09
-169.140731276578 4.03960628067512e-09
-164.7603219772 8.92965598886073e-09
-160.379912677823 9.78009941637127e-09
-155.999503378445 1.4032316553924e-08
-151.619094079067 1.78593119777215e-08
-147.238684779689 2.38124159702953e-08
-142.858275480311 2.55133028253164e-08
-138.477866180934 4.03960628067509e-08
-134.097456881556 5.29401033625315e-08
-129.717047582178 7.16498587677635e-08
-125.3366382828 9.3761387883038e-08
-120.956228983422 1.13534197572658e-07
-116.575819684045 1.62859916368269e-07
-112.195410384667 2.29194503714092e-07
-107.815001085289 2.87024656784809e-07
-103.434591785911 4.02047130355612e-07
-99.0541824865335 5.57465666733163e-07
-94.6737731871557 7.61572089335694e-07
-90.2933638877779 1.07836226608337e-06
-85.9129545884001 1.54929531406734e-06
-81.5325452890224 2.25176158519105e-06
-77.1521359896446 3.24996955823155e-06
-72.7717266902668 4.85433108423019e-06
-68.391317390889 7.32571968457589e-06
-64.0109080915112 1.1021108987966e-05
-59.6304987921334 1.68464338555564e-05
-55.2500894927556 2.59272561636571e-05
-50.8696801933779 4.08210719096493e-05
-46.4892708940001 6.55725900347731e-05
-42.1088615946223 0.000107615749149755
-37.7284522952445 0.0001811244646392
-33.3480429958667 0.000313894842098719
-28.9676336964889 0.000565390311201563
-24.5872243971111 0.00106255464887452
-20.2068150977334 0.00209151848324924
-15.8264057983556 0.00439090171121801
-11.4459964989778 0.00999548229360088
-7.0655871996 0.0249626808894552
-2.68517790022221 0.0608604645726089
1.69523139915557 0.0696738521882776
6.07564069853336 0.0312683532137721
10.4560499979112 0.0122688785217552
14.8364592972889 0.00523246757239963
19.2168685966667 0.00243067128253415
23.5972778960445 0.00120820392898941
27.9776871954223 0.000633299707164278
32.3580964948001 0.000347903011710581
36.7385057941779 0.000199145573479005
41.1189150935556 0.00011727572343199
45.4993243929334 7.10590131965005e-05
49.8797336923112 4.38911726829624e-05
54.260142991689 2.74469985686185e-05
58.6405522910668 1.75314660364161e-05
63.0209615904446 1.12715645773679e-05
67.4013708898224 7.37568323594208e-06
71.7817801892002 4.89982980760201e-06
76.162189488578 3.3026970507372e-06
80.5425987879557 2.27621183373198e-06
84.9230080873335 1.52505767638329e-06
89.3034173867113 1.06156600839004e-06
93.6838266860891 7.1415986825199e-07
98.0642359854668 5.23022707918986e-07
102.444645284845 3.67178949827678e-07
106.825054584222 2.57684358535695e-07
111.2054638836 1.7540395692405e-07
115.585873182978 1.11620699860759e-07
119.966282482356 8.16425690410124e-08
124.346691781734 5.93184290688606e-08
128.727101081111 4.25221713755273e-08
133.107510380489 2.76394113940927e-08
137.487919679867 2.55133028253164e-08
141.868328979245 1.65836468364556e-08
146.248738278623 1.12683754145147e-08
150.629147578 7.016158276962e-09
155.009556877378 7.016158276962e-09
159.389966176756 6.37832570632909e-09
163.770375476134 4.25221713755273e-09
168.150784775511 3.82699542379746e-09
172.531194074889 1.48827599814345e-09
176.911603374267 2.338719425654e-09
181.292012673645 1.06305428438818e-09
185.672421973023 1.48827599814345e-09
190.0528312724 6.37832570632917e-10
194.433240571778 8.50443427510546e-10
198.813649871156 1.27566514126582e-09
203.194059170534 1.27566514126581e-09
};
% \addlegendentry{Comp cutoff 1e-05}
% \addlegendentry{MPS}
\end{axis}

% \draw ({$(current bounding box.south west)!0.5!(current bounding box.south east)$}|-{$(current bounding box.south west)!-0.2!(current bounding box.north west)$}) node[
%   scale=0.6,
%   fill=white,
%   draw=black,
%   line width=0.4pt,
%   inner sep=3.3pt,
%   fill opacity=0.5,
%   anchor=base,
%   text=black,
%   rotate=0.0
% ]{DNS Laplace Fit: $\mu = 0.0146$, $b = 5.3836$};
\end{tikzpicture}

%% file: PDF_Dissipation.tex
% This file was created by tikzplotlib v0.9.6.
\begin{tikzpicture}

\begin{axis}[
width=8cm,         
height=6cm,
legend cell align={left},
legend style={fill opacity=0.8, text opacity=1, draw=none},
log basis y={10},
tick align=inside,               
tick pos=left,
tick style={line width=0.5pt},     
axis line style={line width=0.5pt},
x grid style={white!69.0196078431373!black},
xlabel={\(\displaystyle \varepsilon\)},
xmin=-0.5, xmax=16.5,
xtick={0,2,4,6,8,10,12,14,16}, 
xticklabels={0,,4,,8,,12,,16},
xtick style={color=black},
y grid style={white!69.0196078431373!black},
ylabel={PDF},
ymin=1e-08, ymax=10,
ymode=log,
ytick style={color=black},
ytick={1e-08,1e-06,0.0001,0.01,1,100,10000},
yticklabels={,\(\displaystyle 10^{-6}\),,\(\displaystyle 10^{-2}\),,\(\displaystyle 10^{2}\),}
]
\addplot [very thick, black]
table {%
1.12254841641697e-06 4.35512800171782
0.206031975178144 0.349341421983016
0.412062827807871 0.0864265851935135
0.618093680437598 0.0313622426675007
0.824124533067325 0.0139108405323197
1.03015538569705 0.00697790653017465
1.23618623832678 0.00384195302977585
1.44221709095651 0.00224785794939705
1.64824794358623 0.00138918055222149
1.85427879621596 0.000892290395308842
2.06030964884569 0.000591251552290833
2.26634050147541 0.000402207819782519
2.47237135410514 0.000283346363903747
2.67840220673487 0.000200493668531928
2.8844330593646 0.000147361987512757
3.09046391199432 0.00010813024856726
3.29649476462405 8.10445728256512e-05
3.50252561725378 6.25293979528825e-05
3.70855646988351 4.93210627531195e-05
3.91458732251323 3.73060884338278e-05
4.12061817514296 2.93232273924926e-05
4.32664902777269 2.33338214583084e-05
4.53267988040241 1.88225557056172e-05
4.73871073303214 1.48266048785842e-05
4.94474158566187 1.18884057410598e-05
5.15077243829159 9.91303185937046e-06
5.35680329092132 7.88341430129598e-06
5.56283414355105 6.6538909699012e-06
5.76886499618078 5.4831316212569e-06
5.9748958488105 4.65591555638465e-06
6.18092670144023 3.87390255516666e-06
6.38695755406996 3.23201905127674e-06
6.59298840669968 2.73930565744574e-06
6.79901925932941 2.36412022911571e-06
7.00505011195914 1.92565051166977e-06
7.21108096458887 1.65443212974445e-06
7.41711181721859 1.55498538970517e-06
7.62314266984832 1.31088884597238e-06
7.82917352247805 1.16171873591345e-06
8.03520437510778 9.2214249854609e-07
8.2412352277375 8.67898822161026e-07
8.44726608036723 7.23249018467521e-07
8.65329693299696 6.73525648447879e-07
8.85932778562668 6.14761665697393e-07
9.06535863825641 5.10794619292687e-07
9.27138949088614 4.88193087465577e-07
9.47742034351587 3.39022977406651e-07
9.68345119614559 3.70665121964605e-07
9.88948204877532 3.2094175194496e-07
10.095512901405 3.34502671041229e-07
10.3015437540348 2.16974705540256e-07
10.5075746066645 1.76291948251458e-07
10.7136054592942 1.5821072278977e-07
10.919636311924 2.21495011905678e-07
11.1256671645537 9.94467400392842e-08
11.3316980171834 1.1300765913555e-07
11.5377288698131 1.1300765913555e-07
11.7437597224429 1.1300765913555e-07
11.9497905750726 1.22048271866393e-07
12.1558214277023 8.13655145775961e-08
12.361852280332 6.32842891159081e-08
12.5678831329618 8.58858209430182e-08
12.7739139855915 5.42436763850641e-08
12.9799448382212 3.61624509233761e-08
13.185975690851 4.97233700196421e-08
13.3920065434807 4.52030636542201e-08
13.5980373961104 3.16421445579541e-08
13.8040682487401 1.3560919096266e-08
14.0100991013699 1.3560919096266e-08
14.2161299539996 1.35609190962659e-08
14.4221608066293 1.8081225461688e-08
14.628191659259 1.8081225461688e-08
14.8342225118888 2.7121838192532e-08
15.0402533645185 1.3560919096266e-08
15.2462842171482 4.52030636542201e-09
15.452315069778 0
15.6583459224077 1.3560919096266e-08
15.8643767750374 0
16.0704076276671 0
16.2764384802969 0
16.4824693329266 9.04061273084402e-09
16.6885001855563 9.04061273084402e-09
16.894531038186 0
17.1005618908158 0
17.3065927434455 4.52030636542201e-09
17.5126235960752 0
17.7186544487049 0
17.9246853013347 0
18.1307161539644 9.04061273084402e-09
18.3367470065941 0
18.5427778592239 0
18.7488087118536 0
18.9548395644833 0
19.160870417113 0
19.3669012697428 4.52030636542201e-09
19.5729321223725 4.52030636542201e-09
19.7789629750022 0
19.9849938276319 0
20.1910246802617 0
20.3970555328914 4.52030636542201e-09
};
\addlegendentry{DNS}
\addplot [very thick, red, dotted]
table {%
1.12254841641697e-06 9.65870371117685
0.0206248315203611 10.3697631638584
0.0412485404923057 7.29978377507986
0.0618722494642504 5.1049714876499
0.0824959584361951 3.66334186393509
0.10311966740814 2.70165236715533
0.123743376380084 2.04142687595961
0.144367085352029 1.57513360498616
0.164990794323974 1.23734532784304
0.185614503295918 0.98715702177075
0.206238212267863 0.798230633247491
0.226861921239808 0.653128985862958
0.247485630211752 0.540012435263796
0.268109339183697 0.450657930711301
0.288733048155642 0.379237515777093
0.309356757127586 0.32154519387089
0.329980466099531 0.274495710329244
0.350604175071476 0.235792441398946
0.37122788404342 0.203702919875887
0.391851593015365 0.176904344313898
0.41247530198731 0.154375485181187
0.433099010959254 0.135319904486506
0.453722719931199 0.11911065787026
0.474346428903144 0.105249958295186
0.494970137875088 0.0933394055671883
0.515593846847033 0.0830577738351823
0.536217555818978 0.0741442702162205
0.556841264790922 0.0663857979444887
0.577464973762867 0.059607180936291
0.598088682734812 0.0536635995423117
0.618712391706756 0.048434692245475
0.639336100678701 0.0438199231492905
0.659959809650646 0.0397349188753815
0.68058351862259 0.0361085534492952
0.701207227594535 0.0328806144058414
0.72183093656648 0.0299999235412295
0.742454645538424 0.0274228155482812
0.763078354510369 0.0251119000505346
0.783702063482314 0.0230350493267106
0.804325772454258 0.0211645667374693
0.824949481426203 0.0194765005764441
0.845573190398148 0.0179500755268172
0.866196899370092 0.0165672196696363
0.886820608342037 0.0153121694712601
0.907444317313982 0.0141711386798425
0.928068026285926 0.0131320398127563
0.948691735257871 0.0121842490900482
0.969315444229816 0.0113184073934152
0.98993915320176 0.0105262512047936
1.0105628621737 0.00980046857928204
1.03118657114565 0.00913457609212639
1.05181028011759 0.00852281341403134
1.07243398908954 0.00796005274823491
1.09305769806148 0.00744172083400993
1.11368140703343 0.00696373160605296
1.13430511600537 0.00652242791454579
1.15492882497732 0.00611453096994315
1.17555253394926 0.00573709639041593
1.19617624292121 0.00538747590686606
1.21679995189315 0.00506328392733577
1.2374236608651 0.00476236828493438
1.25804736983704 0.00448278459550872
1.27867107880899 0.00422277373676332
1.29929478778093 0.00398074203228926
1.31991849675287 0.00375524378434584
1.34054220572482 0.00354496585018942
1.36116591469676 0.00334871399983283
1.38178962366871 0.00316540082965036
1.40241333264065 0.00299403503728353
1.4230370416126 0.00283371188973731
1.44366075058454 0.00268360473911753
1.46428445955649 0.00254295745975496
1.48490816852843 0.0024110776969981
1.50553187750038 0.00228733083215656
1.52615558647232 0.0021711345802942
1.54677929544427 0.00206195414810312
1.56740300441621 0.0019592978881845
1.58802671338816 0.0018627133939312
1.6086504223601 0.00177178398602632
1.62927413133204 0.00168612554749242
1.64989784030399 0.00160538366937433
1.67052154927593 0.00152923107362261
1.69114525824788 0.00145736528365646
1.71176896721982 0.00138950651650324
1.73239267619177 0.00132539577340194
1.75301638516371 0.00126479310837971
1.77364009413566 0.00120747605661022
1.7942638031076 0.00115323820638444
1.81488751207955 0.00110188790030349
1.83551122105149 0.00105324705287083
1.85613493002344 0.00100715007304476
1.87675863899538 0.000963442881534205
1.89738234796733 0.000921982013702417
1.91800605693927 0.0008826337999007
1.93862976591121 0.000845273615903921
1.95925347488316 0.000809785196873572
1.9798771838551 0.000776060008944561
2.00050089282705 0.000743996673128529
2.02112460179899 0.00071350043675791
2.04174831077094 0.000684482688169073
2.06237201974288 0.000656860510746045
2.08299572871483 0.000630556272824649
2.10361943768677 0.000605497250295314
2.12424314665872 0.000581615279045929
2.14486685563066 0.00055884643465787
2.16549056460261 0.000537130737012154
2.18611427357455 0.000516411877681747
2.2067379825465 0.000496636968182933
2.22736169151844 0.000477756307335893
2.24798540049038 0.000459723166144267
2.26860910946233 0.000442493588747328
2.28923281843427 0.000426026208128299
2.30985652740622 0.000410282075379515
2.33048023637816 0.000395224501431206
2.35110394535011 0.000380818910246445
2.37172765432205 0.000367032702571667
2.392351363294 0.000353835129410774
2.41297507226594 0.000341197174462118
2.43359878123789 0.000329091444822417
2.45422249020983 0.000317492069320357
2.47484619918178 0.000306374603896089
2.49546990815372 0.00029571594349136
2.51609361712566 0.000285494239959274
2.53671732609761 0.000275688825542885
2.55734103506955 0.000266280141508573
2.5779647440415 0.000257249671553617
2.59858845301344 0.0002485798796379
2.61921216198539 0.000240254151917638
2.63983587095733 0.000232256742484476
2.66045957992928 0.00022457272263665
2.68108328890122 0.000217187933430242
2.70170699787317 0.000210088941278091
2.72233070684511 0.000203262996381855
2.74295441581706 0.000196697993799099
2.763578124789 0.000190382436962375
2.78420183376095 0.00018430540348107
2.80482554273289 0.000178456513069491
2.82544925170484 0.000172825897456339
2.84607296067678 0.000167404172141424
2.86669666964872 0.000162182409875396
2.88732037862067 0.000157152115747274
2.90794408759261 0.000152305203773019
2.92856779656456 0.000147633974886023
2.9491915055365 0.000143131096237551
2.96981521450845 0.000138789581721719
2.99043892348039 0.000134602773645634
3.01106263245234 0.000130564325470925
3.03168634142428 0.00012666818555807
3.05231005039623 0.000122908581849662
3.07293375936817 0.000119280007433232
3.09355746834012 0.000115777206928268
3.11418117731206 0.000112395163645909
3.134804886284 0.00010912908747326
3.15542859525595 0.000105974403437539
3.17605230422789 0.000102926740908264
3.19667601319984 9.99819233984997e-05
3.21729972217178 9.71359589287405e-05
3.23792343114373 9.43850309194569e-05
3.25854714011567 9.17254895805129e-05
3.27917084908762 8.9153843767781e-05
3.29979455805956 8.66667532791851e-05
3.32041826703151 8.42610215641992e-05
3.34104197600345 8.19335888225042e-05
3.3616656849754 7.96815254690521e-05
3.38228939394734 7.75020259442291e-05
3.40291310291929 7.53924028491657e-05
3.42353681189123 7.33500813874858e-05
3.44416052086317 7.13725940959611e-05
3.46478422983512 6.94575758476279e-05
3.48540793880706 6.76027591119382e-05
3.50603164777901 6.5805969457465e-05
3.52665535675095 6.40651212835773e-05
3.5472790657229 6.23782137683107e-05
3.56790277469484 6.0743327020451e-05
3.58852648366679 5.91586184245597e-05
3.60915019263873 5.7622319168346e-05
3.62977390161068 5.613273094243e-05
3.65039761058262 5.46882228031246e-05
3.67102131955457 5.32872281894177e-05
3.69164502852651 5.19282420858621e-05
3.71226873749846 5.06098183235564e-05
3.7328924464704 4.93305670118576e-05
3.75351615544234 4.80891520938974e-05
3.77413986441429 4.68842890193655e-05
3.79476357338623 4.57147425284018e-05
3.81538728235818 4.45793245407927e-05
3.83601099133012 4.34768921449931e-05
3.85663470030207 4.24063456818048e-05
3.87725840927401 4.13666269178364e-05
3.89788211824596 4.03567173041395e-05
3.9185058272179 3.93756363156741e-05
3.93912953618985 3.84224398674961e-05
3.95975324516179 3.74962188037871e-05
3.98037695413374 3.65960974560604e-05
4.00100066310568 3.57212322670758e-05
4.02162437207762 3.48708104771876e-05
4.04224808104957 3.40440488700235e-05
4.06287179002151 3.32401925745632e-05
4.08349549899346 3.24585139208405e-05
4.1041192079654 3.16983113466435e-05
4.12474291693735 3.09589083527238e-05
4.14536662590929 3.02396525041602e-05
4.16599033488124 2.95399144756459e-05
4.18661404385318 2.88590871385854e-05
4.20723775282513 2.81965846879963e-05
4.22786146179707 2.75518418073183e-05
4.24848517076902 2.69243128693281e-05
4.26910887974096 2.63134711714513e-05
4.28973258871291 2.57188082038533e-05
4.31035629768485 2.51398329487692e-05
4.33098000665679 2.45760712096168e-05
4.35160371562874 2.40270649685049e-05
4.37222742460068 2.34923717708236e-05
4.39285113357263 2.29715641356677e-05
4.41347484254457 2.24642289909053e-05
4.43409855151652 2.19699671317671e-05
4.45472226048846 2.14883927018822e-05
4.47534596946041 2.1019132695745e-05
4.49596967843235 2.05618264816425e-05
4.5165933874043 2.0116125344124e-05
4.53721709637624 1.96816920451362e-05
4.55784080534819 1.92582004029898e-05
4.57846451432013 1.88453348883676e-05
4.59908822329208 1.84427902366173e-05
4.61971193226402 1.80502710756122e-05
4.64033564123597 1.76674915684957e-05
4.66095935020791 1.72941750706584e-05
4.68158305917985 1.6930053800328e-05
4.7022067681518 1.65748685221801e-05
4.72283047712374 1.62283682434094e-05
4.74345418609569 1.58903099217209e-05
4.76407789506763 1.55604581847334e-05
4.78470160403958 1.52385850603054e-05
4.80532531301152 1.49244697173189e-05
4.82594902198347 1.46178982164779e-05
4.84657273095541 1.43186632706982e-05
4.86719643992736 1.40265640146845e-05
4.8878201488993 1.37414057833109e-05
4.90844385787125 1.34629998984345e-05
4.92906756684319 1.31911634637946e-05
4.94969127581513 1.29257191676584e-05
4.97031498478708 1.26664950928967e-05
4.99093869375902 1.24133245341813e-05
5.01156240273097 1.21660458220135e-05
5.03218611170291 1.19245021533035e-05
5.05280982067486 1.16885414282354e-05
5.0734335296468 1.14580160931615e-05
5.09405723861875 1.1232782989283e-05
5.11468094759069 1.10127032068842e-05
5.13530465656264 1.07976419448962e-05
5.15592836553458 1.05874683755795e-05
5.17655207450653 1.0382055514118e-05
5.19717578347847 1.01812800929335e-05
5.21779949245041 9.98502244052992e-06
5.23842320142236 9.79316636469251e-06
5.2590469103943 9.60559903986772e-06
5.27967061936625 9.42221089856243e-06
5.30029432833819 9.24289552660445e-06
5.32091803731014 9.06754956211477e-06
5.34154174628208 8.89607259804755e-06
5.36216545525403 8.72836708816026e-06
5.38278916422597 8.56433825628207e-06
5.40341287319792 8.40389400875358e-06
5.42403658216986 8.24694484991738e-06
5.44466029114181 8.09340380054289e-06
5.46528400011375 7.94318631907406e-06
5.4859077090857 7.79621022559327e-06
5.50653141805764 7.65239562839912e-06
5.52715512702959 7.51166485309996e-06
5.54777883600153 7.37394237412848e-06
5.56840254497347 7.23915474858754e-06
5.58902625394542 7.10723055234039e-06
5.60964996291736 6.97810031826195e-06
5.63027367188931 6.85169647657108e-06
5.65089738086125 6.72795329716791e-06
5.6715210898332 6.60680683390182e-06
5.69214479880514 6.4881948706998e-06
5.71276850777709 6.37205686948734e-06
5.73339221674903 6.25833391983668e-06
5.75401592572098 6.14696869027948e-06
5.77463963469292 6.0379053812244e-06
5.79526334366487 5.9310896794212e-06
5.81588705263681 5.8264687139162e-06
5.83651076160876 5.72399101344552e-06
5.8571344705807 5.62360646521505e-06
5.87775817955264 5.52526627501762e-06
5.89838188852459 5.42892292864011e-06
5.91900559749653 5.33453015451471e-06
5.93962930646848 5.24204288757098e-06
5.96025301544042 5.15141723424582e-06
5.98087672441237 5.06261043861147e-06
6.00150043338431 4.97558084958229e-06
6.02212414235626 4.89028788916251e-06
6.0427478513282 4.80669202169937e-06
6.06337156030015 4.72475472410635e-06
6.08399526927209 4.64443845702353e-06
6.10461897824404 4.56570663688244e-06
6.12524268721598 4.48852360884468e-06
6.14586639618792 4.41285462058441e-06
6.16649010515987 4.33866579688576e-06
6.18711381413181 4.26592411502797e-06
6.20773752310376 4.19459738093097e-06
6.2283612320757 4.12465420603606e-06
6.24898494104765 4.05606398489718e-06
6.26960865001959 3.98879687345838e-06
6.29023235899154 3.92282376799487e-06
6.31085606796348 3.85811628469555e-06
6.33147977693543 3.7946467398655e-06
6.35210348590737 3.732388130728e-06
6.37272719487932 3.67131411680603e-06
6.39335090385126 3.61139900186461e-06
6.41397461282321 3.55261771639495e-06
6.43459832179515 3.49494580062311e-06
6.45522203076709 3.43835938802576e-06
6.47584573973904 3.38283518933664e-06
6.49646944871098 3.32835047702775e-06
6.51709315768293 3.2748830702497e-06
6.53771686665487 3.22241132021678e-06
6.55834057562682 3.17091409602197e-06
6.57896428459876 3.12037077086832e-06
6.59958799357071 3.07076120870341e-06
6.62021170254265 3.02206575124377e-06
6.6408354115146 2.97426520537706e-06
6.66145912048654 2.92734083092982e-06
6.68208282945849 2.88127432878938e-06
6.70270653843043 2.83604782936864e-06
6.72333024740238 2.79164388140287e-06
6.74395395637432 2.7480454410682e-06
6.76457766534627 2.70523586141159e-06
6.78520137431821 2.66319888208272e-06
6.80582508329015 2.62191861935817e-06
6.8264487922621 2.58137955644904e-06
6.84707250123404 2.54156653408293e-06
6.86769621020599 2.50246474135216e-06
6.88831991917793 2.46405970681962e-06
6.90894362814988 2.42633728987467e-06
6.92956733712182 2.38928367233121e-06
6.95019104609377 2.3528853502606e-06
6.97081475506571 2.3171291260522e-06
6.99143846403766 2.28200210069474e-06
7.0120621730096 2.24749166627153e-06
7.03268588198155 2.21358549866351e-06
7.05330959095349 2.18027155045333e-06
7.07393329992543 2.14753804402471e-06
7.09455700889738 2.11537346485133e-06
7.11518071786932 2.08376655496919e-06
7.13580442684127 2.05270630662731e-06
7.15642813581321 2.02218195611123e-06
7.17705184478516 1.9921829777344e-06
7.1976755537571 1.96269907799217e-06
7.21829926272905 1.93372018987389e-06
7.23892297170099 1.90523646732824e-06
7.25954668067294 1.87723827987741e-06
7.28017038964488 1.84971620737565e-06
7.30079409861683 1.82266103490806e-06
7.32141780758877 1.79606374782552e-06
7.34204151656072 1.76991552691164e-06
7.36266522553266 1.74420774367823e-06
7.3832889345046 1.71893195578513e-06
7.40391264347655 1.69407990258119e-06
7.42453635244849 1.66964350076271e-06
7.44516006142044 1.64561484014592e-06
7.46578377039238 1.62198617955043e-06
7.48640747936433 1.59874994279029e-06
7.50703118833627 1.57589871476967e-06
7.52765489730822 1.5534252376801e-06
7.54827860628016 1.53132240729661e-06
7.56890231525211 1.50958326936957e-06
7.58952602422405 1.48820101610999e-06
7.610149733196 1.46716898276531e-06
7.63077344216794 1.44648064428318e-06
7.65139715113989 1.42612961206095e-06
7.67202086011183 1.40610963077827e-06
7.69264456908377 1.38641457531054e-06
7.71326827805572 1.36703844772103e-06
7.73389198702766 1.34797537432938e-06
7.75451569599961 1.32921960285435e-06
7.77513940497155 1.31076549962898e-06
7.7957631139435 1.29260754688584e-06
7.81638682291544 1.27474034011073e-06
7.83701053188739 1.2571585854627e-06
7.85763424085933 1.23985709725881e-06
7.87825794983128 1.22283079552171e-06
7.89888165880322 1.20607470358837e-06
7.91950536777517 1.18958394577826e-06
7.94012907674711 1.1733537451194e-06
7.96075278571906 1.15737942113084e-06
7.981376494691 1.14165638765969e-06
8.00200020366294 1.12618015077171e-06
8.02262391263489 1.11094630669351e-06
8.04324762160683 1.09595053980549e-06
8.06387133057878 1.08118862068368e-06
8.08449503955072 1.06665640418959e-06
8.10511874852267 1.05234982760645e-06
8.12574245749461 1.03826490882085e-06
8.14636616646656 1.02439774454844e-06
8.1669898754385 1.01074450860258e-06
8.18761358441045 9.97301450204837e-07
8.20823729338239 9.84064892336082e-07
8.22886100235434 9.71031230127349e-07
8.24948471132628 9.58196929289158e-07
8.27010842029823 9.45558524578551e-07
8.29073212927017 9.3311261830263e-07
8.31135583824211 9.20855878857781e-07
8.33197954721406 9.08785039303656e-07
8.352603256186 8.96896895970927e-07
8.37322696515795 8.85188307102013e-07
8.39385067412989 8.73656191523896e-07
8.41447438310184 8.62297527352228e-07
8.43509809207378 8.51109350725889e-07
8.45572180104573 8.40088754571232e-07
8.47634551001767 8.29232887395246e-07
8.49696921898962 8.18538952106954e-07
8.51759292796156 8.08004204866204e-07
8.53821663693351 7.97625953959308e-07
8.55884034590545 7.87401558700725e-07
8.5794640548774 7.77328428360207e-07
8.60008776384934 7.674040211147e-07
8.62071147282129 7.5762584302446e-07
8.64133518179323 7.47991447032676e-07
8.66195889076517 7.38498431988096e-07
8.68258259973712 7.29144441690033e-07
8.70320630870906 7.19927163955202e-07
8.72383001768101 7.10844329705838e-07
8.74445372665295 7.01893712078599e-07
8.7650774356249 6.9307312555368e-07
8.78570114459684 6.84380425103696e-07
8.80632485356879 6.75813505361795e-07
8.82694856254073 6.67370299808574e-07
8.84757227151268 6.59048779977286e-07
8.86819598048462 6.50846954676945e-07
8.88881968945657 6.42762869232828e-07
8.90944339842851 6.34794604744002e-07
8.93006710740046 6.26940277357425e-07
8.9506908163724 6.19198037558245e-07
8.97131452534435 6.11566069475872e-07
8.99193823431629 6.04042590205469e-07
9.01256194328823 5.96625849144475e-07
9.03318565226018 5.89314127343791e-07
9.05380936123212 5.821057368733e-07
9.07443307020407 5.74999020201346e-07
9.09505677917601 5.67992349587841e-07
9.11568048814796 5.61084126490717e-07
9.1363041971199 5.54272780985334e-07
9.15692790609185 5.4755677119659e-07
9.17755161506379 5.40934582743396e-07
9.19817532403574 5.34404728195267e-07
9.21879903300768 5.27965746540676e-07
9.23942274197963 5.21616202666947e-07
9.26004645095157 5.153546868514e-07
9.28067015992352 5.09179814263466e-07
9.30129386889546 5.03090224477529e-07
9.3219175778674 4.97084580996256e-07
9.34254128683935 4.91161570784132e-07
9.36316499581129 4.8531990381099e-07
9.38378870478324 4.7955831260531e-07
9.40441241375518 4.73875551817025e-07
9.42503612272713 4.68270397789651e-07
9.44565983169907 4.62741648141483e-07
9.46628354067102 4.57288121355704e-07
9.48690724964296 4.51908656379137e-07
9.50753095861491 4.46602112229491e-07
9.52815466758685 4.41367367610879e-07
9.5487783765588 4.36203320537426e-07
9.56940208553074 4.31108887964779e-07
9.59002579450268 4.26083005429341e-07
9.61064950347463 4.21124626695054e-07
9.63127321244657 4.1623272340755e-07
9.65189692141852 4.11406284755515e-07
9.67252063039046 4.06644317139091e-07
9.69314433936241 4.0194584384516e-07
9.71376804833435 3.97309904729366e-07
9.7343917573063 3.92735555904692e-07
9.75501546627824 3.88221869436493e-07
9.77563917525019 3.8376793304378e-07
9.79626288422213 3.79372849806672e-07
9.81688659319408 3.75035737879833e-07
9.83751030216602 3.707557302118e-07
9.85813401113796 3.66531974270019e-07
9.87875772010991 3.62363631771523e-07
9.89938142908185 3.5824987841908e-07
9.9200051380538 3.54189903642703e-07
9.94062884702574 3.5018291034642e-07
9.96125255599769 3.46228114660155e-07
9.98187626496963 3.42324745696637e-07
10.0024999739416 3.38472045313214e-07
10.0231236829135 3.34669267878461e-07
10.0437473918855 3.30915680043487e-07
10.0643711008574 3.27210560517815e-07
10.0849948098294 3.2355319984978e-07
10.1056185188013 3.19942900211291e-07
10.1262422277732 3.16378975186897e-07
10.1468659367452 3.12860749567071e-07
10.1674896457171 3.09387559145581e-07
10.1881133546891 3.05958750520896e-07
10.208737063661 3.02573680901524e-07
10.229360772633 2.99231717915209e-07
10.2499844816049 2.95932239421874e-07
10.2706081905769 2.92674633330273e-07
10.2912318995488 2.89458297418236e-07
10.3118556085207 2.86282639156448e-07
10.3324793174927 2.83147075535683e-07
10.3531030264646 2.80051032897414e-07
10.3737267354366 2.76993946767736e-07
10.3943504444085 2.73975261694525e-07
10.4149741533805 2.70994431087766e-07
10.4355978623524 2.68050917062982e-07
10.4562215713244 2.65144190287704e-07
10.4768452802963 2.62273729830903e-07
10.4974689892682 2.59439023015344e-07
10.5180926982402 2.56639565272776e-07
10.5387164072121 2.53874860001924e-07
10.5593401161841 2.51144418429195e-07
10.579963825156 2.48447759472072e-07
10.600587534128 2.45784409605109e-07
10.6212112430999 2.43153902728507e-07
10.6418349520719 2.4055578003918e-07
10.6624586610438 2.37989589904293e-07
10.6830823700158 2.35454887737192e-07
10.7037060789877 2.32951235875704e-07
10.7243297879596 2.30478203462731e-07
10.7449534969316 2.28035366329113e-07
10.7655772059035 2.25622306878699e-07
10.7862009148755 2.23238613975587e-07
10.8068246238474 2.20883882833484e-07
10.8274483328194 2.18557714907149e-07
10.8480720417913 2.16259717785877e-07
10.8686957507633 2.13989505088963e-07
10.8893194597352 2.11746696363135e-07
10.9099431687071 2.09530916981897e-07
10.9305668776791 2.07341798046746e-07
10.951190586651 2.05178976290227e-07
10.971814295623 2.03042093980794e-07
10.9924380045949 2.00930798829418e-07
11.0130617135669 1.98844743897953e-07
11.0336854225388 1.96783587509165e-07
11.0543091315108 1.94746993158436e-07
11.0749328404827 1.92734629427095e-07
11.0955565494546 1.90746169897335e-07
11.1161802584266 1.88781293068698e-07
11.1368039673985 1.86839682276087e-07
11.1574276763705 1.84921025609279e-07
11.1780513853424 1.83025015833908e-07
11.1986750943144 1.81151350313885e-07
11.2192988032863 1.79299730935241e-07
11.2399225122583 1.77469864031338e-07
11.2605462212302 1.75661460309449e-07
11.2811699302021 1.73874234778664e-07
11.3017936391741 1.72107906679101e-07
11.322417348146 1.70362199412391e-07
11.343041057118 1.68636840473421e-07
11.3636647660899 1.66931561383302e-07
11.3842884750619 1.6524609762354e-07
11.4049121840338 1.63580188571394e-07
11.4255358930058 1.61933577436384e-07
11.4461596019777 1.60306011197941e-07
11.4667833109496 1.58697240544158e-07
11.4874070199216 1.57107019811653e-07
11.5080307288935 1.5553510692648e-07
11.5286544378655 1.53981263346102e-07
11.5492781468374 1.52445254002391e-07
11.5699018558094 1.50926847245638e-07
11.5905255647813 1.49425814789545e-07
11.6111492737533 1.47941931657202e-07
11.6317729827252 1.46474976128e-07
11.6523966916972 1.45024729685494e-07
11.6730204006691 1.43590976966168e-07
11.693644109641 1.42173505709104e-07
11.714267818613 1.40772106706536e-07
11.7348915275849 1.3938657375526e-07
11.7555152365569 1.38016703608903e-07
11.7761389455288 1.36662295931011e-07
11.7967626545008 1.35323153248965e-07
11.8173863634727 1.33999080908687e-07
11.8380100724447 1.32689887030144e-07
11.8586337814166 1.31395382463608e-07
11.8792574903885 1.30115380746685e-07
11.8998811993605 1.28849698062076e-07
11.9205049083324 1.27598153196076e-07
11.9411286173044 1.26360567497778e-07
11.9617523262763 1.25136764838979e-07
11.9823760352483 1.23926571574786e-07
12.0029997442202 1.22729816504875e-07
12.0236234531922 1.21546330835436e-07
12.0442471621641 1.20375948141748e-07
12.064870871136 1.19218504331408e-07
12.085494580108 1.18073837608172e-07
12.1061182890799 1.16941788436425e-07
12.1267419980519 1.15822199506239e-07
12.1473657070238 1.14714915699035e-07
12.1679894159958 1.13619784053821e-07
12.1886131249677 1.12536653734e-07
12.2092368339397 1.11465375994739e-07
12.2298605429116 1.10405804150889e-07
12.2504842518835 1.09357793545441e-07
12.2711079608555 1.08321201518515e-07
12.2917316698274 1.07295887376864e-07
12.3123553787994 1.06281712363898e-07
12.3329790877713 1.052785396302e-07
12.3536027967433 1.04286234204536e-07
12.3742265057152 1.03304662965352e-07
12.3948502146872 1.02333694612743e-07
12.4154739236591 1.01373199640881e-07
12.436097632631 1.00423050310919e-07
12.456721341603 9.94831206243196e-08
12.4773450505749 9.85532862966437e-08
12.4979687595469 9.76334247317663e-08
12.5185924685188 9.67234149965115e-08
12.5392161774908 9.58231377957142e-08
12.5598398864627 9.49324754476883e-08
12.5804635954347 9.40513118600946e-08
12.6010873044066 9.3179532506211e-08
12.6217110133785 9.23170244015825e-08
12.6423347223505 9.14636760810638e-08
12.6629584313224 9.06193775762248e-08
12.6835821402944 8.97840203931354e-08
12.7042058492663 8.89574974905009e-08
12.7248295582383 8.81397032581648e-08
12.7454532672102 8.73305334959492e-08
12.7660769761822 8.65298853928446e-08
12.7867006851541 8.57376575065358e-08
12.8073243941261 8.49537497432563e-08
12.827948103098 8.41780633379703e-08
12.8485718120699 8.34105008348702e-08
12.8691955210419 8.26509660681936e-08
12.8898192300138 8.18993641433425e-08
12.9104429389858 8.11556014183097e-08
12.9310666479577 8.04195854854009e-08
12.9516903569297 7.96912251532476e-08
12.9723140659016 7.89704304291109e-08
12.9929377748736 7.82571125014652e-08
13.0135614838455 7.75511837228589e-08
13.0341851928174 7.6852557593049e-08
13.0548089017894 7.61611487424031e-08
13.0754326107613 7.54768729155638e-08
13.0960563197333 7.47996469553734e-08
13.1166800287052 7.41293887870511e-08
13.1373037376772 7.34660174026212e-08
13.1579274466491 7.28094528455862e-08
13.1785511556211 7.21596161958414e-08
13.199174864593 7.15164295548278e-08
13.2197985735649 7.08798160309154e-08
13.2404222825369 7.02496997250188e-08
13.2610459915088 6.96260057164359e-08
13.2816697004808 6.9008660048909e-08
13.3022934094527 6.83975897169032e-08
13.3229171184247 6.77927226520989e-08
13.3435408273966 6.71939877100945e-08
13.3641645363686 6.6601314657316e-08
13.3847882453405 6.60146341581295e-08
13.4054119543124 6.54338777621537e-08
13.4260356632844 6.48589778917696e-08
13.4466593722563 6.42898678298222e-08
13.4672830812283 6.37264817075116e-08
13.4879067902002 6.31687544924744e-08
13.5085304991722 6.26166219770431e-08
13.5291542081441 6.207002076669e-08
13.5497779171161 6.15288882686492e-08
13.570401626088 6.09931626807083e-08
13.5910253350599 6.04627829801772e-08
13.6116490440319 5.99376889130224e-08
13.6322727530038 5.94178209831682e-08
13.6528964619758 5.89031204419603e-08
13.6735201709477 5.8393529277792e-08
13.6941438799197 5.78889902058855e-08
13.7147675888916 5.73894466582302e-08
13.7353912978636 5.68948427736723e-08
13.7560150068355 5.64051233881568e-08
13.7766387158075 5.59202340251136e-08
13.7972624247794 5.54401208859932e-08
13.8178861337513 5.49647308409404e-08
13.8385098427233 5.4494011419613e-08
13.8591335516952 5.40279108021354e-08
13.8797572606672 5.35663778101895e-08
13.9003809696391 5.31093618982397e-08
13.9210046786111 5.26568131448883e-08
13.941628387583 5.22086822443611e-08
13.962252096555 5.17649204981202e-08
13.9828758055269 5.13254798066004e-08
14.0034995144988 5.08903126610725e-08
14.0241232234708 5.04593721356225e-08
14.0447469324427 5.00326118792564e-08
14.0653706414147 4.96099861081158e-08
14.0859943503866 4.91914495978121e-08
14.1066180593586 4.87769576758763e-08
14.1272417683305 4.83664662143163e-08
14.1478654773025 4.79599316222856e-08
14.1684891862744 4.75573108388617e-08
14.1891128952463 4.71585613259289e-08
14.2097366042183 4.67636410611676e-08
14.2303603131902 4.6372508531144e-08
14.2509840221622 4.59851227245039e-08
14.2716077311341 4.56014431252649e-08
14.2922314401061 4.5221429706207e-08
14.312855149078 4.48450429223592e-08
14.33347885805 4.44722437045827e-08
14.3541025670219 4.41029934532464e-08
14.3747262759938 4.37372540319932e-08
14.3953499849658 4.33749877616013e-08
14.4159736939377 4.30161574139299e-08
14.4365974029097 4.26607262059546e-08
14.4572211118816 4.23086577938904e-08
14.4778448208536 4.19599162673975e-08
14.4984685298255 4.16144661438717e-08
14.5190922387975 4.12722723628188e-08
14.5397159477694 4.09333002803068e-08
14.5603396567413 4.05975156635009e-08
14.5809633657133 4.02648846852749e-08
14.6015870746852 3.99353739189015e-08
14.6222107836572 3.96089503328186e-08
14.6428344926291 3.9285581285469e-08
14.6634582016011 3.8965234520215e-08
14.684081910573 3.86478781603271e-08
14.704705619545 3.83334807040403e-08
14.7253293285169 3.80220110196854e-08
14.7459530374888 3.77134383408857e-08
14.7665767464608 3.74077322618245e-08
14.7872004554327 3.71048627325783e-08
14.8078241644047 3.68048000545171e-08
14.8284478733766 3.65075148757694e-08
14.8490715823486 3.6212978186751e-08
14.8696952913205 3.59211613157571e-08
14.8903190002925 3.56320359246173e-08
14.9109427092644 3.53455740044112e-08
14.9315664182364 3.50617478712435e-08
14.9521901272083 3.47805301620797e-08
14.9728138361802 3.45018938306404e-08
14.9934375451522 3.4225812143351e-08
15.0140612541241 3.39522586753512e-08
15.0346849630961 3.36812073065578e-08
15.055308672068 3.34126322177833e-08
15.07593238104 3.31465078869091e-08
15.0965560900119 3.28828090851118e-08
15.1171797989839 3.26215108731407e-08
15.1378035079558 3.23625885976491e-08
15.1584272169277 3.21060178875754e-08
15.1790509258997 3.18517746505742e-08
15.1996746348716 3.1599835069497e-08
15.2202983438436 3.13501755989222e-08
15.2409220528155 3.11027729617314e-08
15.2615457617875 3.08576041457352e-08
15.2821694707594 3.06146464003432e-08
15.3027931797314 3.03738772332816e-08
15.3234168887033 3.01352744073552e-08
15.3440405976752 2.98988159372527e-08
15.3646643066472 2.96644800863981e-08
15.3852880156191 2.9432245363844e-08
15.4059117245911 2.92020905212063e-08
15.426535433563 2.89739945496431e-08
15.447159142535 2.87479366768727e-08
15.4677828515069 2.85238963642335e-08
15.4884065604789 2.83018533037838e-08
15.5090302694508 2.80817874154401e-08
15.5296539784227 2.78636788441552e-08
15.5502776873947 2.76475079571346e-08
15.5709013963666 2.74332553410895e-08
15.5915251053386 2.72209017995285e-08
15.6121488143105 2.70104283500846e-08
15.6327725232825 2.6801816221879e-08
15.6533962322544 2.65950468529203e-08
15.6740199412264 2.63901018875383e-08
15.6946436501983 2.61869631738531e-08
15.7152673591702 2.59856127612774e-08
15.7358910681422 2.5786032898053e-08
15.7565147771141 2.55882060288191e-08
15.7771384860861 2.53921147922154e-08
15.797762195058 2.51977420185147e-08
15.81838590403 2.50050707272889e-08
15.8390096130019 2.48140841251046e-08
15.8596333219739 2.46247656032516e-08
15.8802570309458 2.44370987354993e-08
15.9008807399178 2.42510672758842e-08
15.9215044488897 2.40666551565266e-08
15.9421281578616 2.38838464854762e-08
15.9627518668336 2.37026255445856e-08
15.9833755758055 2.35229767874133e-08
16.0039992847775 2.33448848371532e-08
16.0246229937494 2.31683344845915e-08
16.0452467027214 2.29933106860916e-08
16.0658704116933 2.28197985616041e-08
16.0864941206653 2.2647783392704e-08
16.1071178296372 2.24772506206534e-08
16.1277415386091 2.23081858444893e-08
16.1483652475811 2.21405748191369e-08
16.168988956553 2.19744034535474e-08
16.189612665525 2.18096578088599e-08
16.2102363744969 2.1646324096588e-08
16.2308600834689 2.14843886768296e-08
16.2514837924408 2.13238380564993e-08
16.2721075014128 2.11646588875855e-08
16.2927312103847 2.10068379654288e-08
16.3133549193566 2.08503622270232e-08
16.3339786283286 2.0695218749339e-08
16.3546023373005 2.05413947476684e-08
16.3752260462725 2.03888775739913e-08
16.3958497552444 2.02376547153629e-08
16.4164734642164 2.00877137923219e-08
16.4370971731883 1.99390425573197e-08
16.4577208821603 1.97916288931687e-08
16.4783445911322 1.96454608115121e-08
16.4989683001041 1.95005264513124e-08
16.5195920090761 1.93568140773587e-08
16.540215718048 1.9214312078795e-08
16.56083942702 1.9073008967666e-08
16.5814631359919 1.89328933774811e-08
16.6020868449639 1.87939540617981e-08
16.6227105539358 1.86561798928244e-08
16.6433342629078 1.85195598600356e-08
16.6639579718797 1.8384083068812e-08
16.6845816808516 1.82497387390928e-08
16.7052053898236 1.8116516204047e-08
16.7258290987955 1.7984404908761e-08
16.7464528077675 1.7853394408944e-08
16.7670765167394 1.77234743696472e-08
16.7877002257114 1.75946345640031e-08
16.8083239346833 1.74668648719769e-08
16.8289476436553 1.73401552791359e-08
16.8495713526272 1.7214495875434e-08
16.8701950615991 1.70898768540109e-08
16.8908187705711 1.69662885100073e-08
16.911442479543 1.68437212393944e-08
16.932066188515 1.67221655378182e-08
16.9526898974869 1.66016119994583e-08
16.9733136064589 1.64820513159009e-08
16.9939373154308 1.63634742750267e-08
17.0145610244028 1.62458717599114e-08
17.0351847333747 1.61292347477403e-08
17.0558084423466 1.60135543087378e-08
17.0764321513186 1.5898821605108e-08
17.0970558602905 1.57850278899904e-08
17.1176795692625 1.56721645064275e-08
17.1383032782344 1.55602228863453e-08
17.1589269872064 1.54491945495472e-08
17.1795506961783 1.53390711027196e-08
17.2001744051503 1.52298442384497e-08
17.2207981141222 1.51215057342566e-08
17.2414218230942 1.50140474516326e-08
17.2620455320661 1.49074613350982e-08
17.282669241038 1.4801739411267e-08
17.30329295001 1.46968737879238e-08
17.3239166589819 1.45928566531124e-08
17.3445403679539 1.44896802742358e-08
17.3651640769258 1.43873369971672e-08
17.3857877858978 1.42858192453712e-08
17.4064114948697 1.41851195190368e-08
17.4270352038417 1.40852303942203e-08
17.4476589128136 1.39861445219987e-08
17.4682826217855 1.38878546276338e-08
17.4889063307575 1.37903535097463e-08
17.5095300397294 1.36936340395001e-08
17.5301537487014 1.35976891597958e-08
17.5507774576733 1.35025118844752e-08
17.5714011666453 1.34080952975345e-08
17.5920248756172 1.33144325523476e-08
17.6126485845892 1.32215168708983e-08
17.6332722935611 1.31293415430224e-08
17.653896002533 1.30378999256583e-08
17.674519711505 1.29471854421074e-08
17.6951434204769 1.28571915813027e-08
17.7157671294489 1.27679118970866e-08
17.7363908384208 1.26793400074971e-08
17.7570145473928 1.25914695940636e-08
17.7776382563647 1.25042944011094e-08
17.7982619653367 1.2417808235064e-08
17.8188856743086 1.23320049637833e-08
17.8395093832805 1.22468785158775e-08
17.8601330922525 1.21624228800474e-08
17.8807568012244 1.20786321044291e-08
17.9013805101964 1.19955002959452e-08
17.9220042191683 1.19130216196652e-08
17.9426279281403 1.18311902981726e-08
17.9632516371122 1.17500006109399e-08
17.9838753460842 1.16694468937107e-08
18.0044990550561 1.15895235378899e-08
18.025122764028 1.15102249899397e-08
18.045746473 1.14315457507847e-08
18.0663701819719 1.13534803752225e-08
18.0869938909439 1.12760234713414e-08
18.1076175999158 1.11991696999463e-08
18.1282413088878 1.11229137739895e-08
18.1488650178597 1.10472504580097e-08
18.1694887268317 1.09721745675768e-08
18.1901124358036 1.08976809687437e-08
18.2107361447756 1.08237645775041e-08
18.2313598537475 1.07504203592571e-08
18.2519835627194 1.06776433282775e-08
18.2726072716914 1.06054285471931e-08
18.2932309806633 1.05337711264677e-08
18.3138546896353 1.04626662238898e-08
18.3344783986072 1.03921090440678e-08
18.3551021075792 1.0322094837931e-08
18.3757258165511 1.0252618902236e-08
18.3963495255231 1.01836765790799e-08
18.416973234495 1.01152632554173e-08
18.4375969434669 1.00473743625853e-08
18.4582206524389 9.98000537583179e-09
18.4788443614108 9.91315181385064e-09
18.4994680703828 9.84680923832182e-09
18.5200917793547 9.78097325345672e-09
18.5407154883267 9.71563950554884e-09
18.5613391972986 9.65080368252988e-09
18.5819629062706 9.58646151353061e-09
18.6025866152425 9.52260876844691e-09
18.6232103242144 9.4592412575109e-09
18.6438340331864 9.39635483086707e-09
18.6644577421583 9.33394537815291e-09
18.6850814511303 9.27200882808493e-09
18.7057051601022 9.21054114804878e-09
18.7263288690742 9.14953834369481e-09
18.7469525780461 9.08899645853731e-09
18.7675762870181 9.02891157355937e-09
18.78819999599 8.96927980682139e-09
18.8088237049619 8.91009731307484e-09
18.8294474139339 8.85136028337946e-09
18.8500711229058 8.79306494472579e-09
18.8706948318778 8.73520755966155e-09
18.8913185408497 8.6777844259221e-09
18.9119422498217 8.62079187606537e-09
18.9325659587936 8.56422627711094e-09
18.9531896677656 8.50808403018295e-09
18.9738133767375 8.45236157015753e-09
18.9944370857094 8.39705536531349e-09
19.0150607946814 8.34216191698785e-09
19.0356845036533 8.28767775923434e-09
19.0563082126253 8.23359945848662e-09
19.0769319215972 8.17992361322464e-09
19.0975556305692 8.12664685364492e-09
19.1181793395411 8.07376584133493e-09
19.1388030485131 8.0212772689506e-09
19.159426757485 7.96917785989762e-09
19.180050466457 7.91746436801646e-09
19.2006741754289 7.86613357727089e-09
19.2212978844008 7.81518230143967e-09
19.2419215933728 7.76460738381222e-09
19.2625453023447 7.71440569688719e-09
19.2831690113167 7.66457414207472e-09
19.3037927202886 7.61510964940176e-09
19.3244164292606 7.56600917722107e-09
19.3450401382325 7.51726971192293e-09
19.3656638472045 7.46888826765047e-09
19.3862875561764 7.42086188601811e-09
19.4069112651483 7.37318763583269e-09
19.4275349741203 7.32586261281853e-09
19.4481586830922 7.27888393934457e-09
19.4687823920642 7.23224876415538e-09
19.4894061010361 7.18595426210445e-09
19.5100298100081 7.13999763389099e-09
19.53065351898 7.09437610579916e-09
19.551277227952 7.04908692944056e-09
19.5719009369239 7.0041273814993e-09
19.5925246458958 6.95949476348002e-09
19.6131483548678 6.91518640145856e-09
19.6337720638397 6.87119964583549e-09
19.6543957728117 6.82753187109222e-09
19.6750194817836 6.78418047554996e-09
19.6956431907556 6.74114288113097e-09
19.7162668997275 6.69841653312296e-09
19.7368906086995 6.65599889994555e-09
19.7575143176714 6.61388747291955e-09
19.7781380266433 6.57207976603864e-09
19.7987617356153 6.53057331574374e-09
19.8193854445872 6.48936568069953e-09
19.8400091535592 6.44845444157361e-09
19.8606328625311 6.40783720081811e-09
19.8812565715031 6.36751158245336e-09
19.901880280475 6.32747523185428e-09
19.922503989447 6.28772581553894e-09
19.9431276984189 6.24826102095921e-09
19.9637514073908 6.20907855629402e-09
19.9843751163628 6.17017615024463e-09
20.0049988253347 6.13155155183215e-09
20.0256225343067 6.09320253019732e-09
20.0462462432786 6.05512687440221e-09
20.0668699522506 6.01732239323451e-09
20.0874936612225 5.97978691501339e-09
20.1081173701945 5.94251828739784e-09
20.1287410791664 5.9055143771969e-09
20.1493647881383 5.86877307018189e-09
20.1699884971103 5.83229227090078e-09
20.1906122060822 5.79606990249454e-09
20.2112359150542 5.76010390651523e-09
20.2318596240261 5.7243922427463e-09
20.2524833329981 5.68893288902479e-09
20.27310704197 5.6537238410651e-09
20.293730750942 5.61876311228516e-09
20.3143544599139 5.58404873363391e-09
20.3349781688859 5.54957875342118e-09
20.3556018778578 5.51535123714868e-09
20.3762255868297 5.48136426734356e-09
20.3968492958017 5.44761594339311e-09
20.4174730047736 5.41410438138155e-09
20.4380967137456 5.38082771392844e-09
20.4587204227175 5.34778409002869e-09
20.4793441316895 5.31497167489466e-09
20.4999678406614 5.28238864979933e-09
20.5205915496334 5.25003321192171e-09
20.5412152586053 5.21790357419335e-09
20.5618389675772 5.18599796514695e-09
20.5824626765492 5.15431462876627e-09
20.6030863855211 5.12285182433758e-09
};
\addlegendentry{Log-normal}
\addplot [very thick, blue, dashed]
table {%
1.18077714175657e-05 5.19648553164734
0.166094520948509 0.569937915267739
0.3321772341256 0.147863912956011
0.498259947302691 0.0541475826612259
0.664342660479782 0.0238708745868856
0.830425373656873 0.0118547722141003
0.996508086833964 0.00641717762756659
1.16259080001106 0.00370544028111184
1.32867351318815 0.00225067066910356
1.49475622636524 0.00141571840501898
1.66083893954233 0.000925615663654204
1.82692165271942 0.000620563156443174
1.99300436589651 0.000426378169821653
2.1590870790736 0.00029784676324474
2.32516979225069 0.000212886275436757
2.49125250542778 0.000153916934153095
2.65733521860487 0.000115544244689031
2.82341793178197 8.49941333051027e-05
2.98950064495906 6.47339364566935e-05
3.15558335813615 4.99747783872188e-05
3.32166607131324 3.87091444352526e-05
3.48774878449033 2.98603786929915e-05
3.65383149766742 2.32210005948691e-05
3.81991421084451 1.8662035735263e-05
3.9859969240216 1.48937400579503e-05
4.15207963719869 1.18880756486652e-05
4.31816235037579 9.05063872497427e-06
4.48424506355288 7.65435059454148e-06
4.65032777672997 6.55526435532527e-06
4.81641048990706 5.19262172201129e-06
4.98249320308415 4.32905396262712e-06
5.14857591626124 3.76829567731273e-06
5.31465862943833 2.80379142657197e-06
5.48074134261542 2.23742555840443e-06
5.64682405579251 2.04116015854439e-06
5.81290676896961 1.92340091862837e-06
5.9789894821467 1.44675637611114e-06
6.14507219532379 1.4299336275517e-06
6.31115490850088 1.05422557639106e-06
6.47723762167797 1.02618766212534e-06
6.64332033485506 8.24314679412159e-07
6.80940304803215 6.67302359524129e-07
6.97548576120924 6.78517525230417e-07
7.14156847438633 6.22441696698977e-07
7.30765118756343 4.20568713985795e-07
7.47373390074052 4.20568713985795e-07
7.63981661391761 3.92530799720076e-07
7.8058993270947 3.25239805482348e-07
7.97198204027179 2.7477155980405e-07
8.13806475344888 2.41126062685191e-07
8.30414746662597 2.01872982713181e-07
8.47023017980306 1.79442651300607e-07
8.63631289298015 1.51404737034886e-07
8.80239560615725 1.96265399860037e-07
8.96847831933434 1.12151657062879e-07
9.13456103251143 1.12151657062878e-07
9.30064374568852 5.60758285314397e-08
9.46672645886561 7.28985770908708e-08
9.6328091720427 5.60758285314397e-08
9.79889188521979 6.72909942377269e-08
9.96497459839688 5.04682456782957e-08
10.131057311574 4.48606628251513e-08
10.2971400247511 3.36454971188638e-08
10.4632227379282 1.68227485594317e-08
10.6293054511052 3.36454971188635e-08
10.7953881642823 1.12151657062879e-08
10.9614708774594 2.80379142657195e-08
11.1275535906365 1.68227485594319e-08
11.2936363038136 5.60758285314391e-09
11.4597190169907 1.12151657062879e-08
11.6258017301678 5.60758285314391e-09
11.7918844433449 1.12151657062879e-08
11.957967156522 1.68227485594317e-08
12.1240498696991 1.12151657062879e-08
12.2901325828762 5.60758285314391e-09
12.4562152960532 5.60758285314391e-09
12.6222980092303 1.12151657062879e-08
12.7883807224074 1.12151657062878e-08
12.9544634355845 5.60758285314397e-09
13.1205461487616 5.60758285314391e-09
13.2866288619387 0
13.4527115751158 5.60758285314391e-09
13.6187942882929 0
13.78487700147 0
13.9509597146471 5.60758285314397e-09
14.1170424278242 0
14.2831251410013 0
14.4492078541783 5.60758285314397e-09
14.6152905673554 5.60758285314391e-09
14.7813732805325 0
14.9474559937096 0
15.1135387068867 0
15.2796214200638 0
15.4457041332409 0
15.611786846418 0
15.7778695595951 0
15.9439522727722 0
16.1100349859493 0
16.2761176991263 0
16.4422004123034 5.60758285314397e-09
};
\addlegendentry{$C\ceq5$}
\end{axis}

% \draw ({$(current bounding box.south west)!0.5!(current bounding box.south east)$}|-{$(current bounding box.south west)!-0.25!(current bounding box.north west)$}) node[
%   scale=0.5,
%   fill=white,
%   draw=black,
%   line width=0.4pt,
%   inner sep=4pt,
%   fill opacity=0.5,
%   anchor=base,
%   text=black,
%   rotate=0.0,
%   align=center
% ]{DNS Log-normal fit:   shape ($\sigma$): 1.073,   loc: -0.009,   scale ($e^\mu$): 0.057
% };
\end{tikzpicture}

%% file: PDF_div.tex
% This file was created by tikzplotlib v0.9.6.
\definecolor{mediumseagreen1}{RGB}{77,175,74}
\definecolor{darkorange1}{RGB}{255,127,0}

\begin{tikzpicture}

\begin{axis}[
width=0.8\linewidth,
height=0.35\linewidth,
legend cell align={left},
legend style={fill opacity=0.8, text opacity=1, draw=none},
log basis y={10},
tick align=inside,               
tick pos=left,
tick style={line width=0.5pt},     
axis line style={line width=0.5pt},
x grid style={white!69.0196078431373!black},
xlabel={\(\displaystyle \nabla \cdot \mathbf{U}\)},
xmin=-150, xmax=150,
xtick style={color=black},
xtick={-150, -100, -50, 0, 50, 100, 150},
xticklabels={, -100, -50, 0, 50, 100,},
y grid style={white!69.0196078431373!black},
ylabel={PDF},
ymin=1e-08, ymax=1,
ymode=log,
ytick style={color=black},
ytick={1e-10,1e-08,1e-06,1e-04,1e-02,1,100,10000},
yticklabels={\(\displaystyle 10^{-10}\),\(\displaystyle 10^{-8}\),\(\displaystyle 10^{-6}\),\(\displaystyle 10^{-4}\),\(\displaystyle 10^{-2}\),\(\displaystyle 10^{0}\),\(\displaystyle 10^{2}\),}
]
\addplot [thick, blue]
table {%
-170.056406756498 5.51001487767951e-10
-166.675934669614 2.75500743883973e-10
-163.29546258273 0
-159.914990495847 5.51001487767947e-10
-156.534518408963 0
-153.154046322079 5.51001487767951e-10
-149.773574235196 1.10200297553589e-09
-146.393102148312 2.75500743883976e-10
-143.012630061428 2.75500743883976e-09
-139.632157974544 3.85701041437563e-09
-136.251685887661 3.58150967049168e-09
-132.871213800777 6.61201785321541e-09
-129.490741713893 1.0193527523707e-08
-126.110269627009 1.26730342186629e-08
-122.729797540126 1.48770401697346e-08
-119.349325453242 2.39685647179058e-08
-115.968853366358 3.55395959610327e-08
-112.588381279475 4.87636316674635e-08
-109.207909192591 7.16301934098337e-08
-105.827437105707 9.39457536644353e-08
-102.446965018823 1.32791358552076e-07
-99.0664929319397 1.86514003609451e-07
-95.686020845056 2.78255751322814e-07
-92.3055487581723 3.87354045900868e-07
-88.9250766712885 5.587155085967e-07
-85.5446045844048 8.03084668421786e-07
-82.1641324975211 1.19429572473703e-06
-78.7836604106373 1.7243591559698e-06
-75.4031883237536 2.52964783034265e-06
-72.0227162368699 3.81017528791538e-06
-68.6422441499862 5.78634212379512e-06
-65.2617720631025 8.68433444871064e-06
-61.8812999762187 1.33529700545685e-05
-58.500827889335 2.03765860191465e-05
-55.1203558024513 3.18456819870365e-05
-51.7398837155676 5.0056556658511e-05
-48.3594116286838 7.97872194347501e-05
-44.9789395418001 0.000128473159892439
-41.5984674549164 0.000208211340194776
-38.2179953680326 0.000340539581996385
-34.8375232811489 0.000564651172123683
-31.4570511942652 0.00094119264032933
-28.0765791073815 0.00158113100273717
-24.6961070204978 0.00267141665613494
-21.315634933614 0.0045034649203778
-17.9351628467303 0.00753346563073133
-14.5546907598466 0.0123059735575564
-11.1742186729629 0.0192281953303585
-7.79374658607914 0.0279202132193153
-4.41327449919541 0.0363864014955057
-1.03280241231168 0.0412924749379736
2.34766967457205 0.0401200335852456
5.72814176145577 0.0334910313866708
9.10861384833949 0.0245747074040615
12.4890859352232 0.0163813924211602
15.8695580221069 0.0102475923406884
19.2500301089907 0.0061789758659518
22.6305021958744 0.0036603375963362
26.0109742827581 0.00215631126229061
29.3914463696418 0.00126999919664519
32.7719184565256 0.000753325377065922
36.1523905434093 0.000449421057489003
39.532862630293 0.000270309758867711
42.9133347171767 0.000164324071694061
46.2938068040605 0.000100669624819668
49.6742788909442 6.22967792085318e-05
53.0547509778279 3.90458939284442e-05
56.4352230647116 2.43705203032324e-05
59.8156951515954 1.56567072749263e-05
63.1961672384791 1.00188600520847e-05
66.5766393253628 6.45966594184752e-06
69.9571114122465 4.16116323562357e-06
73.3375834991302 2.77291498719221e-06
76.718055586014 1.85770151600963e-06
80.0985276728977 1.22597831028369e-06
83.4789997597814 8.54603307528092e-07
86.8594718466651 5.75796554717504e-07
90.2399439335489 3.9038455408359e-07
93.6204160204326 2.66409219335807e-07
97.0008881073163 1.93126021462665e-07
100.3813601942 1.35270865247031e-07
103.761832281084 9.00887432500608e-08
107.142304367967 7.30076971292529e-08
110.522776454851 4.93146331552312e-08
113.903248541735 3.38865914977293e-08
117.283720628619 2.0938056535182e-08
120.664192715502 1.70810461208063e-08
124.044664802386 1.2673034218663e-08
127.42513688927 5.78551562156344e-09
130.805608976154 9.36702529205509e-09
134.186081063037 3.30600892660768e-09
137.566553149921 3.30600892660774e-09
140.947025236805 1.92850520718781e-09
144.327497323688 3.03050818272371e-09
147.707969410572 8.26502231651934e-10
151.088441497456 1.10200297553589e-09
154.46891358434 1.10200297553589e-09
157.849385671223 2.75500743883978e-10
161.229857758107 0
164.610329844991 2.75500743883973e-10
};
\addlegendentry{$C\ceq3$}
\addplot [thick, mediumseagreen1]
table {%
-136.462264579658 3.49884698913321e-10
-133.800466057151 0
-131.138667534643 0
-128.476869012136 0
-125.815070489628 3.49884698913319e-10
-123.15327196712 3.49884698913319e-10
-120.491473444613 3.49884698913321e-10
-117.829674922105 0
-115.167876399598 0
-112.50607787709 0
-109.844279354583 3.49884698913319e-10
-107.182480832075 0
-104.520682309568 0
-101.85888378706 0
-99.1970852645524 3.49884698913319e-10
-96.5352867420448 2.09930819347993e-09
-93.8734882195373 1.39953879565328e-09
-91.2116896970297 2.79907759130655e-09
-88.5498911745222 4.19861638695983e-09
-85.8880926520146 4.54850108587318e-09
-83.226294129507 5.59815518261311e-09
-80.5644956069995 1.22459644619662e-08
-77.9026970844919 1.01466562684863e-08
-75.2408985619844 1.92436584402326e-08
-72.5791000394768 2.23926207304524e-08
-69.9173015169692 3.49884698913321e-08
-67.2555029944617 4.79342037511248e-08
-64.5937044719541 7.20762479761438e-08
-61.9319059494465 1.04615524975083e-07
-59.270107426939 1.5604857571534e-07
-56.6083089044314 2.35122517669752e-07
-53.9465103819239 3.45336197827446e-07
-51.2847118594163 5.64364019347184e-07
-48.6229133369087 8.64565091014817e-07
-45.9611148144012 1.36490021046086e-06
-43.2993162918936 2.19902533267022e-06
-40.6375177693861 3.61605836326916e-06
-37.9757192468785 6.06875010265152e-06
-35.3139207243709 1.01931909334418e-05
-32.6521222018634 1.78507674538586e-05
-29.9903236793558 3.19080850020991e-05
-27.3285251568482 5.85287124342204e-05
-24.6667266343407 0.000110103116592839
-22.0049281118331 0.000215949186054
-19.3431295893256 0.000438684035958727
-16.681331066818 0.000933700208211891
-14.0195325443104 0.00207978042866418
-11.3577340218029 0.00488544879804415
-8.69593549929532 0.0119873737780016
-6.03413697678775 0.02950456815761
-3.37233845428018 0.0641666075209361
-0.71053993177263 0.0971232878686764
1.95125859073494 0.0849338938144836
4.6130571132425 0.0457913181181648
7.27485563575006 0.0194327186372905
9.93665415825762 0.00787330877795754
12.5984526807652 0.00328859847447861
15.2602512032727 0.00144471695763107
17.9220497257803 0.000666461074259492
20.5838482482879 0.000321735775116343
23.2456467707954 0.000162216343187785
25.907445293303 8.43043683184658e-05
28.5692438158106 4.56452580508336e-05
31.2310423383181 2.53159073898734e-05
33.8928408608257 1.45139170803224e-05
36.5546393833332 8.28142093857931e-06
39.2164379058408 4.93442390877457e-06
41.8782364283484 3.02965160789045e-06
44.5400349508559 1.85089005725145e-06
47.2018334733635 1.150420890027e-06
49.863631995871 7.41405676997328e-07
52.5254305183786 4.57299301479711e-07
55.1872290408862 2.9215372359262e-07
57.8490275633937 2.05732202961033e-07
60.5108260859013 1.19660567028356e-07
63.1726246084089 8.74711747283294e-08
65.8344231309164 5.91305141163513e-08
68.496221653424 3.53383545902454e-08
71.1580201759315 2.93903147087187e-08
73.8198186984391 1.60946961500128e-08
76.4816172209467 1.50450420532728e-08
79.1434157434542 1.32956185587061e-08
81.8052142659618 5.94803988152646e-09
84.4670127884693 4.8983857847865e-09
87.1288113109769 2.09930819347991e-09
89.7906098334845 2.44919289239325e-09
92.452408355992 2.09930819347993e-09
95.1142068784996 3.14896229021986e-09
97.7760054010072 1.39953879565328e-09
100.437803923515 3.49884698913321e-10
103.099602446022 3.49884698913321e-10
105.76140096853 3.49884698913318e-10
108.423199491037 3.49884698913321e-10
111.084998013545 0
113.746796536053 3.49884698913318e-10
116.40859505856 0
119.070393581068 0
121.732192103575 0
124.393990626083 0
127.05578914859 3.49884698913314e-10
};
\addlegendentry{$C\ceq4$}
\addplot [thick, darkorange1]
table {%
-67.8029443019002 6.47413538534956e-10
-66.3644161909142 0
-64.9258880799281 0
-63.4873599689421 0
-62.048831857956 6.47413538534956e-10
-60.61030374697 6.47413538534952e-10
-59.1717756359839 0
-57.7332475249979 6.47413538534952e-10
-56.2947194140118 1.2948270770699e-09
-54.8561913030258 1.2948270770699e-09
-53.4176631920397 6.47413538534952e-10
-51.9791350810537 6.47413538534956e-10
-50.5406069700676 3.23706769267475e-09
-49.1020788590816 3.23706769267478e-09
-47.6635507480955 4.53189476974467e-09
-46.2250226371095 8.41637600095438e-09
-44.7864945261234 9.71120307802429e-09
-43.3479664151374 1.48905113863039e-08
-41.9094383041513 1.4890511386304e-08
-40.4709101931653 1.68327520019087e-08
-39.0323820821792 2.9133609234073e-08
-37.5938539711932 4.01396393891671e-08
-36.1553258602072 5.69723913910758e-08
-34.7167977492211 7.96318652397992e-08
-33.2782696382351 1.13297369243617e-07
-31.839741527249 1.67032692942018e-07
-30.401213416263 2.38248182180863e-07
-28.9626853052769 3.65141235733715e-07
-27.5241571942909 5.34116169291336e-07
-26.0856290833048 7.38698847468381e-07
-24.6471009723188 1.17764522659508e-06
-23.2085728613327 1.8134053214364e-06
-21.7700447503467 2.80459544893341e-06
-20.3315166393606 4.52347839374371e-06
-18.8929885283746 7.46662033992361e-06
-17.4544604173885 1.25876614297351e-05
-16.0159323064025 2.19272491366404e-05
-14.5774041954164 3.84854977982102e-05
-13.1388760844304 7.12277900960769e-05
-11.7003479734443 0.000136439166178549
-10.2618198624583 0.000276689655858452
-8.82329175147224 0.000595630814068773
-7.3847636404862 0.0013879918275057
-5.94623552950015 0.00359431501513314
-4.5077074185141 0.0107426241302193
-3.06917930752805 0.0380699411754039
-1.630651196542 0.133478098646313
-0.192123085555956 0.256168865373092
1.24640502543009 0.173554771242137
2.68493313641613 0.0542545914381858
4.12346124740218 0.0148085261728516
5.56198935838823 0.00472400114162058
7.00051746937428 0.00176484088967029
8.43904558036033 0.000739916632334361
9.87757369134638 0.000337006585589601
11.3161018023324 0.000164139401838306
12.7546299133185 8.41242677836928e-05
14.1931580243045 4.54303028260749e-05
15.6316861352906 2.49370746772892e-05
17.0702142462766 1.4482640857027e-05
18.5087423572627 8.61836902497724e-06
19.9472704682487 5.33662979814364e-06
21.3857985792348 3.26814354252442e-06
22.8243266902208 2.01021903715104e-06
24.2628548012068 1.30518569368647e-06
25.7013829121929 8.61060006251482e-07
27.1399110231789 5.65839432679551e-07
28.578439134165 3.62551581579572e-07
30.016967245151 2.42780076950608e-07
31.4554953561371 1.70917174173227e-07
32.8940234671231 1.01643925549988e-07
34.3325515781092 1.04233579704127e-07
35.7710796890952 6.92732486232403e-08
37.2096078000813 4.46715341589119e-08
38.6481359110673 3.23706769267475e-08
40.0866640220534 2.13646467716535e-08
41.5251921330394 1.2948270770699e-08
42.9637202440255 1.23008572321642e-08
44.4022483550115 8.41637600095434e-09
45.8407764659976 4.53189476974469e-09
47.2793045769836 3.88448123120973e-09
48.7178326879697 3.8844812312097e-09
50.1563607989557 3.23706769267478e-09
51.5948889099418 6.47413538534949e-10
53.0334170209278 1.29482707706991e-09
54.4719451319139 1.2948270770699e-09
55.9104732428999 6.47413538534956e-10
57.3490013538859 6.47413538534956e-10
58.787529464872 0
60.2260575758581 6.47413538534949e-10
61.6645856868441 6.47413538534962e-10
63.1031137978301 0
64.5416419088162 6.47413538534949e-10
65.9801700198022 0
67.4186981307883 0
68.8572262417743 0
70.2957543527604 0
71.7342824637464 0
73.1728105747325 0
74.6113386857185 6.47413538534943e-10
};
\addlegendentry{$C\ceq5$}
\end{axis}

\end{tikzpicture}

%% file: PDF_Strain_rate_eigen_values_alpha.tex
% This file was created by tikzplotlib v0.9.6.
\definecolor{mediumseagreen1}{RGB}{77,175,74}
\definecolor{darkorange1}{RGB}{255,127,0}

\begin{tikzpicture}

\begin{axis}[
legend cell align={left},
legend style={fill opacity=0.8, text opacity=1, draw=none},
log basis y={10},
tick pos=left,
x grid style={white!69.0196078431373!black},
xlabel={\(\displaystyle \alpha\)},
xlabel style={font=\LARGE},
xmin=-5, xmax=125,
xtick style={color=black},
y grid style={white!69.0196078431373!black},
ylabel={PDF},
ymin=1e-08, ymax=1,
ymode=log,
ytick style={color=black},
ytick={1e-10,1e-08,1e-06,0.0001,0.01,1,100},
yticklabels={\(\displaystyle 10^{-10}\),,\(\displaystyle 10^{-6}\),,\(\displaystyle 10^{-2}\),,\(\displaystyle 10^{2}\)}
]
\addplot [very thick, black]
table {%
0.731304434360096 0.00297945630186607
2.12231785031099 0.0389361752144771
3.51333126626189 0.093622308785235
4.90434468221278 0.117063795954004
6.29535809816368 0.10999535744548
7.68637151411457 0.0903751771560904
9.07738493006547 0.0695749448800484
10.4683983460164 0.0518425968301697
11.8594117619673 0.0380188186050677
13.2504251779182 0.0277379788987224
14.641438593869 0.0202338195346414
16.0324520098199 0.0147950311773319
17.4234654257708 0.0108684450208126
18.8144788417217 0.00802725698481212
20.2054922576726 0.00595526568415123
21.5965056736235 0.00444622968176345
22.9875190895744 0.00334133892003324
24.3785325055253 0.00252257701917604
25.7695459214762 0.0019163240079642
27.1605593374271 0.00146515044377398
28.551572753378 0.00112314678453404
29.9425861693289 0.00086712258531621
31.3335995852798 0.000671469735209827
32.7246130012307 0.000522579405978897
34.1156264171816 0.0004075397481434
35.5066398331325 0.000320180390699544
36.8976532490834 0.000252105451226821
38.2886666650343 0.00019987087715975
39.6796800809852 0.000158507430912603
41.0706934969361 0.000126158511053796
42.461706912887 0.000100743893765104
43.8527203288379 8.09908061213751e-05
45.2437337447887 6.47915736521342e-05
46.6347471607396 5.22573380225966e-05
48.0257605766905 4.20851975429447e-05
49.4167739926414 3.42885427752524e-05
50.8077874085923 2.78965579691421e-05
52.1988008245432 2.27880585378419e-05
53.5898142404941 1.89302376042703e-05
54.980827656445 1.52799703828484e-05
56.3718410723959 1.24887077184853e-05
57.7628544883468 1.02210160312226e-05
59.1538679042977 8.5331356817721e-06
60.5448813202486 7.07624252810116e-06
61.9358947361995 5.82154686175037e-06
63.3269081521504 4.78511735720873e-06
64.7179215681013 3.91740893480177e-06
66.1089349840522 3.30880788853023e-06
67.4999484000031 2.66673043707323e-06
68.890961815954 2.27907366502568e-06
70.2819752319049 1.93292763540809e-06
71.6729886478558 1.6015112240721e-06
73.0640020638066 1.37521072501843e-06
74.4550154797575 1.06053251627517e-06
75.8460288957084 9.5072990726688e-07
77.2370423116593 7.83347881339612e-07
78.6280557276102 6.72875744227616e-07
80.0190691435611 5.57047382285941e-07
81.410082559512 4.95450796744712e-07
82.8010959754629 4.37201851722023e-07
84.1921093914138 3.61545176002898e-07
85.5831228073647 3.026267028765e-07
86.9741362233156 2.79193219246682e-07
88.3651496392665 2.42369173542683e-07
89.7561630552174 1.98849846801594e-07
91.1471764711683 1.72068722653231e-07
92.5381898871192 1.39931373675196e-07
93.9292033030701 1.38592317467778e-07
95.320216719021 1.07124496593451e-07
96.7112301349719 8.23519567562157e-08
98.1022435509228 8.83777096895964e-08
99.4932569668737 7.29785633042887e-08
100.884270382825 5.22231920893075e-08
102.275283798775 4.82060234670531e-08
103.666297214726 4.41888548447986e-08
105.057310630677 3.48154613928717e-08
106.448324046628 3.54849894965807e-08
107.839337462579 1.9416315007563e-08
109.23035087853 2.74506522520719e-08
110.621364294481 2.14248993186902e-08
112.012377710432 1.80772588001449e-08
113.403391126383 1.40600901778905e-08
114.794404542333 9.37339345192698e-09
116.185417958284 1.0042921555636e-08
117.576431374235 1.00429215556361e-08
118.967444790186 1.00429215556361e-08
120.358458206137 7.36480914079977e-09
121.749471622088 1.13819777630542e-08
123.140485038039 4.68669672596349e-09
124.53149845399 2.67811241483628e-09
125.922511869941 2.00858431112721e-09
127.313525285892 2.67811241483628e-09
128.704538701842 6.69528103709063e-10
130.095552117793 1.33905620741813e-09
131.486565533744 2.00858431112723e-09
132.877578949695 1.33905620741813e-09
134.268592365646 1.33905620741815e-09
135.659605781597 6.69528103709063e-10
137.050619197548 1.33905620741815e-09
138.441632613499 6.69528103709063e-10
};
\addlegendentry{DNS}
\addplot [very thick, blue]
table {%
0.700173356076152 0.000466045996440052
1.94790253641514 0.00900899200989862
3.19563171675413 0.0355939080582343
4.44336089709311 0.0700942949205583
5.6910900774321 0.0952665671691782
6.93881925777109 0.104069190626119
8.18654843811008 0.0991457495897083
9.43427761844907 0.086492178316016
10.6820067987881 0.0712114236996318
11.929735979127 0.0563949498104901
13.177465159466 0.043530070962157
14.425194339805 0.0330241657541762
15.672923520144 0.0248002908419543
16.920652700483 0.0184836254598292
18.168381880822 0.0137229049227329
19.416111061161 0.0101769359101991
20.6638402415 0.00753994692630285
21.9115694218389 0.00559668037331942
23.1592986021779 0.0041584330793055
24.4070277825169 0.00309914989527356
25.6547569628559 0.00231530543352987
26.9024861431949 0.00173558281849244
28.1502153235339 0.00130608199971936
29.3979445038729 0.000982845161741916
30.6456736842119 0.000745857212157191
31.8934028645508 0.000567069404281716
33.1411320448898 0.000432322263833053
34.3888612252288 0.000331315277100203
35.6365904055678 0.000254702593925528
36.8843195859068 0.000196945822154274
38.1320487662458 0.00015206469252177
39.3797779465848 0.000118158834889736
40.6275071269238 9.21925833695772e-05
41.8752363072627 7.21521128893905e-05
43.1229654876017 5.66976102527086e-05
44.3706946679407 4.45900281580645e-05
45.6184238482797 3.5227010477575e-05
46.8661530286187 2.77755591459163e-05
48.1138822089577 2.21057981206453e-05
49.3616113892967 1.77445009022861e-05
50.6093405696357 1.39288323449947e-05
51.8570697499746 1.11141050113591e-05
53.1047989303136 8.99130349004912e-06
54.3525281106526 7.34993902262271e-06
55.6002572909916 6.02356127882255e-06
56.8479864713306 4.73823630705893e-06
58.0957156516696 3.78357275369907e-06
59.3434448320086 3.18047021177979e-06
60.5911740123476 2.53706131186093e-06
61.8389031926865 2.07055253871794e-06
63.0866323730255 1.64061805338934e-06
64.3343615533645 1.36071278950352e-06
65.5820907337035 1.12857802398758e-06
66.8298199140425 1.00168763769267e-06
68.0775490943815 7.78509840621051e-07
69.3252782747205 6.31466275326382e-07
70.5730074550595 5.92652745400877e-07
71.8207366353984 5.01590232883367e-07
73.0684658157374 3.97092267699328e-07
74.3161949960764 3.14986723626159e-07
75.5639241764154 2.60498498923058e-07
76.8116533567544 2.20938554960528e-07
78.0593825370934 1.76153712738802e-07
79.3071117174324 1.47789979331705e-07
80.5548408977714 1.28383214368956e-07
81.8025700781103 1.02258723072948e-07
83.0502992584493 9.25553405915727e-08
84.2980284387883 7.01629194807091e-08
85.5457576191273 6.49380212215067e-08
86.7934867994663 6.04595369993338e-08
88.0412159798053 4.17991860736139e-08
89.2889451601443 4.25456001106423e-08
90.5366743404833 3.88135299254987e-08
91.7844035208223 2.46316632219508e-08
93.0321327011612 2.53780772589796e-08
94.2798618815002 2.16460070738358e-08
95.5275910618392 1.94067649627491e-08
96.7753202421782 1.86603509257203e-08
98.0230494225172 1.86603509257205e-08
99.2707786028562 1.04497965184034e-08
100.518507783195 1.49282807405764e-08
101.766236963534 1.49282807405762e-08
103.013966143873 5.9713122962305e-09
104.261695324212 2.23924211108646e-09
105.509424504551 8.21055440731693e-09
106.75715368489 5.97131229623056e-09
108.004882865229 4.47848422217287e-09
109.252612045568 1.49282807405762e-09
110.500341225907 4.47848422217292e-09
111.748070406246 1.49282807405762e-09
112.995799586585 2.98565614811528e-09
114.243528766924 2.23924211108644e-09
115.491257947263 0
116.738987127602 7.46414037028821e-10
117.986716307941 7.46414037028812e-10
119.23444548828 7.46414037028821e-10
120.482174668619 1.49282807405762e-09
121.729903848958 1.49282807405762e-09
122.977633029297 1.49282807405764e-09
124.225362209636 7.46414037028812e-10
};
% \addlegendentry{$\alpha_{comp}$ cutoff 0.001}
\addlegendentry{$C\ceq3$}
\addplot [very thick, mediumseagreen1]
table {%
0.626634798382566 0.000787398627420345
1.78262528736115 0.0146615452120976
2.93861577633974 0.0527769747536599
4.09460626531833 0.0932656585155761
5.25059675429692 0.114319070579071
6.40658724327551 0.114392824521883
7.5625777322541 0.101935389940778
8.71856822123269 0.084879090153622
9.87455871021128 0.0678396920767124
11.0305491991899 0.0528666805465551
12.1865396881685 0.040552015829341
13.342530177147 0.0308390269635282
14.4985206661256 0.0233134089539093
15.6545111551042 0.0175814781969297
16.8105016440828 0.0132530137684109
17.9664921330614 0.00998941348087476
19.12248262204 0.00754090842187757
20.2784731110186 0.00570157068028667
21.4344635999972 0.004319580537061
22.5904540889757 0.00328361260854915
23.7464445779543 0.00250156912999248
24.9024350669329 0.00191067070917604
26.0584255559115 0.00146370873199352
27.2144160448901 0.00112433312148089
28.3704065338687 0.00086696774467199
29.5263970228473 0.000668327734339802
30.6823875118259 0.000520189831201726
31.8383780008045 0.000404842647771835
32.994368489783 0.000314337651074505
34.1503589787616 0.000245966249191624
35.3063494677402 0.000193877819128161
36.4623399567188 0.000151453148748728
37.6183304456974 0.000119444714213798
38.774320934676 9.46951771155587e-05
39.9303114236546 7.50091439451329e-05
41.0863019126332 5.99427021067808e-05
42.2422924016118 4.73205993164776e-05
43.3982828905903 3.80193816797857e-05
44.5542733795689 3.04865635851067e-05
45.7102638685475 2.43644368790676e-05
46.8662543575261 1.97883505926744e-05
48.0222448465047 1.58116671578791e-05
49.1782353354833 1.26857490608359e-05
50.3342258244619 1.0240604363793e-05
51.4902163134405 8.30543548000489e-06
52.6462068024191 6.83673736379101e-06
53.8021972913976 5.61215089278437e-06
54.9581877803762 4.4649067252097e-06
56.1141782693548 3.6697311680495e-06
57.2701687583334 3.09046954130359e-06
58.426159247312 2.46448027290086e-06
59.5821497362906 2.04795974295979e-06
60.7381402252692 1.67897250366963e-06
61.8941307142478 1.45258513633222e-06
63.0501212032264 1.18188707432022e-06
64.2061116922049 9.45831919053817e-07
65.3621021811835 7.50059213491566e-07
66.5180926701621 6.48547440237068e-07
67.6740831591407 5.69593838816903e-07
68.8300736481193 4.79361151479572e-07
69.9860641370979 3.85905868165908e-07
71.1420546260765 3.08563564733909e-07
72.2980451150551 2.73920657988327e-07
73.4540356040336 1.97384003550412e-07
74.6100260930122 2.07051791479412e-07
75.7660165819908 1.64352394792996e-07
76.9220070709694 1.16013455147997e-07
78.077997559948 1.2890383905333e-07
79.2339880489266 9.90948262722477e-08
80.3899785379052 8.21761973964981e-08
81.5459690268838 6.36462705325818e-08
82.7019595158623 7.25084094674983e-08
83.8579500048409 4.02824497041657e-08
85.0139404938195 4.75332906509156e-08
86.1699309827981 4.59219926627489e-08
87.3259214717767 2.8197714792916e-08
88.4819119607553 2.65864168047494e-08
89.6379024497339 1.69186288757496e-08
90.7938929387125 2.49751188165827e-08
91.949883427691 1.45016818934997e-08
93.1058739166697 1.20847349112497e-08
94.2618644056482 9.66778792899977e-09
95.4178548946268 1.20847349112497e-08
96.5738453836054 5.6395429585832e-09
97.729835872584 7.25084094674983e-09
98.8858263615626 4.83389396449989e-09
100.041816850541 6.44519195266652e-09
101.19780733952 3.22259597633326e-09
102.353797828498 3.22259597633326e-09
103.509788317477 8.05648994083314e-10
104.665778806456 8.05648994083314e-10
105.821769295434 8.05648994083314e-10
106.977759784413 0
108.133750273391 8.05648994083314e-10
109.28974076237 0
110.445731251348 8.05648994083305e-10
111.601721740327 0
112.757712229306 0
113.913702718284 0
115.069693207263 8.05648994083324e-10
};
% \addlegendentry{$\beta_{comp}$ cutoff 0.0001}
\addlegendentry{$C\ceq4$}
\addplot [very thick, darkorange1]
table {%
0.80938279041503 0.00320429862559094
2.29660760263063 0.0416416664246113
3.78383241484622 0.0982529932064435
5.27105722706182 0.117928475004103
6.75828203927742 0.106214774136255
8.24550685149301 0.084122131013926
9.73273166370861 0.062683044753266
11.2199564759242 0.0453039950264812
12.7071812881398 0.0322843808895647
14.1944061003554 0.0228900898017785
15.681630912571 0.0162282918844201
17.1688557247866 0.0115357443246031
18.6560805370022 0.00822415752525848
20.1433053492178 0.00589357356367152
21.6305301614334 0.00424501545279936
23.117754973649 0.00307402961157527
24.6049797858646 0.00223705999484621
26.0922045980802 0.00163476948147193
27.5794294102958 0.00120272304666633
29.0666542225114 0.00088792536235446
30.553879034727 0.000659628642888456
32.0411038469426 0.000490719028612068
33.5283286591582 0.000369009122032811
35.0155534713737 0.000278105239101975
36.5027782835894 0.000209979928968864
37.990003095805 0.00015934731031429
39.4772279080205 0.000120977861174432
40.9644527202361 9.29641304773523e-05
42.4516775324517 7.13265215184559e-05
43.9389023446673 5.5308566505589e-05
45.4261271568829 4.26302164062686e-05
46.9133519690985 3.29746064002717e-05
48.4005767813141 2.56009240719355e-05
49.8878015935297 2.02467853112327e-05
51.3750264057453 1.57424203865855e-05
52.8622512179609 1.23333056014081e-05
54.3494760301765 9.66249837165416e-06
55.8367008423921 7.64671209437902e-06
57.3239256546077 6.04485397158636e-06
58.8111504668233 4.70913724917945e-06
60.2983752790389 3.73599904635701e-06
61.7856000912545 3.03902168487606e-06
63.2728249034701 2.38337451733737e-06
64.7600497156857 1.93249967433082e-06
66.2472745279013 1.53735797164038e-06
67.7344993401169 1.15787164544319e-06
69.2217241523325 9.86914934136544e-07
70.7089489645481 7.66487233111131e-07
72.1961737767637 6.43749081403786e-07
73.6833985889793 5.46685747145434e-07
75.1706234011949 4.17059229781052e-07
76.6578482134105 3.58821229226044e-07
78.1450730256261 2.80544346759627e-07
79.6322978378417 2.36709292578435e-07
81.1195226500573 1.99762604054288e-07
82.6067474622729 1.74714001665037e-07
84.0939722744885 1.41524603499276e-07
85.5811970867041 1.10840065572442e-07
87.0684218989196 1.00820624616742e-07
88.5556467111352 8.45390330637268e-08
90.0428715233508 7.20147318691006e-08
91.5300963355664 4.94709897187735e-08
93.017321147782 5.32282800771618e-08
94.5045459599976 3.50680433449534e-08
95.9917707722132 2.81796776879089e-08
97.4789955844288 2.25437421503274e-08
98.9662203966444 1.94126668516706e-08
100.45344520886 2.06650969711332e-08
101.940670021076 1.25243011946262e-08
103.427894833291 1.06456560154324e-08
104.915119645507 6.2621505973131e-09
106.402344457722 6.2621505973131e-09
107.889569269938 4.38350541811921e-09
109.376794082154 3.75729035838786e-09
110.864018894369 3.13107529865655e-09
112.351243706585 5.00972047785048e-09
113.8384685188 1.87864517919395e-09
115.325693331016 3.75729035838786e-09
116.812918143232 1.87864517919393e-09
118.300142955447 1.87864517919393e-09
119.787367767663 1.25243011946263e-09
121.274592579878 0
122.761817392094 6.2621505973131e-10
124.24904220431 6.26215059731316e-10
125.736267016525 1.87864517919393e-09
127.223491828741 0
128.710716640956 0
130.197941453172 0
131.685166265388 6.26215059731322e-10
133.172391077603 6.2621505973131e-10
134.659615889819 0
136.146840702034 0
137.63406551425 0
139.121290326466 0
140.608515138681 0
142.095739950897 0
143.582964763112 0
145.070189575328 0
146.557414387544 0
148.044639199759 6.2621505973131e-10
};
% \addlegendentry{$\alpha_{comp}$ cutoff 1e-05}
\addlegendentry{$C\ceq5$}
\end{axis}
\end{tikzpicture}

%% file: PDF_Strain_rate_eigen_values_beta.tex
% This file was created by tikzplotlib v0.9.6.
\definecolor{mediumseagreen1}{RGB}{77,175,74}
\definecolor{darkorange1}{RGB}{255,127,0}

\begin{tikzpicture}

\begin{axis}[
legend cell align={left},
legend style={fill opacity=0.8, text opacity=1, draw=none},
log basis y={10},
tick pos=left,
x grid style={white!69.0196078431373!black},
xlabel={\(\displaystyle \beta\)},
xlabel style={font=\LARGE},
xmin=-60, xmax=60,
xtick style={color=black},
y grid style={white!69.0196078431373!black},
ymin=1e-08, ymax=1,
ymode=log,
ytick style={color=black},
ytick={1e-10,1e-08,1e-06,0.0001,0.01,1,100},
% yticklabels={\(\displaystyle 10^{-10}\),,\(\displaystyle 10^{-6}\),,\(\displaystyle 10^{-2}\),,\(\displaystyle 10^{2}\)}
% ]
yticklabels={,,,,,,}
]
\addplot [very thick, black]
table {%
-52.5526829344895 7.20099715731492e-10
-51.2593584764047 0
-49.9660340183198 0
-48.672709560235 7.20099715731492e-10
-47.3793851021501 0
-46.0860606440653 7.20099715731488e-10
-44.7927361859805 7.20099715731492e-10
-43.4994117278956 2.16029914719448e-09
-42.2060872698108 2.16029914719448e-09
-40.9127628117259 2.16029914719448e-09
-39.6194383536411 5.04069801012041e-09
-38.3261138955563 7.20099715731492e-09
-37.0327894374714 7.20099715731492e-09
-35.7394649793866 1.00813960202408e-08
-34.4461405213017 1.72823931775558e-08
-33.1528160632169 2.37632906191392e-08
-31.859491605132 3.31245869236486e-08
-30.5661671470472 4.68064815225468e-08
-29.2728426889624 6.04883761214451e-08
-27.9795182308775 8.9292364750705e-08
-26.6861937727927 1.53381239450808e-07
-25.3928693147078 2.30431909034077e-07
-24.099544856623 3.39887065825263e-07
-22.8062203985382 5.3575418850423e-07
-21.5128959404533 8.79961852623883e-07
-20.2195714823685 1.36602916074264e-06
-18.9262470242836 2.25319201052382e-06
-17.6329225661988 3.6494653593272e-06
-16.339598108114 6.23606353823472e-06
-15.0462736500291 1.09404749811085e-05
-13.7529491919443 1.94448526238975e-05
-12.4596247338594 3.60993188493354e-05
-11.1663002757746 7.02126026826834e-05
-9.87297581768975 0.000139707265948782
-8.57965135960491 0.000294292512124292
-7.28632690152006 0.000658257552144471
-5.99300244343522 0.00158810071207707
-4.69967798535038 0.0041774971984799
-3.40635352726554 0.0121292191923367
-2.1130290691807 0.0375740765862561
-0.819704611095858 0.103724756634635
0.473619846988981 0.173731077392451
1.76694430507382 0.164925521644055
3.06026876315867 0.114388580826555
4.35359322124351 0.0689694095716496
5.64691767932835 0.0392642845576248
6.94024213741319 0.0219891787891924
8.23356659549803 0.0123443821069984
9.52689105358287 0.00700450210768383
10.8202155116677 0.00403222788232683
12.1135399697526 0.00236729901147865
13.4068644278374 0.00141149481659449
14.7001888859222 0.000856234566990521
15.9935133440071 0.00052839548950917
17.2868378020919 0.000331924203168705
18.5801622601768 0.000209888184243974
19.8734867182616 0.00013590513944972
21.1668111763464 8.78723281112825e-05
22.4601356344313 5.84195296381487e-05
23.7534600925161 3.89703564159564e-05
25.046784550601 2.63232451085647e-05
26.3401090086858 1.7976569303521e-05
27.6334334667707 1.23129850392928e-05
28.9267579248555 8.69664426688923e-06
30.2200823829403 6.09348379451988e-06
31.5134068410252 4.31843799524176e-06
32.80673129911 3.02081830749361e-06
34.1000557571949 2.18334233809788e-06
35.3933802152797 1.60582236608121e-06
36.6867046733645 1.18024343408391e-06
37.9800291314494 8.36755869679993e-07
39.2733535895342 6.35848048990907e-07
40.5666780476191 4.60143718352423e-07
41.8600025057039 3.19724273784782e-07
43.1533269637887 2.69317293683578e-07
44.4466514218736 1.97307322110427e-07
45.7399758799584 1.55541538598002e-07
47.0333003380433 1.04414458781066e-07
48.3266247961281 7.63305698675381e-08
49.619949254213 6.40888747001028e-08
50.9132737122978 5.83280769742508e-08
52.2065981703826 3.09642877764541e-08
53.4999226284675 3.74451852180376e-08
54.7932470865523 1.58421937460928e-08
56.0865715446372 1.87225926090186e-08
57.379896002722 1.08014957359724e-08
58.6732204608068 1.51220940303613e-08
59.9665449188917 5.76079772585193e-09
61.2598693769765 6.48089744158343e-09
62.5531938350614 6.48089744158343e-09
63.8465182931462 3.60049857865746e-09
65.139842751231 2.88039886292594e-09
66.4331672093159 2.16029914719448e-09
67.7264916674007 2.88039886292597e-09
69.0198161254856 3.60049857865746e-09
70.3131405835704 7.20099715731492e-10
71.6064650416553 7.20099715731492e-10
72.8997894997401 7.20099715731492e-10
74.1931139578249 0
75.4864384159098 7.200997157315e-10
};
% \addlegendentry{DNS}
\addplot [very thick, blue]
table {%
-49.2775188327613 8.79077132020225e-10
-48.2180866843272 0
-47.1586545358931 8.79077132020225e-10
-46.099222387459 2.63723139606069e-09
-45.0397902390249 0
-43.9803580905908 8.79077132020231e-10
-42.9209259421567 1.75815426404045e-09
-41.8614937937226 0
-40.8020616452885 3.5163085280809e-09
-39.7426294968544 5.27446279212139e-09
-38.6831973484203 3.5163085280809e-09
-37.6237651999862 1.1428002716263e-08
-36.5643330515521 1.05489255842427e-08
-35.504900903118 1.9339696904445e-08
-34.4454687546839 3.16467767527283e-08
-33.3860366062498 3.95584709409104e-08
-32.3266044578157 5.62609364492944e-08
-31.2671723093816 8.43914046739419e-08
-30.2077401609475 1.26587107010913e-07
-29.1483080125134 1.87243429120309e-07
-28.0888758640792 2.83062836510513e-07
-27.0294437156452 4.11408097785467e-07
-25.970011567211 5.4063243619244e-07
-24.9105794187769 8.17541732778809e-07
-23.8511472703428 1.3256483150865e-06
-22.7917151219087 2.05264510326723e-06
-21.7322829734746 3.03984872252595e-06
-20.6728508250405 4.95535779319802e-06
-19.6134186766064 7.80356770094356e-06
-18.5539865281723 1.25409143654006e-05
-17.4945543797382 2.03857986915492e-05
-16.4351222313041 3.32299946674965e-05
-15.37569008287 5.51893414253617e-05
-14.3162579344359 9.34177686655259e-05
-13.2568257860018 0.000159042634725099
-12.1973936375677 0.000274974447818796
-11.1379614891336 0.000483880095626345
-10.0785293406995 0.000864901134189273
-9.01909719226542 0.00157602882198276
-7.95966504383131 0.00292001644419749
-6.90023289539721 0.00551707841071048
-5.84080074696311 0.0105682811045356
-4.78136859852901 0.0203548674835818
-3.72193645009491 0.0386118639826855
-2.66250430166081 0.0695196357536092
-1.60307215322671 0.111831734935477
-0.543640004792604 0.149536998760215
0.515792143641502 0.158228887968299
1.5752242920756 0.133420264103051
2.6346564405097 0.0946012552640529
3.69408858894381 0.0600879249010658
4.75352073737791 0.0358974487353561
5.81295288581201 0.0207997007311582
6.87238503424611 0.0119157806398926
7.93181718268021 0.00682039289218092
8.99124933111431 0.00392987212299408
10.0506814795484 0.00228514078591255
11.1101136279825 0.0013474028368726
12.1695457764166 0.000804562158924536
13.2289779248507 0.000488084721548797
14.2884100732848 0.000299095445244299
15.3478422217189 0.00018561186196328
16.407274370153 0.000117009561657551
17.4667065185871 7.46556254368181e-05
18.5261386670212 4.81189240525234e-05
19.5855708154553 3.10903209281591e-05
20.6450029638894 2.0511506721428e-05
21.7044351123235 1.37628315789087e-05
22.7638672607576 9.23646342613657e-06
23.8232994091917 6.28540149394457e-06
24.8827315576258 4.19583515113256e-06
25.9421637060599 2.98182963181262e-06
27.001595854494 2.03594263775885e-06
28.0610280029281 1.37575571161164e-06
29.1204601513623 9.99510699107003e-07
30.1798922997964 6.94470934295982e-07
31.2393244482304 5.17776430759916e-07
32.2987565966646 3.27016693111522e-07
33.3581887450987 2.21527437269098e-07
34.4176208935328 1.84606197724248e-07
35.4770530419669 1.23949875614851e-07
36.536485190401 7.9996019013841e-08
37.5959173388351 4.92283193931329e-08
38.6553494872692 2.46141596965665e-08
39.7147816357033 2.37350825645459e-08
40.7742137841374 1.49443112443439e-08
41.8336459325715 1.40652341123237e-08
42.8930780810056 1.23070798482832e-08
43.9525102294397 8.79077132020219e-09
45.0119423778738 5.27446279212139e-09
46.0713745263079 3.51630852808092e-09
47.130806674742 2.63723139606069e-09
48.1902388231761 1.75815426404044e-09
49.2496709716102 8.79077132020231e-10
50.3091031200443 2.63723139606069e-09
51.3685352684784 0
52.4279674169125 0
53.4873995653466 0
54.5468317137807 0
55.6062638622148 8.79077132020231e-10
};
% \addlegendentry{C3}

\addplot [very thick, mediumseagreen1]
table {%
-43.6640815868891 9.54730099093582e-10
-42.6885990128091 9.54730099093589e-10
-41.7131164387291 0
-40.7376338646491 2.86419029728075e-09
-39.7621512905691 9.54730099093589e-10
-38.7866687164891 3.81892039637433e-09
-37.811186142409 3.81892039637433e-09
-36.835703568329 7.63784079274871e-09
-35.860220994249 3.81892039637433e-09
-34.884738420169 4.77365049546794e-09
-33.909255846089 1.05020310900294e-08
-32.933773272009 1.43209514864037e-08
-31.9582906979289 2.1004062180059e-08
-30.9828081238489 3.34155534682754e-08
-30.0073255497689 2.38682524773396e-08
-29.0318429756889 4.96459651528664e-08
-28.0563604016089 7.16047574320186e-08
-27.0808778275289 7.92425982247676e-08
-26.1053952534489 1.2793383327854e-07
-25.1299126793688 1.79489258629593e-07
-24.1544301052888 2.7687172873714e-07
-23.1789475312088 3.76163659042873e-07
-22.2034649571288 5.74747519654338e-07
-21.2279823830488 8.77396961067002e-07
-20.2524998089688 1.29365928427181e-06
-19.2770172348887 1.96674400413279e-06
-18.3015346608087 3.03604171511759e-06
-17.3260520867287 4.7621937342788e-06
-16.3505695126487 7.5318657517493e-06
-15.3750869385687 1.19646776018408e-05
-14.3996043644887 1.94927244331937e-05
-13.4241217904086 3.23472104873898e-05
-12.4486392163286 5.43231879083259e-05
-11.4731566422486 9.318547659193e-05
-10.4976740681686 0.000162440643250081
-9.52219149408858 0.000288528983247071
-8.54670892000856 0.000528410649024928
-7.57122634592855 0.000992208029133508
-6.59574377184853 0.00191931205807971
-5.62026119776851 0.00383684640871491
-4.6447786236885 0.00795025234942461
-3.66929604960848 0.0169874525051759
-2.69381347552847 0.0367257411576972
-1.71833090144845 0.0755492238816942
-0.742848327368435 0.131408427400486
0.23263424671158 0.17289247654096
1.2081168207916 0.17091266610069
2.18359939487161 0.13730168415478
3.15908196895163 0.097270052547838
4.13456454303164 0.0641179501672961
5.11004711711166 0.0405816812136639
6.08552969119168 0.0251417106274448
7.06101226527169 0.0154296910072423
8.03649483935171 0.0094469407176492
9.01197741343173 0.00579628863704986
9.98745998751174 0.00357825489718511
10.9629425615918 0.00222167889938304
11.9384251356718 0.0013869956082194
12.9139077097518 0.000877678606446234
13.8893902838318 0.000556671614688202
14.8648728579118 0.000356903888753856
15.8403554319918 0.000230658018290514
16.8158380060719 0.000150515109582302
17.7913205801519 9.82436366569278e-05
18.7668031542319 6.57045254196203e-05
19.7422857283119 4.36607621616489e-05
20.7177683023919 2.93130782324702e-05
21.6932508764719 2.02078172774148e-05
22.668733450552 1.34340072243458e-05
23.644216024632 9.27615764279338e-06
24.619698598712 6.39860112412519e-06
25.595181172792 4.4920051162353e-06
26.570663746872 3.0494079365049e-06
27.546146320952 2.22547586098714e-06
28.5216288950321 1.52279450805426e-06
29.4971114691121 1.10939637514676e-06
30.4725940431921 7.97199632743141e-07
31.4480766172721 5.53743457474278e-07
32.4235591913521 4.3440219508758e-07
33.3990417654321 3.31291344385473e-07
34.3745243395121 2.37727794674305e-07
35.3500069135922 1.64213577044096e-07
36.3254894876722 1.040655808012e-07
37.3009720617522 9.73824701075454e-08
38.2764546358322 7.82878681256737e-08
39.2519372099122 5.82385360447085e-08
40.2274197839922 3.91439340628374e-08
41.2029023580723 2.19587922791524e-08
42.1783849321523 2.86419029728075e-08
43.1538675062323 1.62304116845909e-08
44.1293500803123 1.05020310900294e-08
45.1048326543923 9.54730099093596e-09
46.0803152284723 1.24114912882166e-08
47.0557978025523 1.90946019818716e-09
48.0312803766324 3.81892039637433e-09
49.0067629507124 2.86419029728075e-09
49.9822455247924 9.54730099093582e-10
50.9577280988724 0
51.9332106729524 1.90946019818716e-09
52.9086932470324 1.90946019818718e-09
};
% \addlegendentry{C4}
\addplot [very thick, darkorange1]
table {%
-45.1817054418913 1.69931381221004e-09
-44.0855893885483 0
-42.9894733352052 0
-41.8933572818622 3.39862762442008e-09
-40.7972412285192 1.69931381221003e-09
-39.7011251751762 8.49656906105019e-10
-38.6050091218331 2.54897071831506e-09
-37.5088930684901 5.09794143663012e-09
-36.4127770151471 1.10455397793653e-08
-35.3166609618041 8.49656906105019e-09
-34.220544908461 1.18951966854703e-08
-33.124428855118 1.78427950282054e-08
-32.028312801775 2.37903933709405e-08
-30.932196748432 3.48359331503058e-08
-29.8360806950889 3.22869624319906e-08
-28.7399646417459 4.5881472929671e-08
-27.6438485884029 7.47698077372417e-08
-26.5477325350599 1.15553339230283e-07
-25.4516164817168 1.80127264094264e-07
-24.3555004283738 2.67641925423081e-07
-23.2593843750308 3.98489088963254e-07
-22.1632683216878 6.01557089522354e-07
-21.0671522683447 8.9893700665911e-07
-19.9710362150017 1.40108423816718e-06
-18.8749201616587 2.11904432382591e-06
-17.7788041083157 3.43771184210091e-06
-16.6826880549726 5.49558086868726e-06
-15.5865720016296 8.75401510360001e-06
-14.4904559482866 1.44441674037853e-05
-13.3943398949436 2.4449727130078e-05
-12.2982238416005 4.16527305079864e-05
-11.2021077882575 7.35361059095772e-05
-10.1059917349145 0.000133921222226461
-9.00987568157146 0.000251328512825865
-7.91375962822843 0.000487376795852337
-6.81764357488541 0.00098983670110185
-5.72152752154238 0.00212153211924699
-4.62541146819936 0.00482951187925504
-3.52929541485633 0.0117645789292947
-2.4331793615133 0.0302700825538687
-1.33706330817028 0.075533107214724
-0.240947254827255 0.144953078281353
0.85516879851577 0.179486643178827
1.95128485185879 0.157184116303129
3.04740090520182 0.113492204122549
4.14351695854484 0.0740627742393955
5.23963301188787 0.0458947972492726
6.33574906523089 0.0277800253461091
7.43186511857392 0.0167028536897864
8.52798117191695 0.0100500596570931
9.62409722525997 0.00608661920571204
10.720213278603 0.00371802470096473
11.816329331946 0.00229653850116816
12.912445385289 0.001434141839413
14.0085614386321 0.000907504097243367
15.1046774919751 0.000581142383039368
16.2007935453181 0.000374326545867439
17.2969095986611 0.000244600079786419
18.3930256520042 0.000162565705501978
19.4891417053472 0.000107614145099637
20.5852577586902 7.22599212366075e-05
21.6813738120333 4.88612196993813e-05
22.7774898653763 3.33473342508098e-05
23.8736059187193 2.30681850007513e-05
24.9697219720623 1.61604743541175e-05
26.0658380254053 1.10990681644499e-05
27.1619540787484 7.94089344445751e-06
28.2580701320914 5.53806371399252e-06
29.3541861854344 4.01632819515843e-06
30.4503022387774 2.91942112937685e-06
31.5464182921205 2.18021962106548e-06
32.6425343454635 1.53023208789514e-06
33.7386503988065 1.1504354508662e-06
34.8347664521495 8.35212738701234e-07
35.9308825054926 5.78616353057511e-07
37.0269985588356 4.4522021879903e-07
38.1231146121786 3.06726143103912e-07
39.2192306655217 2.43851532052141e-07
40.3153467188647 1.79277607188159e-07
41.4114627722077 1.42742360225643e-07
42.5075788255507 1.18951966854703e-07
43.6036948788938 7.47698077372417e-08
44.6998109322368 5.60773558029313e-08
45.7959269855798 3.48359331503058e-08
46.8920430389228 2.63393640892556e-08
47.9881590922659 2.71890209953606e-08
49.0842751456089 2.29407364648355e-08
50.1803911989519 1.10455397793653e-08
51.2765072522949 1.27448535915753e-08
52.372623305638 8.49656906105019e-09
53.468739358981 7.64691215494517e-09
54.564855412324 3.39862762442008e-09
55.660971465667 3.39862762442008e-09
56.75708751901 3.39862762442008e-09
57.8532035723531 8.49656906105019e-10
58.9493196256961 0
60.0454356790391 0
61.1415517323822 2.54897071831506e-09
62.2376677857252 2.54897071831506e-09
63.3337838390682 1.69931381221007e-09
};
% \addlegendentry{C5}
\end{axis}

\end{tikzpicture}

%% file: PDF_Strain_rate_eigen_values_gamma.tex
% This file was created by tikzplotlib v0.9.6.
\definecolor{mediumseagreen1}{RGB}{77,175,74}
\definecolor{darkorange1}{RGB}{255,127,0}

\begin{tikzpicture}

\begin{axis}[
legend cell align={left},
legend style={fill opacity=0.8, text opacity=1, draw=none},
log basis y={10},
tick pos=left,
x grid style={white!69.0196078431373!black},
xlabel={\(\displaystyle \gamma\)},
xlabel style={font=\LARGE},
xmin=-160, xmax=10,
xtick style={color=black},
xtick={-160,-140,-120,-100,-80,-60,-40,-20,0,20},
xticklabels={$-160$,,$-120$,,$-80$,,$-40$,,$0$,},
y grid style={white!69.0196078431373!black},
ymin=1e-08, ymax=1,
ymode=log,
ytick style={color=black},
ytick={1e-10,1e-08,1e-06,0.0001,0.01,1,100},
% yticklabels={\(\displaystyle 10^{-10}\),,\(\displaystyle 10^{-6}\),,\(\displaystyle 10^{-2}\),,\(\displaystyle 10^{2}\)}
% ]
yticklabels={,,,,,,}
]
\addplot [very thick, black]
table {%
-190.390230947391 4.86825332110279e-10
-188.477178127033 0
-186.564125306674 4.86825332110279e-10
-184.651072486316 4.86825332110279e-10
-182.738019665958 0
-180.824966845599 0
-178.911914025241 4.86825332110279e-10
-176.998861204882 0
-175.085808384524 0
-173.172755564166 4.86825332110279e-10
-171.259702743807 4.86825332110286e-10
-169.346649923449 9.73650664220558e-10
-167.433597103091 4.86825332110279e-10
-165.520544282732 4.86825332110286e-10
-163.607491462374 1.46047599633084e-09
-161.694438642015 0
-159.781385821657 2.43412666055143e-09
-157.868333001299 2.4341266605514e-09
-155.95528018094 4.86825332110279e-09
-154.042227360582 3.40777732477201e-09
-152.129174540223 3.89460265688223e-09
-150.216121719865 8.76285597798503e-09
-148.303068899507 5.84190398532335e-09
-146.390016079148 1.11969826385366e-08
-144.47696325879 1.11969826385364e-08
-142.563910438432 1.31442839669775e-08
-140.650857618073 1.65520612917497e-08
-138.737804797715 2.33676159412934e-08
-136.824751977356 2.14203146128523e-08
-134.911699156998 2.48280919376242e-08
-132.99864633664 3.94328519009332e-08
-131.085593516281 3.26172972513887e-08
-129.172540695923 4.81957078789176e-08
-127.259487875565 5.20903105358006e-08
-125.346435055206 7.88657038018652e-08
-123.433382234848 6.52345945027779e-08
-121.520329414489 8.86022104440715e-08
-119.607276594131 1.0515427173582e-07
-117.694223773773 1.28521887677115e-07
-115.781170953414 1.29008713009225e-07
-113.868118133056 1.75257119559701e-07
-111.955065312698 1.99111560833106e-07
-110.042012492339 2.15663622124855e-07
-108.128959671981 2.794377406313e-07
-106.215906851622 3.24712496517559e-07
-104.302854031264 3.63658523086379e-07
-102.389801210906 4.6540501749743e-07
-100.476748390547 5.56441354602053e-07
-98.5636955701889 6.57701023680987e-07
-96.6506427498305 8.12024653959952e-07
-94.7375899294721 9.43954318961838e-07
-92.8245371091137 1.14744730778393e-06
-90.9114842887553 1.44635806169965e-06
-88.998431468397 1.65958755716395e-06
-87.0853786480386 2.08555972276044e-06
-85.1723258276802 2.50228220704685e-06
-83.2592730073218 3.06651276696267e-06
-81.3462201869634 3.82303933306202e-06
-79.4331673666051 4.60585446709539e-06
-77.5201145462467 5.64814750314346e-06
-75.6070617258883 6.88760479869628e-06
-73.6940089055299 8.36317238032255e-06
-71.7809560851716 1.04901122563123e-05
-69.8679032648132 1.30240381099464e-05
-67.9548504444548 1.61114843661898e-05
-66.0417976240964 2.01540819240334e-05
-64.128744803738 2.51464757048244e-05
-62.2156919833797 3.17541559375577e-05
-60.3026391630213 3.95384929980006e-05
-58.3895863426629 4.98630846413953e-05
-56.4765335223045 6.32566231784143e-05
-54.5634807019461 8.04873189831246e-05
-52.6504278815878 0.000102917309334774
-50.7373750612294 0.000130757876427498
-48.824322240871 0.000168169429549507
-46.9112694205126 0.000216421609166949
-44.9982166001543 0.000279308245092963
-43.0851637797959 0.000362314398344422
-41.1721109594375 0.000472500009887602
-39.2590581390791 0.000619216994352347
-37.3460053187207 0.000814358494602082
-35.4329524983624 0.00107575059382272
-33.519899678004 0.00142862544897754
-31.6068468576456 0.00190713580441536
-29.6937940372872 0.00255647949762001
-27.7807412169288 0.00344698575019484
-25.8676883965705 0.00468046483066924
-23.9546355762121 0.00637870145094942
-22.0415827558537 0.00873167578911416
-20.1285299354953 0.0120042578245563
-18.2154771151369 0.0165424065717631
-16.3024242947786 0.0228159026343168
-14.3893714744202 0.0313902223581586
-12.4763186540618 0.0427720063268903
-10.5632658337034 0.0571655167133601
-8.65021301334505 0.0734289124903092
-6.73716019298668 0.0870403856896253
-4.82410737262829 0.0861641911281639
-2.91105455226992 0.0527726619601731
-0.998001731911545 0.00685949063576687
};
% \addlegendentry{DNS}
\addplot [very thick, blue]
table {%
-130.465467669085 7.1068597197846e-10
-129.155011840018 0
-127.844556010951 7.1068597197846e-10
-126.534100181883 0
-125.223644352816 7.10685971978468e-10
-123.913188523749 1.42137194395694e-09
-122.602732694682 1.42137194395694e-09
-121.292276865615 7.1068597197846e-10
-119.981821036547 1.42137194395694e-09
-118.67136520748 0
-117.360909378413 4.26411583187081e-09
-116.050453549346 5.68548777582774e-09
-114.739997720279 4.26411583187076e-09
-113.429541891211 7.81754569176323e-09
-112.119086062144 5.68548777582768e-09
-110.808630233077 5.68548777582774e-09
-109.49817440401 4.26411583187081e-09
-108.187718574943 9.23891763571998e-09
-106.877262745875 1.27923474956126e-08
-105.566806916808 1.27923474956123e-08
-104.256351087741 1.56350913835263e-08
-102.945895258674 1.98992072153971e-08
-101.635439429606 1.84778352714402e-08
-100.324983600539 2.2741951103311e-08
-99.014527771472 3.48236126269446e-08
-97.7040719424048 2.70060669351818e-08
-96.3936161133376 4.26411583187081e-08
-95.0831602842704 4.47732162346435e-08
-93.7727044552032 5.33014478983851e-08
-92.462248626136 6.1829679562126e-08
-91.1517927970688 7.81754569176315e-08
-89.8413369680016 9.30998623291793e-08
-88.5308811389343 1.0660289579677e-07
-87.2204253098672 1.07313581768749e-07
-85.9099694807999 1.47111996199541e-07
-84.5995136517327 1.89042468546273e-07
-83.2890578226655 1.74828749106703e-07
-81.9786019935983 2.43054602416636e-07
-80.6681461645311 2.70060669351818e-07
-79.3576903354639 3.63871217652972e-07
-78.0472345063967 4.15751293607408e-07
-76.7367786773295 5.2164350343219e-07
-75.4263228482622 5.8915867077015e-07
-74.115867019195 7.01447054342748e-07
-72.8054111901278 8.8622540705714e-07
-71.4949553610606 1.06034347019189e-06
-70.1844995319934 1.30481944455245e-06
-68.8740437029262 1.58056560168011e-06
-67.563587873859 1.92809104197758e-06
-66.2531320447918 2.3303393021174e-06
-64.9426762157245 2.81360576306272e-06
-63.6322203866573 3.46814754325492e-06
-62.3217645575901 4.18380831703724e-06
-61.0113087285229 5.14394506518015e-06
-59.7008528994557 6.18652138607256e-06
-58.3903970703885 7.68962221680702e-06
-57.0799412413213 9.43080284815417e-06
-55.7694854122541 1.17234757937568e-05
-54.4590295831869 1.45612448798668e-05
-53.1485737541196 1.77678599854337e-05
-51.8381179250524 2.21073085303342e-05
-50.5276620959852 2.76570552855138e-05
-49.217206266918 3.44803513024793e-05
-47.9067504378508 4.30924439109144e-05
-46.5962946087836 5.43703196002407e-05
-45.2858387797164 6.8389311083488e-05
-43.9753829506492 8.6017165932441e-05
-42.6649271215819 0.00010886571993155
-41.3544712925147 0.000138934843405959
-40.0440154634475 0.000176911769690572
-38.7335596343803 0.000227236864738311
-37.4231038053131 0.000290986817796721
-36.1126479762459 0.000375240771832687
-34.8021921471787 0.000485434053159893
-33.4917363181115 0.000631411083862186
-32.1812804890443 0.000822289964960051
-30.8708246599771 0.0010733205960402
-29.5603688309098 0.00141053327219828
-28.2499130018426 0.00185685259083243
-26.9394571727754 0.00245296887726825
-25.6290013437082 0.0032459529954878
-24.318545514641 0.00431386242853736
-23.0080896855738 0.00573014728230681
-21.6976338565066 0.00763333801988425
-20.3871780274394 0.0101689191665932
-19.0767221983721 0.0135365231226484
-17.7662663693049 0.0179681227641055
-16.4558105402377 0.0237626269252909
-15.1453547111705 0.0312054577528403
-13.8348988821033 0.0405672828419266
-12.5244430530361 0.0519500650196346
-11.2139872239689 0.0650160116900533
-9.90353139490167 0.0787166485494089
-8.59307556583445 0.0906753152053014
-7.28261973676724 0.0967755948486754
-5.97216390770003 0.0915284162498916
-4.66170807863282 0.0707684460844414
-3.35125224956561 0.0382343523787554
-2.0407964204984 0.0102647992324446
-0.73034059143119 0.000547786797597404
};
% \addlegendentry{C3}
% \addlegendentry{MPS - Eq. DNS 130}
\addplot [very thick, mediumseagreen1]
table {%
-141.195740908847 6.56564855986595e-10
-139.777263257898 0
-138.358785606949 0
-136.940307956 0
-135.521830305051 0
-134.103352654102 0
-132.684875003153 6.56564855986595e-10
-131.266397352204 1.31312971197319e-09
-129.847919701255 0
-128.429442050306 0
-127.010964399357 6.56564855986595e-10
-125.592486748409 0
-124.17400909746 2.62625942394638e-09
-122.755531446511 0
-121.337053795562 1.31312971197319e-09
-119.918576144613 1.31312971197318e-09
-118.500098493664 5.25251884789276e-09
-117.081620842715 1.96969456795979e-09
-115.663143191766 5.25251884789271e-09
-114.244665540817 7.22221341585255e-09
-112.826187889868 5.90908370387936e-09
-111.407710238919 5.25251884789276e-09
-109.98923258797 7.22221341585248e-09
-108.570754937021 9.84847283979893e-09
-107.152277286073 8.53534312782565e-09
-105.733799635124 9.84847283979893e-09
-104.315321984175 2.1010075391571e-08
-102.896844333226 3.02019833753834e-08
-101.478366682277 3.41413725113029e-08
-100.059889031328 2.69191590954501e-08
-98.6414113803791 4.13635859271555e-08
-97.2229337294301 4.26767156391283e-08
-95.8044560784812 5.12120587669544e-08
-94.3859784275323 7.68180881504316e-08
-92.9675007765834 8.33837367102976e-08
-91.5490231256345 1.05706941813842e-07
-90.1305454746855 1.28030146917385e-07
-88.7120678237366 1.54292741156848e-07
-87.2935901727877 1.90403808236113e-07
-85.8751125218388 2.04848235067818e-07
-84.4566348708899 2.42272431859054e-07
-83.0381572199409 3.03989528321794e-07
-81.619679568992 3.88029829888074e-07
-80.2012019180431 4.43837842646938e-07
-78.7827242670942 5.27221579357231e-07
-77.3642466161453 6.64443634258434e-07
-75.9457689651963 8.59443396486453e-07
-74.5272913142474 9.92069497395745e-07
-73.1088136632985 1.21924093756709e-06
-71.6903360123496 1.48514970424168e-06
-70.2718583614007 1.80161396482722e-06
-68.8533807104517 2.23888615891429e-06
-67.4349030595028 2.83110765901417e-06
-66.0164254085539 3.51853106323216e-06
-64.597947757605 4.40292392424611e-06
-63.1794701066561 5.3070137309396e-06
-61.7609924557071 6.85847648563597e-06
-60.3425148047582 8.21296978353632e-06
-58.9240371538093 1.05030680012175e-05
-57.5055595028604 1.33085696308483e-05
-56.0870818519115 1.64029597971131e-05
-54.6686042009625 2.06325505993788e-05
-53.2501265500136 2.63033012605347e-05
-51.8316488990647 3.31814746918506e-05
-50.4131712481158 4.20286861262699e-05
-48.9946935971669 5.30990261630594e-05
-47.5762159462179 6.78244627531273e-05
-46.157738295269 8.56331279069077e-05
-44.7392606443201 0.000109839361017421
-43.3207829933712 0.000141053110835879
-41.9023053424223 0.000181374728336585
-40.4838276914734 0.000232265070324106
-39.0653500405244 0.000299985139265128
-37.6468723895755 0.000387689729292677
-36.2283947386266 0.00050285054846787
-34.8099170876777 0.000653890010763299
-33.3914394367287 0.000849921892944936
-31.9729617857798 0.00111072586786421
-30.5544841348309 0.00145169378544003
-29.136006483882 0.00190001727918305
-27.7175288329331 0.00249354862617067
-26.2990511819842 0.00328132140297762
-24.8805735310352 0.00431878973390537
-23.4620958800863 0.00569455048062088
-22.0436182291374 0.00751548977541921
-20.6251405781885 0.00992085123639308
-19.2066629272395 0.0130853198525617
-17.7881852762906 0.0172265291464333
-16.3697076253417 0.0226096952087594
-14.9512299743928 0.0295073259395819
-13.5327523234439 0.0381927400965404
-12.1142746724949 0.0488276863712966
-10.695797021546 0.061253631926853
-9.27731937059711 0.0746953555367548
-7.8588417196482 0.0870751012496548
-6.44036406869928 0.094143726418441
-5.02188641775037 0.0887609496990745
-3.60340876680144 0.0631537463776335
-2.18493111585251 0.0233929593547003
-0.766453464903592 0.00159804077870973
};
% \addlegendentry{C4}
% \addlegendentry{MPS - Eq. DNS 300}
\addplot [very thick, darkorange1]
table {%
-159.891409904871 5.79821239899628e-10
-158.285186278471 0
-156.678962652071 0
-155.07273902567 0
-153.46651539927 0
-151.86029177287 0
-150.254068146469 1.15964247979926e-09
-148.647844520069 2.31928495959851e-09
-147.041620893669 1.15964247979924e-09
-145.435397267269 1.15964247979926e-09
-143.829173640868 5.79821239899618e-10
-142.222950014468 1.15964247979926e-09
-140.616726388068 1.15964247979926e-09
-139.010502761668 2.31928495959847e-09
-137.404279135267 2.89910619949814e-09
-135.798055508867 1.73946371969885e-09
-134.191831882467 4.63856991919702e-09
-132.585608256066 5.21839115909656e-09
-130.979384629666 3.47892743939777e-09
-129.373161003266 7.53767611869503e-09
-127.766937376866 5.21839115909665e-09
-126.160713750465 1.15964247979926e-08
-124.554490124065 1.68148159570889e-08
-122.948266497665 1.79744584368885e-08
-121.342042871264 1.56551734772898e-08
-119.735819244864 2.72515982752823e-08
-118.129595618464 3.13103469545796e-08
-116.523371992064 2.66717770353827e-08
-114.917148365663 4.0587486792974e-08
-113.310924739263 3.82682018333748e-08
-111.704701112863 5.4503196550565e-08
-110.098477486463 6.95785487879547e-08
-108.492253860062 7.88556886263487e-08
-106.886030233662 8.81328284647435e-08
-105.279806607262 1.15384426740024e-07
-103.673582980861 1.28720315257717e-07
-102.067359354461 1.45535131214805e-07
-100.461135728061 1.57131556012798e-07
-98.8549121016606 2.19172428682057e-07
-97.2486884752603 2.55701166795734e-07
-95.64246484886 2.75994910192221e-07
-94.0362412224597 3.32237570462487e-07
-92.4300175960595 4.34865929924717e-07
-90.8237939696592 4.95167338874278e-07
-89.2175703432589 5.9953516205621e-07
-87.6113467168586 7.27675656074027e-07
-86.0051230904583 8.65093289930237e-07
-84.3988994640581 1.07382893629411e-06
-82.7926758376578 1.26748923042058e-06
-81.1864522112575 1.60494519204216e-06
-79.5802285848573 1.92848544390615e-06
-77.974004958457 2.32798227819699e-06
-76.3677813320567 2.88982905965972e-06
-74.7615577056564 3.44239870128406e-06
-73.1553340792561 4.30923145493404e-06
-71.5491104528559 5.28507060168506e-06
-69.9428868264556 6.40238613097164e-06
-68.3366632000553 7.86295583427879e-06
-66.730439573655 9.60647830265695e-06
-65.1242159472548 1.1980266458806e-05
-63.5179923208545 1.48324071378724e-05
-61.9117686944542 1.80330203821182e-05
-60.3055450680539 2.27974115103734e-05
-58.6993214416536 2.85138691145438e-05
-57.0930978152534 3.55499998607257e-05
-55.4868741888531 4.42218063246644e-05
-53.8806505624528 5.61939552861119e-05
-52.2744269360525 7.00516829197135e-05
-50.6682033096523 8.85346445839931e-05
-49.061979683252 0.000111917675546665
-47.4557560568517 0.000141901971325594
-45.8495324304514 0.000179629779763383
-44.2433088040511 0.000229295527709465
-42.6370851776509 0.000291800257370647
-41.0308615512506 0.000372525869495669
-39.4246379248503 0.000479470418267715
-37.81841429845 0.000616301852850388
-36.2121906720498 0.00079662451899545
-34.6059670456495 0.00103006113000028
-32.9997434192492 0.00133919108458904
-31.3935197928489 0.00174138640693039
-29.7872961664487 0.00227425662192911
-28.1810725400484 0.0029753782046795
-26.5748489136481 0.00390693698316068
-24.9686252872478 0.00514558764457228
-23.3624016608476 0.00678933853273731
-21.7561780344473 0.00897055819815448
-20.149954408047 0.0118647022874817
-18.5437307816467 0.0156960404379203
-16.9375071552464 0.0207352848083467
-15.3312835288462 0.0273231418382525
-13.7250599024459 0.0357664100210205
-12.1188362760456 0.0463048999425478
-10.5126126496453 0.058900764351218
-8.90638902324505 0.072662887579858
-7.30016539684479 0.0850853947149392
-5.69394177044451 0.0902624892658648
-4.08771814404423 0.0775547622646663
-2.48149451764395 0.038756959248442
-0.875270891243669 0.00382839439800385
};
% \addlegendentry{C5}
% \addlegendentry{MPS - Eq. DNS 500}
\end{axis}

\end{tikzpicture}

%% file: PDF_Strain_rate_eigen_values_stilde.tex
% This file was created by tikzplotlib v0.9.6.
\definecolor{mediumseagreen1}{RGB}{77,175,74}
\definecolor{darkorange1}{RGB}{255,127,0}

\begin{tikzpicture}

\begin{axis}[
legend cell align={left},
legend style={fill opacity=0.8, text opacity=1, draw=none},
tick pos=left,
x grid style={white!69.0196078431373!black},
xlabel={\(\displaystyle s^*\)},
xlabel style={font=\LARGE},
xmin=-1.1, xmax=1.1,
xtick style={color=black},
xtick={-1.25,-1,-0.75,-0.5,-0.25,0,0.25,0.5,0.75,1,1.25},
xticklabels={,-1,,-0.5,,0,,0.5,,1,},
y grid style={white!69.0196078431373!black},
ymin=0, ymax=1.6,
ytick style={color=black},
ytick={0,0.2,0.4,0.6,0.8,1,1.2,1.4,1.6},
yticklabels={0.0,,0.4,,0.8,,1.2,,1.6}
]
\addplot [very thick, black]
table {%
-0.989999981451225 0.156583265937171
-0.969999981638611 0.158484188462029
-0.949999981825996 0.160710608229765
-0.929999982013382 0.162989973686364
-0.909999982200767 0.164944400693635
-0.889999982388152 0.167236339005108
-0.869999982575538 0.169380383290348
-0.849999982762923 0.171876560644348
-0.829999982950309 0.174036483979633
-0.809999983137694 0.17671119587286
-0.78999998332508 0.178961178101954
-0.769999983512465 0.181591698738602
-0.749999983699851 0.18426580527215
-0.729999983887236 0.186707593387284
-0.709999984074622 0.189621841448945
-0.689999984262007 0.192510850668599
-0.669999984449392 0.195204421844139
-0.649999984636778 0.198325796048337
-0.629999984824163 0.201284328498837
-0.609999985011549 0.20423825090255
-0.589999985198934 0.207573270507366
-0.569999985386319 0.210685936845409
-0.549999985573705 0.214144915486068
-0.52999998576109 0.217703964738306
-0.509999985948476 0.221154095814424
-0.489999986135861 0.224787324910425
-0.469999986323247 0.228722628486335
-0.449999986510632 0.232199581652851
-0.429999986698018 0.236504853124992
-0.409999986885403 0.240539668518931
-0.389999987072788 0.244761120958571
-0.369999987260174 0.248858241833335
-0.349999987447559 0.253467031335387
-0.329999987634945 0.258148929660231
-0.30999998782233 0.262917953212747
-0.289999988009715 0.267668210615208
-0.269999988197101 0.272793325357465
-0.249999988384486 0.277753735700859
-0.229999988571872 0.283101389974403
-0.209999988759257 0.288614028043587
-0.189999988946643 0.294305060953611
-0.169999989134028 0.300353256074346
-0.149999989321414 0.306300030165755
-0.129999989508799 0.31260541001788
-0.109999989696184 0.319307719555025
-0.0899999898835698 0.325913870035439
-0.0699999900709553 0.333065263322271
-0.0499999902583407 0.340163431523465
-0.0299999904457262 0.347801627124612
-0.00999999063311163 0.355277167236574
0.010000009179503 0.363533062541117
0.0300000089921176 0.371938295421904
0.0500000088047322 0.38058269194209
0.0700000086173467 0.389545880182558
0.0900000084299611 0.39915466123779
0.110000008242576 0.408709984376743
0.13000000805519 0.419203195923249
0.150000007867805 0.42954967759675
0.170000007680419 0.440230923172833
0.190000007493034 0.45174942775134
0.210000007305649 0.463999346520674
0.230000007118263 0.475919996190667
0.250000006930878 0.489036929718963
0.270000006743492 0.502899386975223
0.290000006556107 0.516624008489148
0.310000006368721 0.531273992121882
0.330000006181336 0.545987305690283
0.350000005993951 0.562041025286757
0.370000005806565 0.578582525086257
0.39000000561918 0.5957483369795
0.410000005431794 0.613645982158732
0.430000005244409 0.632017251437722
0.450000005057024 0.651563477287856
0.470000004869638 0.67148870486328
0.490000004682253 0.691954751462859
0.510000004494867 0.713241479522862
0.530000004307482 0.735492824948569
0.550000004120096 0.758492068974828
0.570000003932711 0.78171250211463
0.590000003745325 0.806120371548193
0.61000000355794 0.830960328094297
0.630000003370555 0.856325224614501
0.650000003183169 0.882651618569747
0.670000002995784 0.909059968912327
0.690000002808398 0.936413704809257
0.710000002621013 0.964371729364387
0.730000002433627 0.991912195439277
0.750000002246242 1.02073891113402
0.770000002058857 1.04964362509512
0.790000001871471 1.07842628923167
0.810000001684086 1.1073409683444
0.8300000014967 1.13702091522388
0.850000001309315 1.16629801851992
0.870000001121929 1.19607151675289
0.890000000934544 1.22559518770287
0.910000000747159 1.25556365831866
0.930000000559773 1.28506865625087
0.950000000372388 1.31448136011503
0.970000000185002 1.34397941969398
0.989999999997617 1.37317889725269
};
% \addlegendentry{$\tilde{s}^*$ DNS}
% \addlegendentry{DNS 1024}
\addplot [very thick, blue]
table {%
-0.989999998363741 0.399444439559426
-0.96999999841593 0.400502422006948
-0.949999998468119 0.402072585305721
-0.929999998520308 0.403353153849158
-0.909999998572497 0.404766994653369
-0.889999998624686 0.406072476075745
-0.869999998676875 0.407474069988068
-0.849999998729064 0.408828492411868
-0.829999998781253 0.410625525890407
-0.809999998833442 0.411905116545136
-0.789999998885631 0.413255627414113
-0.769999998937821 0.414799108321048
-0.74999999899001 0.416140119699739
-0.729999999042199 0.41787903864947
-0.709999999094388 0.419503125525451
-0.689999999146577 0.421076036225821
-0.669999999198766 0.42253588436534
-0.649999999250955 0.424441975717653
-0.629999999303144 0.42591733037808
-0.609999999355333 0.427524746579932
-0.589999999407522 0.4293460410116
-0.569999999459711 0.431290689118603
-0.5499999995119 0.432495820533298
-0.529999999564089 0.434501982496514
-0.509999999616279 0.436132355799889
-0.489999999668468 0.437900238881766
-0.469999999720657 0.439708960458648
-0.449999999772846 0.44178175855225
-0.429999999825035 0.443527057061631
-0.409999999877224 0.445205812572887
-0.389999999929413 0.447214256276413
-0.369999999981602 0.44934088481497
-0.350000000033791 0.451017591416555
-0.33000000008598 0.453051413660305
-0.310000000138169 0.454841741616291
-0.290000000190358 0.456868159845553
-0.270000000242548 0.458943984737533
-0.250000000294737 0.460950239833005
-0.230000000346926 0.462880918101347
-0.210000000399115 0.465273998029198
-0.190000000451304 0.467045513269125
-0.170000000503493 0.469132933127184
-0.150000000555682 0.471291645728521
-0.130000000607871 0.473939628645706
-0.11000000066006 0.475902763505858
-0.0900000007122493 0.478017098586399
-0.0700000007644384 0.480033458531829
-0.0500000008166275 0.482661231749095
-0.0300000008688166 0.484790235160222
-0.0100000009210057 0.486797561276654
0.00999999902680532 0.489111152822551
0.0299999989746162 0.491860323937735
0.0499999989224271 0.494011073731036
0.069999998870238 0.496622455670943
0.0899999988180489 0.499021030401996
0.10999999876586 0.501759956968829
0.129999998713671 0.503972686279627
0.149999998661482 0.506411075050853
0.169999998609293 0.508950512321684
0.189999998557103 0.511570788392202
0.209999998504914 0.514322893173501
0.229999998452725 0.51678386651721
0.249999998400536 0.519547333433954
0.269999998348347 0.52219326744147
0.289999998296158 0.525136945276867
0.309999998243969 0.527842064834155
0.32999999819178 0.530648698552334
0.349999998139591 0.533367415419251
0.369999998087402 0.536403806417194
0.389999998035213 0.539309439725319
0.409999997983024 0.542563620507903
0.429999997930835 0.54539912522597
0.449999997878646 0.54833321043883
0.469999997826456 0.551332162269175
0.489999997774267 0.554721384825431
0.509999997722078 0.557671535352739
0.529999997669889 0.560959988806158
0.5499999976177 0.564373612013926
0.569999997565511 0.56764553449161
0.589999997513322 0.570958388596548
0.609999997461133 0.574498531976414
0.629999997408944 0.578021585586992
0.649999997356755 0.581182308148963
0.669999997304566 0.584981452337512
0.689999997252376 0.588498964578756
0.709999997200187 0.59195603398475
0.729999997147998 0.595516480196795
0.749999997095809 0.599576953500457
0.76999999704362 0.603247109512072
0.789999996991431 0.607622556111297
0.809999996939242 0.611201954737786
0.829999996887053 0.615039003755214
0.849999996834864 0.619188371798855
0.869999996782675 0.623461652311365
0.889999996730486 0.627567806987438
0.909999996678297 0.632002113173525
0.929999996626107 0.636701287500525
0.949999996573918 0.640989934835563
0.969999996521729 0.645609108553187
0.98999999646954 0.650327328426894
};
% \addlegendentry{C3}
% \addlegendentry{MPS - Eq. DNS 130}
\addplot [very thick, mediumseagreen1]
table {%
-0.989999999306546 0.225315941651719
-0.969999999320219 0.227835495180677
-0.949999999333891 0.230352161609651
-0.929999999347564 0.232810853208318
-0.909999999361237 0.235533761021583
-0.88999999937491 0.238104118197021
-0.869999999388583 0.240993872449643
-0.849999999402255 0.243727490472523
-0.829999999415928 0.246501201932266
-0.809999999429601 0.24939430894616
-0.789999999443274 0.252268603244032
-0.769999999456947 0.25536054762774
-0.749999999470619 0.258467113775878
-0.729999999484292 0.26154019887744
-0.709999999497965 0.264699151920437
-0.689999999511638 0.26801335643673
-0.669999999525311 0.271427957526634
-0.649999999538983 0.274778716454005
-0.629999999552656 0.278165983226326
-0.609999999566329 0.28178114463281
-0.589999999580002 0.285519520016002
-0.569999999593675 0.289034377981133
-0.549999999607347 0.292650051615033
-0.52999999962102 0.296551594713409
-0.509999999634693 0.30044354518926
-0.489999999648366 0.304520875423713
-0.469999999662039 0.308593502479162
-0.449999999675711 0.31288745834462
-0.429999999689384 0.316969631456465
-0.409999999703057 0.32129939667494
-0.38999999971673 0.326008955239605
-0.369999999730403 0.330521213116702
-0.349999999744075 0.335268536811793
-0.329999999757748 0.339782983296941
-0.309999999771421 0.344924768668778
-0.289999999785094 0.349505106226127
-0.269999999798767 0.354929082338267
-0.249999999812439 0.359833473585674
-0.229999999826112 0.3652820366138
-0.209999999839785 0.370708154767861
-0.189999999853458 0.376354065049686
-0.169999999867131 0.381722860299242
-0.149999999880803 0.387710845362955
-0.129999999894476 0.393740413979652
-0.109999999908149 0.399633683537462
-0.0899999999218219 0.405887700894793
-0.0699999999354947 0.412331801189737
-0.0499999999491675 0.419138511929354
-0.0299999999628403 0.425634300627223
-0.00999999997651313 0.432151556310451
0.0100000000098141 0.439254008533837
0.0299999999961413 0.446263514764217
0.0499999999824685 0.453475723353099
0.0699999999687957 0.46108760902515
0.0899999999551229 0.468662940286122
0.10999999994145 0.476327585382061
0.129999999927777 0.484541990194172
0.149999999914104 0.492395833471273
0.169999999900432 0.500946166336272
0.189999999886759 0.50929747556695
0.209999999873086 0.517935818415315
0.229999999859413 0.527158053952234
0.249999999845741 0.535968086117347
0.269999999832068 0.545349484682377
0.289999999818395 0.554898567877082
0.309999999804722 0.56439368092855
0.329999999791049 0.574545096998799
0.349999999777376 0.584425172236748
0.369999999763704 0.594860408861804
0.389999999750031 0.605413364660366
0.409999999736358 0.616477523420498
0.429999999722685 0.626864890065186
0.449999999709012 0.638327934284291
0.46999999969534 0.649506040361498
0.489999999681667 0.660930481259859
0.509999999667994 0.672787428362167
0.529999999654321 0.68462500395491
0.549999999640648 0.696854573219576
0.569999999626976 0.709534343817119
0.589999999613303 0.722513208659697
0.60999999959963 0.734990882258418
0.629999999585957 0.748490030457668
0.649999999572284 0.761468802167996
0.669999999558612 0.774827646621284
0.689999999544939 0.788854063096514
0.709999999531266 0.802738266614502
0.729999999517593 0.816878444142298
0.74999999950392 0.830755103947426
0.769999999490248 0.84540941810289
0.789999999476575 0.860023918218263
0.809999999462902 0.874783704655364
0.829999999449229 0.890097302061859
0.849999999435556 0.905209128432416
0.869999999421884 0.920423539984637
0.889999999408211 0.935987942560637
0.909999999394538 0.951384101713418
0.929999999380865 0.967181148668717
0.949999999367192 0.983241527082168
0.96999999935352 0.999431405899708
0.989999999339847 1.01558966631581
};
% \addlegendentry{C4}
% \addlegendentry{MPS - Eq. DNS 300}
\addplot [very thick, darkorange1]
table {%
-0.98999999516975 0.166204898162199
-0.969999995226951 0.168085610974756
-0.949999995284153 0.170447864125896
-0.929999995341354 0.172678754227511
-0.909999995398556 0.174787128844084
-0.889999995455758 0.1771183689534
-0.86999999551296 0.179657154299063
-0.849999995570161 0.181864435373344
-0.829999995627363 0.184484432047745
-0.809999995684564 0.187093066586704
-0.789999995741766 0.189526565915006
-0.769999995798968 0.192290824456628
-0.74999999585617 0.194989331624294
-0.729999995913371 0.197730400233742
-0.709999995970573 0.20046327320451
-0.689999996027775 0.203530630678908
-0.669999996084976 0.206628814930613
-0.649999996142178 0.209634984511684
-0.62999999619938 0.212699873981252
-0.609999996256581 0.215971517062976
-0.589999996313783 0.219076220572722
-0.569999996370985 0.222554198171442
-0.549999996428186 0.225992780825144
-0.529999996485388 0.229625870199109
-0.50999999654259 0.233418681395079
-0.489999996599791 0.237175403841178
-0.469999996656993 0.240962999630715
-0.449999996714195 0.245054020315929
-0.429999996771396 0.249087671530383
-0.409999996828598 0.253198063725204
-0.3899999968858 0.257455744585622
-0.369999996943001 0.262036734078743
-0.349999997000203 0.26650885196258
-0.329999997057405 0.271368539736951
-0.309999997114606 0.276175933748608
-0.289999997171808 0.281248708128661
-0.26999999722901 0.286030071674283
-0.249999997286211 0.291554164975286
-0.229999997343413 0.297240401536992
-0.209999997400615 0.302562118441952
-0.189999997457816 0.308251335235906
-0.169999997515018 0.314554247042158
-0.14999999757222 0.320547773488238
-0.129999997629421 0.326867588799268
-0.109999997686623 0.333657070086019
-0.0899999977438247 0.340325386305467
-0.0699999978010263 0.347624114777474
-0.049999997858228 0.354689360679358
-0.0299999979154297 0.362230698682747
-0.00999999797263135 0.37008132891405
0.010000001970167 0.377942622788862
0.0300000019129654 0.386531279596534
0.0500000018557636 0.395382755635511
0.0700000017985619 0.404259656780842
0.0900000017413604 0.413282263343394
0.110000001684159 0.423027855622691
0.130000001626957 0.433104812474981
0.150000001569755 0.44325371399636
0.170000001512554 0.454084762739472
0.190000001455352 0.465202193175102
0.21000000139815 0.476638229064415
0.230000001340949 0.488848380279957
0.250000001283747 0.501256111580273
0.270000001226545 0.51422524413018
0.290000001169344 0.527143060806072
0.310000001112142 0.541366964264335
0.33000000105494 0.555418712744194
0.350000000997739 0.570538129426652
0.370000000940537 0.585944952256452
0.390000000883335 0.601923607148694
0.410000000826134 0.618400985281077
0.430000000768932 0.635619603052722
0.45000000071173 0.653088979428294
0.470000000654529 0.671569958387298
0.490000000597327 0.690306864790546
0.510000000540125 0.709974209014633
0.530000000482924 0.729572542235735
0.550000000425722 0.750247670620679
0.57000000036852 0.770752972333665
0.590000000311319 0.79316017444785
0.610000000254117 0.815251844472844
0.630000000196915 0.838344267173245
0.650000000139714 0.861391427596398
0.670000000082512 0.885192307381353
0.69000000002531 0.909631655359153
0.709999999968108 0.934508166154717
0.729999999910907 0.959465934845159
0.749999999853705 0.985076188133253
0.769999999796503 1.01108155502475
0.789999999739302 1.03704892395508
0.8099999996821 1.06369918516521
0.829999999624899 1.09037068053011
0.849999999567697 1.11740944295336
0.869999999510495 1.14515228524501
0.889999999453294 1.17272339165343
0.909999999396092 1.20021407835731
0.92999999933889 1.22816232384339
0.949999999281688 1.25579480440964
0.969999999224487 1.28355622658672
0.989999999167285 1.310812173435
};
% \addlegendentry{C5}
% \addlegendentry{MPS - Eq. DNS 500}
\end{axis}

\end{tikzpicture}

%% file: scalar_PDF.tex
\definecolor{mediumseagreen1}{RGB}{77,175,74}
\definecolor{darkorange1}{RGB}{255,127,0}
\begin{tikzpicture}

\begin{axis}[
legend cell align={left},
legend style={fill opacity=0.8, text opacity=1, at={(0.5,0.09)}, anchor=south, draw=none},
log basis y={10},
tick pos=left,
x grid style={white!69.0196078431373!black},
xlabel={$\phi$},
xmin=-0.12, xmax=0.12,
xtick style={color=black},
xtick={-0.15,-0.1,-0.05,0,0.05,0.1,0.15},
xticklabels={-0.15,-0.1,-0.05,0,0.05,0.1,0.15},
y grid style={white!69.0196078431373!black},
ylabel={PDF},
ymin=1e-06, ymax=100,
ymode=log,
ytick style={color=black},
ytick={1e-07,1e-06,1e-05,0.0001,0.001,0.01,0.1,1,10,100,1000},
yticklabels={\(\displaystyle 10^{-7}\),,\(\displaystyle 10^{-5}\),,\(\displaystyle 10^{-3}\),,\(\displaystyle 10^{-1}\),,\(\displaystyle 10^{1}\),,\(\displaystyle 10^{3}\)}
]
\addplot [thick, black]
table {%
-0.104321690701141 2.52596771653188e-05
-0.102293844451618 7.89938994988156e-05
-0.100265998202095 0.000185543810450705
-0.0982381519525725 0.000352257679741811
-0.0962103057030496 0.000529534686756592
-0.0941824594535267 0.000773864654973857
-0.0921546132040038 0.00152614376764282
-0.0901267669544809 0.00235374264495016
-0.088098920704958 0.00357998515461202
-0.086071074455435 0.00527651692640632
-0.0840432282059121 0.00756412514386915
-0.0820153819563892 0.00997435761229806
-0.0799875357068663 0.0137614721196657
-0.0779596894573434 0.020354247859814
-0.0759318432078205 0.0321647543686018
-0.0739039969582976 0.0456998079274947
-0.0718761507087747 0.0586254143664175
-0.0698483044592518 0.0749022910648915
-0.0678204582097289 0.0877277773288666
-0.0657926119602059 0.0964295064788914
-0.063764765710683 0.112969543086742
-0.0617369194611601 0.135341809518636
-0.0597090732116372 0.16734490195338
-0.0576812269621143 0.215491683697908
-0.0556533807125914 0.294954494658282
-0.0536255344630685 0.435123198849783
-0.0515976882135456 0.660682930598929
-0.0495698419640227 1.00361060412959
-0.0475419957144998 1.54186814631165
-0.0455141494649768 2.32398490818198
-0.0434863032154539 3.25817273412414
-0.041458456965931 4.2315433712851
-0.0394306107164081 5.32621189672765
-0.0374027644668852 6.46405931431664
-0.0353749182173623 7.51705809850667
-0.0333470719678394 8.03317760035424
-0.0313192257183165 8.18302535356858
-0.0292913794687936 8.47381980903116
-0.0272635332192706 8.81185950539696
-0.0252356869697477 9.09791524542327
-0.0232078407202248 9.50770528441059
-0.0211799944707019 10.162276750936
-0.019152148221179 10.9165950084525
-0.0171243019716561 11.7876754347654
-0.0150964557221332 12.7795311036759
-0.0130686094726103 13.7529995646777
-0.0110407632230874 14.8294967376342
-0.00901291697356446 15.7884895665528
-0.00698507072404155 16.3760774211712
-0.00495722447451864 17.0305005445017
-0.00292937822499573 17.5191774738101
-0.000901531975472818 17.8623967817952
0.00112631427405009 18.1378609095528
0.003154160523573 18.1750054944577
0.00518200677309592 18.1351663908995
0.00720985302261883 18.3101144556797
0.00923769927214173 17.6106836395759
0.0112655455216646 16.9988005681931
0.0132933917711876 16.2679774845751
0.0153212380207105 14.9604975568153
0.0173490842702334 13.3991767034853
0.0193769305197563 11.7775734009668
0.0214047767692792 10.2622605236254
0.0234326230188021 9.01146790409332
0.025460469268325 7.93982655962152
0.0274883155178479 7.00867265054027
0.0295161617673708 6.36528433199624
0.0315440080168937 5.81455861316537
0.0335718542664167 5.06651072605504
0.0355997005159396 4.40051729278218
0.0376275467654625 3.91443200635114
0.0396553930149854 3.50688732303045
0.0416832392645083 3.11214378483024
0.0437110855140312 2.76538398864022
0.0457389317635541 2.44127477459773
0.047766778013077 2.14692008245537
0.0497946242626 1.88236905376701
0.0518224705121229 1.61833194469222
0.0538503167616458 1.37534073936472
0.0558781630111687 1.1794211720682
0.0579060092606916 0.985197677276644
0.0599338555102145 0.817990555380529
0.0619617017597374 0.671370529640997
0.0639895480092603 0.530491798887732
0.0660173942587832 0.413516530269428
0.0680452405083061 0.317434688801698
0.0700730867578291 0.24032516121942
0.072100933007352 0.171679462554949
0.0741287792568749 0.117286651547724
0.0761566255063978 0.0812332846962344
0.0781844717559207 0.0528308444246934
0.0802123180054436 0.0347430785068948
0.0822401642549665 0.0223286360804265
0.0842680105044894 0.0133715545576047
0.0862958567540123 0.00784565572754807
0.0883237030035352 0.00434833860729527
0.0903515492530581 0.00222009598940273
0.0923793955025811 0.0014094899858248
0.094407241752104 0.000799583598996733
0.0964350880016269 0.000127216919541697
};
\addlegendentry{DNS}
\addplot [very thick, red, dashed]
table {%
-0.105335613825903 0.00034680632707817
-0.10328728428093 0.000526427775512057
-0.101238954735957 0.000792558626797618
-0.0991906251909848 0.00118349061387402
-0.0971422956460122 0.00175282701003913
-0.0950939661010395 0.00257486292632545
-0.0930456365560669 0.00375154352976538
-0.0909973070110943 0.00542134049091264
-0.0889489774661216 0.00777041469506113
-0.086900647921149 0.0110464448400607
-0.0848523183761763 0.0155754885526038
-0.0828039888312037 0.0217821952239916
-0.0807556592862311 0.0302135970414868
-0.0787073297412584 0.0415665555521939
-0.0766590001962858 0.0567187252277069
-0.0746106706513132 0.0767626048143382
-0.0725623411063405 0.103041877647225
-0.0705140115613679 0.137188795425567
-0.0684656820163952 0.181160846054832
-0.0664173524714226 0.237274384853115
-0.06436902292645 0.308232330894631
-0.0623206933814773 0.397142479845105
-0.0602723638365047 0.507522516364354
-0.0582240342915321 0.64328748790229
-0.0561757047465594 0.808715398709142
-0.0541273752015868 1.00838677032838
-0.0520790456566141 1.24709455893662
-0.0500307161116415 1.52972177246337
-0.0479823865666689 1.86108551983744
-0.0459340570216962 2.24574804695343
-0.0438857274767236 2.68779752493669
-0.041837397931751 3.19060386658984
-0.0397890683867783 3.75655752018534
-0.0377407388418057 4.38680184583573
-0.0356924092968331 5.08097210289638
-0.0336440797518604 5.83695603007309
-0.0315957502068878 6.6506922433164
-0.0295474206619151 7.51602299041805
-0.0274990911169425 8.42461701031227
-0.0254507615719699 9.36597624282856
-0.0234024320269972 10.3275369036322
-0.0213541024820246 11.2948710654959
-0.019305772937052 12.2519895663226
-0.0172574433920793 13.1817410964283
-0.0152091138471067 14.0662960908328
-0.013160784302134 14.8876980166039
-0.0111124547571614 15.6284592762311
-0.00906412521218877 16.2721747051003
-0.00701579566721613 16.8041229241535
-0.00496746612224351 17.2118249168217
-0.00291913657727087 17.4855302970335
-0.000870807032298235 17.6186048307093
0.0011775225126744 17.6077977104682
0.00322585205764704 17.4533735498531
0.00527418160261968 17.1591016142224
0.00732251114759232 16.7321029002261
0.00937084069256496 16.1825637259763
0.0114191702375376 15.5233319150692
0.0134674997825102 14.7694179206615
0.0155158293274829 13.9374279136176
0.0175641588724555 13.0449586611308
0.0196124884174281 12.1099848174981
0.0216608179624008 11.1502680693799
0.0237091475073734 10.1828146104794
0.025757477052346 9.22340298164602
0.0278058065973187 8.28619881338087
0.0298541361422913 7.38346691330934
0.031902465687264 6.52538492568122
0.0339507952322366 5.71995689613761
0.0359991247772092 4.97301987941298
0.0380474543221819 4.28833251536065
0.0400957838671545 3.66773144882546
0.0421441134121271 3.11133965150066
0.0441924429570998 2.61781008713099
0.0462407725020724 2.18458862697115
0.048289102047045 1.80818148685608
0.0503374315920177 1.48441449627792
0.0523857611369903 1.20867398271685
0.0544340906819629 0.976121726110286
0.0564824202269356 0.781879097691714
0.0585307497719082 0.621177970459462
0.0605790793168809 0.489478145859769
0.0626274088618535 0.382552800457919
0.0646757384068261 0.296544779742205
0.0667240679517988 0.227997455127392
0.0687723974967714 0.173864347165412
0.0708207270417441 0.131501857477054
0.0728690565867167 0.0986493115892861
0.0749173861316893 0.0734001671670958
0.076965715676662 0.0541677576281046
0.0790140452216346 0.0396483834203414
0.0810623747666072 0.028783985686986
0.0831107043115799 0.0207260811381006
0.0851590338565525 0.0148021321237575
0.0872073634015252 0.0104850903535341
0.0892556929464978 0.00736649496966974
0.0913040224914704 0.00513322658050976
0.093352352036443 0.00354781357159229
0.0954006815814157 0.00243204701068063
0.0974490111263883 0.00165357523221084
};
\addlegendentry{Gaussian}
\addplot [very thick, blue]
table {%
-0.107958944933655 5.1498601292784e-06
-0.105788813975758 3.0040850754124e-05
-0.103618683017861 6.99522667560316e-05
-0.101448552059965 0.000156641578932218
-0.0992784211020678 0.000251484836313095
-0.0971082901441709 0.000386239509695883
-0.0949381591862741 0.000646736601235213
-0.0927680282283772 0.00118618444977713
-0.0905978972704804 0.00212045490823038
-0.0884277663125835 0.00346928910709055
-0.0862576353546867 0.00498077305503376
-0.0840875043967898 0.0069209828587394
-0.081917373438893 0.0099555379399167
-0.0797472424809961 0.0147504868752857
-0.0775771115230993 0.0225804200468428
-0.0754069805652024 0.0335444722620762
-0.0732368496073056 0.0459440479883468
-0.0710667186494087 0.0611623138253746
-0.0688965876915119 0.0741193619106399
-0.066726456733615 0.0861151027717719
-0.0645563257757182 0.100799499775398
-0.0623861948178213 0.122720308570683
-0.0602160638599245 0.152837548916724
-0.0580459329020276 0.199205172900704
-0.0558758019441308 0.275726944561653
-0.0537056709862339 0.412385349572227
-0.0515355400283371 0.648129638400182
-0.0493654090704402 1.01938133849982
-0.0471952781125434 1.6066542214423
-0.0450251471546465 2.43382303878695
-0.0428550161967497 3.41545118183404
-0.0406848852388528 4.51985156120812
-0.038514754280956 5.69810780131622
-0.0363446233230591 6.85799148625811
-0.0341744923651623 7.67875773999704
-0.0320043614072654 8.08209135208204
-0.0298342304493686 8.36524267833006
-0.0276640994914717 8.69928363491064
-0.0254939685335749 9.03079171861753
-0.023323837575678 9.47886001987052
-0.0211537066177812 10.1085909246488
-0.0189835756598843 10.9062128486568
-0.0168134447019875 11.8106986618973
-0.0146433137440906 12.7727307288425
-0.0124731827861938 13.8686982122549
-0.0103030518282969 15.1462557719958
-0.00813292087040007 16.3146414468561
-0.00596278991250321 17.0968137904762
-0.00379265895460637 17.6484517870992
-0.00162252799670953 18.1243808316515
0.000547602961187325 18.5105761383813
0.00271773391908418 18.8998815314741
0.00488786487698103 18.9422884838636
0.00705799583487788 18.5435270892634
0.00922812679277473 17.9562087158898
0.0113982577506716 17.2839069215078
0.0135683887085684 16.3056776861461
0.0157385196664653 14.6726205609761
0.0179086506243621 12.661455555254
0.020078781582259 10.7985262692685
0.0222489125401558 9.32012201871061
0.0244190434980527 8.21825236643518
0.0265891744559495 7.26382750121088
0.0287593054138464 6.48043120750015
0.0309294363717432 5.84809018194631
0.0330995673296401 5.13597310228476
0.0352696982875369 4.42926981308425
0.0374398292454338 3.89635284549623
0.0396099602033306 3.46405255764912
0.0417800911612275 3.07555225935644
0.0439502221191243 2.70682356156518
0.0461203530770212 2.35905565281003
0.048290484034918 2.05039878596176
0.0504606149928149 1.77600393855352
0.0526307459507117 1.50351110615308
0.0548008769086086 1.25577193477401
0.0569710078665054 1.03506652498856
0.0591411388244023 0.840930960230134
0.0613112697822991 0.676371471349141
0.063481400740196 0.53247880032197
0.0656515316980928 0.408971421461523
0.0678216626559897 0.310818520637568
0.0699917936138865 0.226611011888679
0.0721619245717834 0.159076604463346
0.0743320555296802 0.106530006634252
0.0765021864875771 0.0692124035174591
0.0786723174454739 0.0437532116583496
0.0808424484033708 0.0276719150946558
0.0830125793612676 0.0174146811821658
0.0851827103191645 0.0107052717437374
0.0873528412770613 0.00651414390852645
0.0895229722349582 0.00362378491096888
0.091693103192855 0.00184879978641096
0.0938632341507519 0.00100121864013387
0.0960333651086487 0.000548460103768153
0.0982034960665456 0.000256205541431599
0.100373627024442 0.000102997202585569
0.102543757982339 3.90531059803615e-05
0.104713888940236 1.33038053339691e-05
0.106884019898133 2.5749300646392e-06
};
\addlegendentry{$C=3$}
\addplot [very thick, mediumseagreen1]
table {%
-0.104943018694834 2.05274237906653e-05
-0.102856013580642 7.05072382375031e-05
-0.100769008466449 0.000181176827369785
-0.0986820033522571 0.000306126363486878
-0.0965949982380648 0.000501583137841474
-0.0945079931238725 0.000730954786285
-0.0924209880096801 0.00147931325795773
-0.0903339828954878 0.00225623162360008
-0.0882469777812955 0.00352223817347223
-0.0861599726671031 0.00521664313288864
-0.0840729675529108 0.00760407176941167
-0.0819859624387185 0.0101909734153788
-0.0798989573245261 0.0141108188627093
-0.0778119522103338 0.0210754167562074
-0.0757249470961414 0.0338595392943592
-0.0736379419819491 0.0470305591461306
-0.0715509368677568 0.0611034468996619
-0.0694639317535644 0.0768573521623106
-0.0673769266393721 0.0887172943815391
-0.0652899215251798 0.0996387763348603
-0.0632029164109874 0.11825536471962
-0.0611159112967951 0.143369775230812
-0.0590289061826028 0.180826076171971
-0.0569419010684104 0.239504163325932
-0.0548548959542181 0.342894103234363
-0.0527678908400258 0.519753477882961
-0.0506808857258334 0.79924149276636
-0.0485938806116411 1.23549967438079
-0.0465068754974488 1.91339151641526
-0.0444198703832564 2.80891510547836
-0.0423328652690641 3.78661934319073
-0.0402458601548718 4.85893224591251
-0.0381588550406794 6.02552314362141
-0.0360718499264871 7.17093866494525
-0.0339848448122948 7.89435587046394
-0.0318978396981024 8.14081169048141
-0.0298108345839101 8.38179427456861
-0.0277238294697177 8.73328044340446
-0.0256368243555254 9.03474217295902
-0.0235498192413331 9.42494707929909
-0.0214628141271407 10.0574742072529
-0.0193758090129484 10.8306922816328
-0.0172888038987561 11.702596838383
-0.0152017987845637 12.7124644254369
-0.0131147936703714 13.7025850770527
-0.0110277885561791 14.815395463175
-0.00894078344198673 15.837342546633
-0.00685377832779439 16.4886424498918
-0.00476677321360206 17.0811718925894
-0.00267976809940972 17.5475023038088
-0.000592762985217384 17.9098653297635
0.00149424212897495 18.15342589053
0.00358124724316728 18.2302783339604
0.00566825235735962 18.3389085681639
0.00775525747155196 18.2531298191228
0.00984226258574429 17.4312572878757
0.0119292676999366 16.848005794483
0.014016272814129 15.9127152305572
0.0161032779283213 14.3732936595054
0.0181902830425136 12.7193732422874
0.020277288156706 11.0826436499978
0.0223642932708983 9.58311221835126
0.0244512983850906 8.38310178221435
0.026538303499283 7.37612677692165
0.0286253086134753 6.60434115972365
0.0307123137276676 6.03749687916813
0.03279931884186 5.35358782447301
0.0348863239560523 4.60609507458623
0.0369733290702446 4.04807357002564
0.039060334184437 3.6220090393167
0.0411473392986293 3.21285508957608
0.0432343444128217 2.84276795104759
0.045321349527014 2.50136610227314
0.0474083546412063 2.1931575440165
0.0494953597553987 1.91576377260812
0.051582364869591 1.65125051338001
0.0536693699837833 1.39176156424882
0.0557563750979757 1.18682379759301
0.057843380212168 0.986808150163511
0.0599303853263603 0.813342940413855
0.0620173904405527 0.663139764302477
0.064104395554745 0.52095611716809
0.0661914006689373 0.403091665997173
0.0682784057831297 0.306880969435359
0.070365410897322 0.22821229524769
0.0724524160115144 0.160440113106125
0.0745394211257067 0.109429464986321
0.076626426239899 0.0735194145546106
0.0787134313540914 0.0475258948071666
0.0808004364682837 0.0306657399015378
0.082887441582476 0.0194885576475202
0.0849744466966684 0.0116430655243968
0.0870614518108607 0.00678252856944183
0.089148456925053 0.00339594989232526
0.0912354620392454 0.00185550061133885
0.0933224671534377 0.00117631063287379
0.09540947226763 0.000492211922632689
0.0974964773818224 0.000120933301027616
0.0995834824960147 1.56186920146366e-05
0.101670487610207 2.677490059652e-06
};
\addlegendentry{$C=4$}
\addplot [very thick, darkorange1]
table {%
-0.10423905833333 2.64796495569275e-05
-0.102199125394959 7.98954943527978e-05
-0.100159192456588 0.000200880100087036
-0.0981192595182167 0.00036751927402287
-0.0960793265798456 0.000541463178870965
-0.0940393936414746 0.00079986803489201
-0.0919994607031036 0.00159973606978403
-0.0899595277647326 0.00242608375423297
-0.0879195948263616 0.0037208473084303
-0.0858796618879905 0.00549954928556459
-0.0838397289496195 0.00775534150040563
-0.0817997960112485 0.0103786529892695
-0.0797598630728775 0.0143035762175609
-0.0777199301345064 0.0213508153651598
-0.0756799971961354 0.0341605741111643
-0.0736400642577644 0.0470543372626601
-0.0716001313193934 0.0608178199383962
-0.0695601983810224 0.0770283874697374
-0.0675202654426513 0.0889611219605893
-0.0654803325042803 0.0984759905194852
-0.0634403995659093 0.11591238320704
-0.0614004666275383 0.139973253740646
-0.0593605336891673 0.174168068777962
-0.0573206007507962 0.226948401984466
-0.0552806678124252 0.314077862668808
-0.0532407348740542 0.472594632375897
-0.0512008019356832 0.716255345596241
-0.0491608689973122 1.09301419113176
-0.0471209360589411 1.68656786268369
-0.0450810031205701 2.51500698290195
-0.0430410701821991 3.46715263421061
-0.0410011372438281 4.4643588877886
-0.0389612043054571 5.5891833558605
-0.036921271367086 6.74014095531664
-0.034881338428715 7.6916990323326
-0.032841405490344 8.08074854417361
-0.030801472551973 8.23712046226665
-0.028761539613602 8.56780426102737
-0.0267216066752309 8.88480082024136
-0.0246816737368599 9.1884146558783
-0.0226417407984889 9.66542225881663
-0.0206018078601179 10.3810238309962
-0.0185618749217469 11.1376049435935
-0.0165219419833758 12.0946753531742
-0.0144820090450048 13.0482025117165
-0.0124420761066338 14.0970979543211
-0.0104021431682628 15.1453213616815
-0.00836221022989175 16.022555627848
-0.00632227729152074 16.5771216677929
-0.00428234435314972 17.2192412993605
-0.0022424114147787 17.6477276837599
-0.000202478476407676 17.9530846108107
0.00183745446196334 18.1908823643828
0.00387738740033437 18.1471735938778
0.00591732033870539 18.2624454434931
0.0079572532770764 18.1666023519218
0.00999718621544742 17.3169429884033
0.0120371191538184 16.7778031705082
0.0140770520921895 15.8210587345445
0.0161169850305605 14.3610658194775
0.0181569179689315 12.7471088082443
0.0201968509073025 11.1328641732314
0.0222367838456736 9.70746555067348
0.0242767167840446 8.5517836164998
0.0263166497224156 7.52165037524347
0.0283565826607866 6.69639885697705
0.0303965155991576 6.14408906579267
0.0324364485375287 5.4978313041632
0.0344763814758997 4.74348315649659
0.0365163144142707 4.15794734021221
0.0385562473526417 3.72947830454879
0.0405961802910127 3.32157528222324
0.0426361132293838 2.94204383476084
0.0446760461677548 2.60780674095694
0.0467159791061258 2.29536102910134
0.0487559120444968 2.00955749706543
0.0507958449828678 1.7598731201166
0.0528357779212388 1.484041915413
0.0548757108596099 1.27800467530191
0.0569156437979809 1.07724373363786
0.0589555767363519 0.895771847039023
0.0609955096747229 0.738224780360535
0.0630354426130939 0.594726080863361
0.065075375551465 0.46439689810962
0.067115308489836 0.359463068918013
0.069155241428207 0.273617871237188
0.071195174366578 0.200650457608982
0.073235107304949 0.139162428609386
0.0752750402433201 0.0956864964023675
0.0773149731816911 0.0637438212144358
0.0793549061200621 0.0415333303300407
0.0813948390584331 0.0269334559648514
0.0834347719968041 0.0168145774686489
0.0854747049351752 0.00972853193807861
0.0875146378735462 0.00563970880994353
0.0895545708119172 0.00291276145126202
0.0915945037502882 0.00161799789706467
0.0936344366886592 0.00108475254046999
0.0956743696270303 0.000310907609452885
0.0977143025654013 2.00880100087037e-05
};
\addlegendentry{$C=5$}
\end{axis}

% \draw ({$(current bounding box.south west)!0.5!(current bounding box.south east)$}|-{$(current bounding box.south west)!-0.2!(current bounding box.north west)$}) node[
%   scale=0.6,
%   fill=white,
%   draw=black,
%   line width=0.4pt,
%   inner sep=3.3pt,
%   fill opacity=0.5,
%   anchor=base,
%   text=black,
%   rotate=0.0
% ]{DNS Normal Fit: $\mu = 0.0000$, $\sigma = 0.0226$};
\end{tikzpicture}